\documentclass[twocolumn]{aastex62}

\newcommand{\Lsun}{$L_{\odot}$}
\newcommand{\Msun}{$M_{\odot}$}

\newcommand{\Mdot}{$\dot{M}$}
\newcommand{\Av}{A$_V$}

\newcommand{\Teff}{$T_{\rm eff}$}
\newcommand{\mic}{$\mu$m}

\newcommand{\rstar}{$R_{*}$}
\newcommand{\mstar}{$M_{*}$}

\newcommand{\brgamma}{Br$\gamma$}
\newcommand{\Lacc}{L$_{acc}$}
\newcommand{\LaccL}{L$_{acc}$/L$_{*}$}
\newcommand{\Lbrg}{L$_{Br\gamma}$}
\newcommand{\Fbrg}{F$_{Br\gamma}$}
\newcommand{\angstrom}{\mbox{\normalfont\AA}}
\newcommand{\Lstar}{L$_{*}$}
\newcommand{\LNIR}{L$_{NIR}$}
\newcommand{\LIR}{L$_{IR}$}

\usepackage{placeins}
\usepackage{amsmath}

\graphicspath{{./}{figures/}}

\shorttitle{Accretion onto Herbig Ae/Be Stars}
\shortauthors{Grant et al.}

\begin{document}

\title{Tracing Accretion onto Herbig Ae/Be Stars Using the \brgamma\ Line}

\correspondingauthor{Sierra Grant}
\email{sierrag@bu.edu}

\author[0000-0002-4022-4899]{Sierra L. Grant}
\affil{Department of Astronomy and Institute for Astrophysical Research, Boston University, 725 Commonwealth Avenue, Boston, MA, USA}

\author[0000-0001-9227-5949]{Catherine C. Espaillat}
\affil{Department of Astronomy and Institute for Astrophysical Research, Boston University, 725 Commonwealth Avenue, Boston, MA, USA}

\author[0000-0001-5638-1330]{Sean Brittain}
\affil{Clemson University, 118 Kinard Laboratory, Clemson, SC, USA}

\author{Caleb Scott-Joseph}
\affil{Department of Astronomy and Institute for Astrophysical Research, Boston University, 725 Commonwealth Avenue, Boston, MA, USA}

\author[0000-0002-3950-5386]{Nuria Calvet}
\affil{Department of Astronomy, University of Michigan, Ann Arbor, MI 48109, USA}

\begin{abstract}
Accretion plays an important role in protoplanetary disk evolution, and it is thought that the accretion mechanism changes between low- and high-mass stars. Here, we characterize accretion in intermediate-mass, pre-main-sequence Herbig Ae/Be (HAeBe) stars to search for correlations between accretion and system properties. We present new high-resolution, near-infrared spectra from the Immersion GRating INfrared Spectrograph for 102 HAeBes and analyze the accretion-tracing Br$\gamma$ line at 2.166 $\mu$m. We also include the samples of Fairlamb et al. and Donehew \& Brittain, for a total of 155 targets. We find a positive correlation between the Br$\gamma$ and stellar luminosity, with a change in the slope between the Herbig Aes and Herbig Bes. We use $L_{Br\gamma}$ to determine the accretion luminosity and accretion rate. We find that the accretion luminosity and rate depend on stellar mass and age; however, the trend disappears when normalizing the accretion luminosity by the stellar luminosity. We classify the objects into flared (Group I) or flat (Group II) disks and find that there is no trend with accretion luminosity or rate, indicating that the disk dust structure is not impacting accretion. We test for Br$\gamma$ variability in objects that are common to our sample and previous studies. We find that the Br$\gamma$ line equivalent width is largely consistent between the literature observations and those that we present here, except in a few cases where we may be seeing changes in the accretion rate. 
\end{abstract}


\section{Introduction} \label{sec:intro}

Accretion of material onto a star is an important mechanism in protoplanetary disk dispersal and heating. Understanding how this accretion takes place and under what conditions has important implications for disk evolution. The accretion paradigm is fairly well-understood for low-mass T Tauri stars (e.g., \citealt{hartmann16}); however, the pre-main-sequence evolution of high-mass objects proceeds very quickly, such that the system is still visibly obscured by the envelope until it reaches the main sequence (e.g., \citealt{zinnecker_yorke07}). Thus, while the picture of accretion in low-mass systems is relatively clear, we do not have such an understanding for high-mass systems. Intermediate-mass, pre-main-sequence stars known as Herbig Ae/Be stars (HAeBes, with masses from 2 to tens of solar masses; \citealt{herbig60}) provide the link between the two mass regimes \citep{mendigutia20}.

The classical magnetospheric accretion paradigm for low-mass T Tauri stars is as follows: The disk is truncated by the stellar magnetic field and the ionized material travels from the disk along the stellar magnetic field lines, falls onto the star, and creates a shock on the stellar surface (e.g., \citealt{calvet&gullbring98, hartmann16}). This paradigm holds as long as the stellar magnetic field is strong enough to truncate the disk. However, as the stellar mass increases, the star will change from having a convective envelope to possessing a radiative envelope, decreasing the dynamo process that generates magnetic fields \citep{gregory12}. \citet{villebrun19} finds that the detection of magnetic fields in intermediate-mass, pre-main-sequence stars drops quickly as the stars evolve from having radiative cores with convective envelope to being fully radiative. Thus, the classical accretion paradigm may break down at certain masses; currently, that break is thought to happen somewhere between $\sim$2 and 8 \Msun\ (e.g., \citealt{vink02,donehew-brittain11,cauley15}). It has been proposed that stars with weak magnetic fields accrete via boundary layer accretion (e.g., \citealt{lynden-bell&pringle74}). Thus, HAeBe objects are important diagnostics in understanding the accretion paradigm and evolution.

There are several observational tracers of accretion. The shock that forms on the stellar surface as material falls onto the star produces ultraviolet continuum that can be used to measure the amount of infalling material (e.g., \citealt{calvet&gullbring98,robinson19}). Spectroscopic tracers are generated in the accretion flows as material travels from the disk to the star; these include the H$\alpha$, He I 1.0830 \mic, and \brgamma\ lines (e.g., \citealt{donehew-brittain11, cauley14, reiter18}).  Spectroscopic line tracers have the additional benefit of allowing for kinematic characterization of the gas closest to the star based on the line shape, which can give insight to the geometry of the accretion region. For instance, redshifted absorption can indicate material falling onto the star, away from the observer, while blueshifted absorption can indicate a wind propagating outward from the system (e.g., \citealt{edwards87,muzerolle01}).

Here, we trace accretion using the \brgamma\ line at 2.166 \mic, which was shown to correlate with the accretion luminosity, \Lacc, for low-mass pre-main-sequence stars (i.e., T Tauri stars) by \citet{muzerolle98}. The relationship between \Lbrg\ and \Lacc\ was extended to slightly higher mass ($\sim1.5$--4 \Msun) objects by \citet{calvet04} and even higher (up to 34.5 \Msun) by \citet{garcialopez06}, \citet{donehew-brittain11} and \citet{fairlamb17}. Previous surveys find that there is a difference in accretion mechanism between HAe stars and HBe stars, with the former appearing more similar to T Tauri stars than their higher-mass HBe counterparts (e.g., \citealt{vink02, donehew-brittain11, mendigutia11b, cauley14, wichittanakom20}). For instance, \citet{fairlamb15} found that observations of several early HBes cannot be explained by magnetospheric accretion and an alternative mechanism is boundary layer accretion. 

In this work, we present new high-resolution, NIR observations of 102 HAeBes along with analysis of another 53 targets from \citet{donehew-brittain11} and \citet{fairlamb15,fairlamb17}. We investigate accretion signatures in this large sample, testing for correlations with stellar mass, spectral energy distribution (SED) shape, and age. We also search for variability in 56 objects that overlap between our sample and those of \citet{fairlamb15,fairlamb17} and/or \citet{donehew-brittain11}.

In Section~\ref{sec: observations}, we describe the observations and data reduction for our new sample of 102 objects and the sample that we utilize from \citet{donehew-brittain11} and \citet{fairlamb15,fairlamb17} to expand the sample (an additional 53 objects) and to test for variability (56 objects). Our analysis determining the \brgamma\ equivalent widths and accretion properties is described in Section~\ref{sec: analysis}. We test for correlations between the accretion and system properties and present the results in Section~\ref{sec: results}. We then discuss the results in Section~\ref{sec: discussion} and offer a summary and conclusions in Section~\ref{sec: summary and conclusions}.  

\section{Observations and Data Reduction} \label{sec: observations}
\subsection{IGRINS Sample}\label{subsec: igrins samp}
We present observations from the Immersion GRating INfrared Spectrograph (IGRINS, \citealt{park14}) with a resolution of R$\,\sim45,000$ in the  H and K bands (1.496--1.780 \mic\ and 2.080--2.460 \mic, respectively). Observations for 65 of the objects were obtained using the Lowell Discovery Telescope (LDT) in Happy Jack, AZ between 2017 November and 2018 December. The other 37 objects were observed using the Gemini South Observatory between 2020 February and December. The targets, observing dates, exposure times, and the continuum SNR from 2.159 to 2.162 \mic\ are listed in Table~\ref{tab: obs log}. 

We reduced the spectra using the standard IGRINS pipeline \citep{igrinsplp}. This publicly available pipeline performs automatic dark and sky subtraction and flat fielding using nodding with the number of AB pairs needed to achieve the desired signal. A one-dimensional spectrum is returned. Then we removed telluric absorption features using the \texttt{xtellcor\_general} Spextool reduction package \citep{vacca03} and A0V standard stars that were taken throughout each observing night.

\begin{figure*}
    \centering
    \includegraphics[scale=0.42]{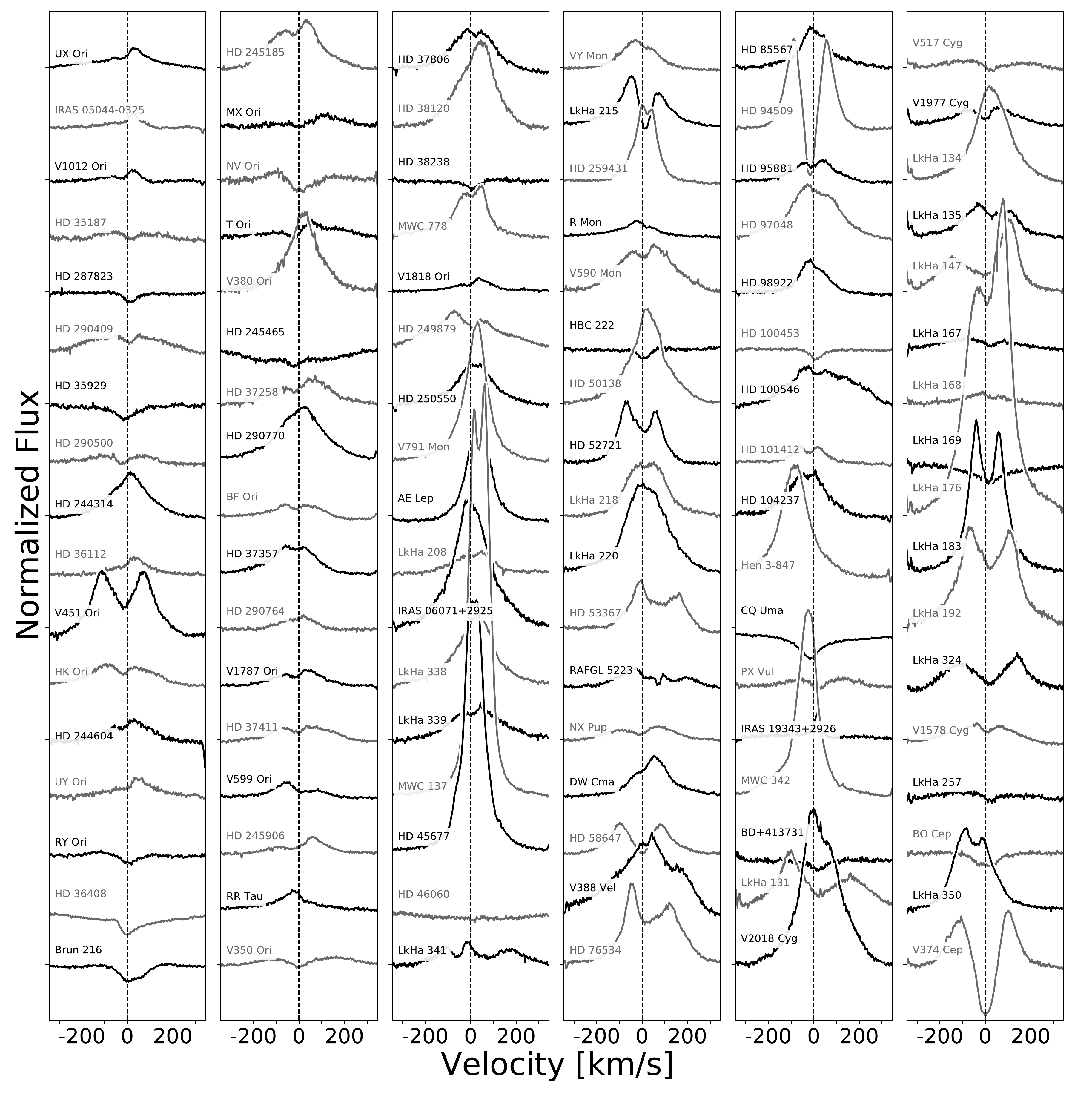}
    \caption{Normalized \brgamma\ line profiles for the IGRINS sample of 102 Herbig Ae/Bes presented here. The data has been corrected for the barycentric radial velocity at the time of the observations. The profiles have alternating colors and are vertically offset for clarity.}
    \label{fig: All}
\end{figure*}

\begin{figure*}
    \centering
    \includegraphics[scale=0.32]{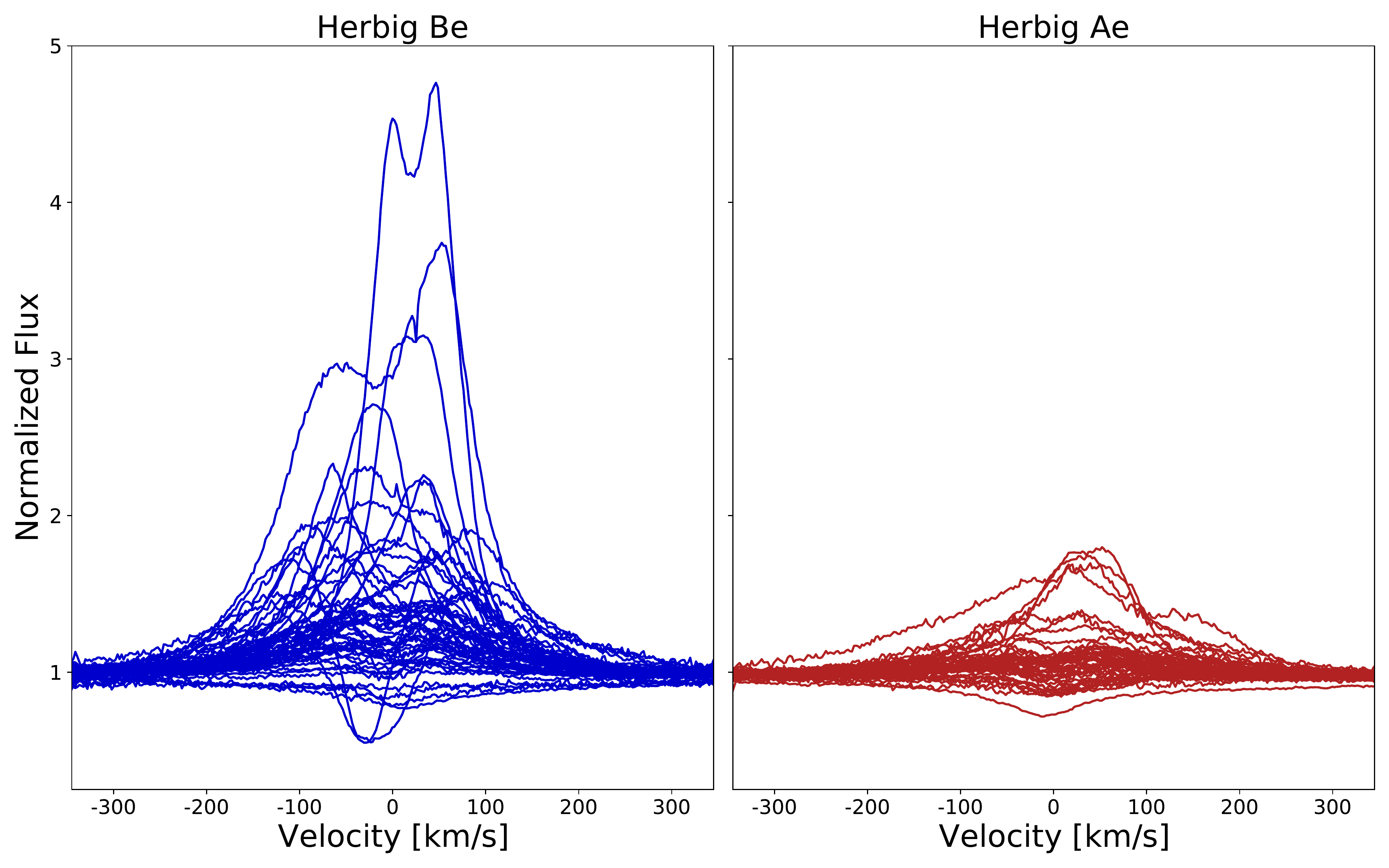}
    \caption{\brgamma\ emission for Herbig Be stars (left, blue) and Herbig Ae stars (right, red) in the IGRINS sample with the local continuum near the line normalized to 1. The data has been corrected for the barycentric radial velocity at the time of the observations.}
    \label{fig: BrBvsA}
\end{figure*}

We show \brgamma\ line profiles from IGRINS for the whole sample in Figure~\ref{fig: All} and separated into HBes and HAes in Figure~\ref{fig: BrBvsA} (the adopted effective temperature, \Teff, for each object is listed in Table~\ref{tab: stellar properties}). We use a \Teff\ cutoff of 10,000 K to distinguish between hotter HBes and cooler HAes. Figure~\ref{fig: BrBvsA} shows that the HBes exhibit higher \brgamma\ emission than the HAes. 

A majority of the targets observed with IGRINS are included in the analysis of \citet{vioque18} (66 out of our 102 sources, with 59 in their ``High-Quality'' sample and seven in their ``Low-Quality'' sample). \citet{vioque18} provides stellar properties derived for these objects homogeneously and thus we primarily adopt these values for our analysis, including effective temperatures, visual extinctions (\Av), distances, and ages. For objects that are not included in \citet{vioque18}, we collect stellar properties from the literature. All distances are either from \citet{vioque18} or \citet{bailer-jones18} using data from the Gaia Data Release 2 \citep{gaia16,gaia18}, with the exception of R Mon, HK Ori, and HD 52721 which do not have Gaia DR2 data and we adopt the distances from \citet{fairlamb15}. The stellar properties and references are listed in Table~\ref{tab: stellar properties}.

While the objects presented here have been included in previous Herbig Ae/Be studies, the evolutionary state of these objects can be difficult to determine. In the Appendix, we note objects that have conflicting classifications in the literature. While we treat them as HAeBes in this work, it is possible that some objects are misclassified. 

\subsection{Auxiliary Sample}\label{subsec: aux samp}
\begin{figure}
    \centering
    \includegraphics[scale=0.55]{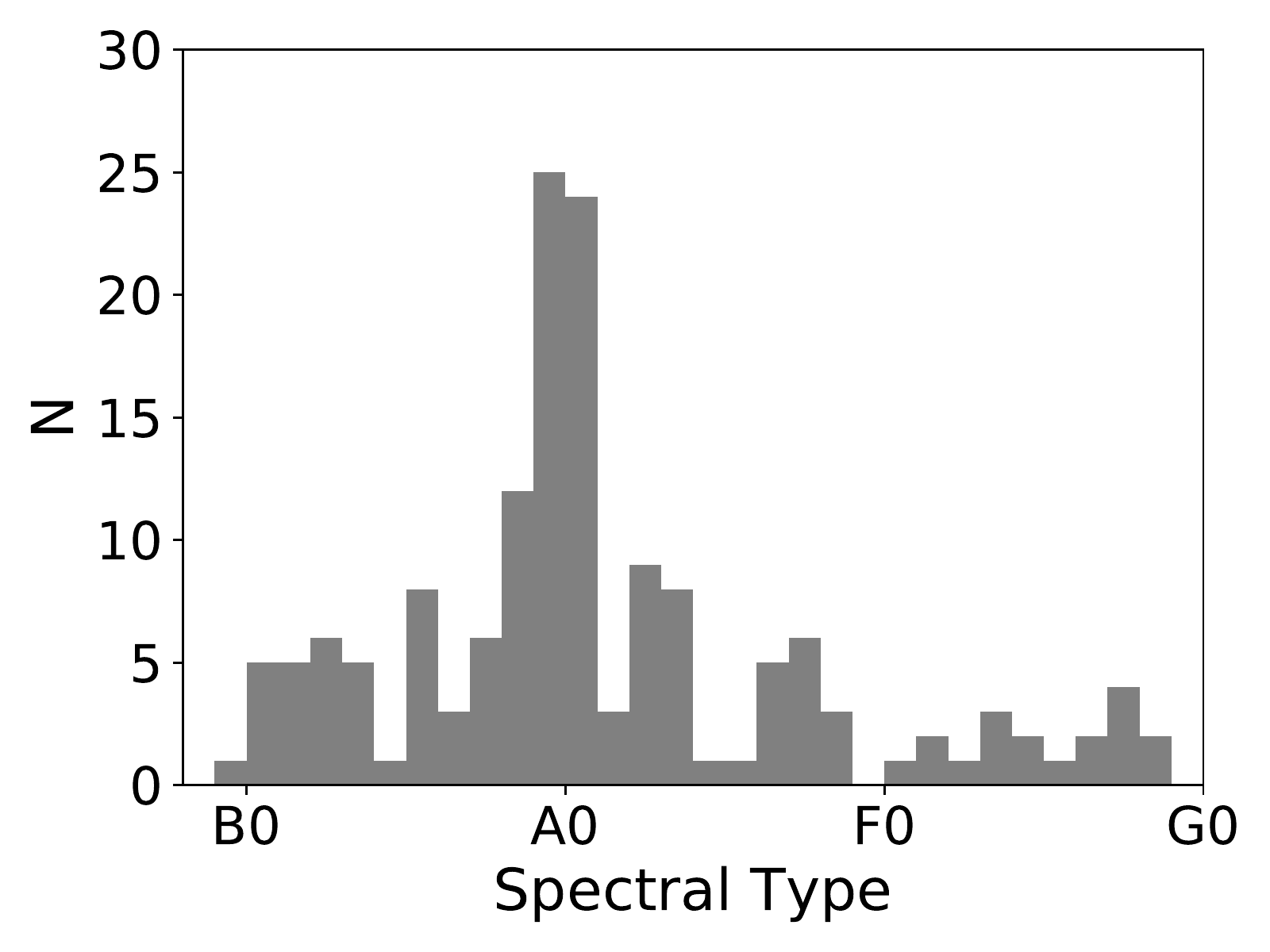}
    \caption{The distribution of spectral types for our IGRINS Sample and the Auxiliary Sample from \citet{donehew-brittain11} and \citet{fairlamb15}. The bin widths correspond to one subclass.} 
    \label{fig: spt hist}
\end{figure}

The works of \citet{fairlamb15,fairlamb17} and \citet{donehew-brittain11} also provide analysis of the \brgamma\ line for a large number of HAeBes. \citet{fairlamb15} presented 91 HAeBes observed between 2009 October and 2010 April with X-Shooter on the VLT. The first paper in the series \citep{fairlamb15} presented the stellar properties along with the mass accretion rates determined via UV continuum excess. The subsequent paper \citep{fairlamb17} provided spectral line analysis for the sample, including data for the \brgamma\ line. In general, we refer to the first paper when discussing stellar parameters taken from that sample and the sample as a whole. We refer to the second paper when discussing \brgamma-specific properties. The \citet{donehew-brittain11} sample was observed between 2006 November and 2008 May with Flamingos on the 2.1 m telescope at Kitt Peak National Observatory. We analyze 45 objects from \citet{fairlamb15,fairlamb17} and eight objects from \citet{donehew-brittain11}. We call these 53 targets the Auxiliary Sample. Table~\ref{tab: Aux Sample} contains stellar properties for the Auxiliary Sample collected from \citet{fairlamb15} and \citet{donehew-brittain11}. For our sample objects, as well as the \citet{donehew-brittain11} and \citet{fairlamb17} samples, the spectral type distribution is shown in Figure~\ref{fig: spt hist}. We note that we calculate effective temperatures from spectral types using conversions in \citet{pecaut-mamajek13}.

\subsection{Variability Sample}
Due to the variability seen in young stars, we checked our overlap sample for changes in the \brgamma\ equivalent width between observations. The \citet{fairlamb15} sample overlaps with 46 of the IGRINS objects, the \citet{donehew-brittain11} sample overlaps with 20, and 10 of our objects overlap with both, totaling 56 that have both IGRINS observations and previous measurements. We call this the Variability Sample. We note which objects in our IGRINS sample are part of the Variability Sample in Table~\ref{tab: results}. We present the variability results in Section~\ref{subsect: variability results} and discuss these results in Section~\ref{subsec: line shape}.

\startlongtable
\begin{deluxetable*}{ccccccc}
\tablewidth{0pt}
\tablecaption{Observing Log for IGRINS Sample\label{tab: obs log}}
\tablehead{\colhead{Num.} & \colhead{Target} & \colhead{RA} & \colhead{Dec} & \colhead{UT Date} & \colhead{Exposure Time} & \colhead{SNR} \\
\colhead{} & \colhead{} & \colhead{(J2000)} & \colhead{(J2000)} & \colhead{} & \colhead{(s)} & \colhead{}}
\startdata
1 & UX Ori & 05:04:29.99 & -03:47:14.29 & 2020/02/06 & 4$\times$35 & 102   \\ 
2 & IRAS 05044-0325 & 05:06:55.52 & -03:21:13.323 & 2020/02/06 & 4$\times$85 & 77   \\ 
3 & V1012 Ori & 05:11:36.55 & -02:22:48.45 & 2020/02/06 & 4$\times$110 & 80   \\ 
4 & HD 35187 & 05:24:01.17 & +24:57:37.58 & 2018/11/14 & 6$\times$45 & 44   \\ 
5 & HD 287823 & 05:24:08.05 & +02:27:46.88 & 2018/11/18 & 2$\times$275 & 72   \\ 
6 & HD 290409 & 05:27:05.47 & +00:25:07.62 & 2018/11/19 & 2$\times$325 & 46   \\ 
7 & HD 35929 & 05:27:42.79 & -08:19:38.45 & 2020/02/07 & 6$\times$4 & 49   \\ 
8 & HD 290500 & 05:29:48.05 & -00:23:43.51 & 2018/11/19 & 4$\times$450 & 52   \\ 
9 & HD 244314 & 05:30:19.02 & +11:20:20.00 & 2018/11/17 & 6$\times$250 & 74   \\ 
10 & HD 36112 & 05:30:27.53 & +25:19:57.08 & 2017/11/16 & 4$\times$35 & 57   \\ 
11 & V451 Ori & 05:31:26.77 & +11:01:22.55 & 2018/11/17 & 4$\times$300 & 46   \\ 
12 & HK Ori & 05:31:28.05 & +12:09:10.15 & 2018/11/16 & 2$\times$175 & 66   \\ 
13 & HD 244604 & 05:31:57.25 & +11:17:41.37 & 2017/11/16 & 4$\times$45 & 46   \\ 
14 & UY Ori & 05:32:00.31 & -04:55:53.91 & 2020/02/06 & 4$\times$160 & 44   \\ 
15 & RY Ori & 05:32:09.94 & -02:49:46.77 & 2020/02/08 & 4$\times$50 & 58   \\ 
16 & HD 36408 & 05:32:14.14 & +17:03:29.25 & 2018/11/16 & 2$\times$80 & 88   \\ 
17 & Brun 216 & 05:34:14.16 & -05:36:54.19 & 2020/02/07 & 6$\times$30 & 78   \\ 
18 & HD 245185 & 05:35:09.60 & +10:01:51.47 & 2018/11/16 & 2$\times$275 & 56   \\ 
19 & MX Ori & 05:35:21.26 & -05:09:16.21 & 2020/02/08 & 4$\times$6 & 40   \\ 
20 & NV Ori & 05:35:31.37 & -05:33:08.87 & 2020/12/29 & 8$\times$10 & 39   \\ 
21 & T Ori & 05:35:50.45 & -05:28:34.92 & 2018/01/12 & 2$\times$35 & 47   \\ 
22 & V380 Ori & 05:36:25.43 & -06:42:57.68 & 2020/12/29 & 4$\times$5 & 40   \\ 
23 & HD 245465 & 05:36:29.35 & +06:50:02.17 & 2018/11/17 & 4$\times$300 & 49   \\ 
24 & HD 37258 & 05:36:59.25 & -06:09:16.32 & 2020/12/29 & 12$\times$10 & 34   \\ 
25 & HD 290770 & 05:37:02.45 & -01:37:21.36 & 2020/12/29 & 12$\times$5 & 72   \\ 
26 & BF Ori & 05:37:13.26 & -06:35:00.57 & 2020/12/29 & 8$\times$15 & 82   \\ 
27 & HD 37357 & 05:37:47.08 & -06:42:30.20 & 2020/12/29 & 4$\times$10 & 63   \\ 
28 & HD 290764 & 05:38:05.25 & -01:15:21.70 & 2018/11/18 & 2$\times$150 & 60   \\ 
29 & V1787 Ori & 05:38:09.30 & -06:49:16.60 & 2020/12/30 & 4$\times$30 & 84   \\ 
30 & HD 37411 & 05:38:14.51 & -05:25:13.32 & 2020/12/30 & 4$\times$25 & 82   \\ 
31 & V599 Ori & 05:38:58.64 & -07:16:45.63 & 2020/12/29 & 8$\times$10 & 80   \\ 
32 & HD 245906 & 05:39:30.48 & +26:19:55.16 & 2018/11/16 & 2$\times$450 & 79   \\ 
33 & RR Tau & 05:39:30.51 & +26:22:26.96 & 2018/11/14 & 4$\times$130 & 69   \\ 
34 & V350 Ori & 05:40:11.76 & -09:42:11.09 & 2020/12/28 & 4$\times$40 & 75   \\ 
35 & HD 37806 & 05:41:02.29 & -02:43:00.73 & 2017/11/16 & 4$\times$25 & 39   \\ 
36 & HD 38120 & 05:43:11.89 & -04:59:49.88 & 2018/01/05 & 2$\times$60 & 36   \\ 
37 & HD 38238 & 05:44:18.80 & +00:08:40.41 & 2018/11/18 & 2$\times$60 & 54   \\ 
38 & MWC 778 & 05:50:13.90 & +23:52:17.70 & 2018/11/15 & 4$\times$225 & 63   \\ 
39 & V1818 Ori & 05:53:42.56 & -10:24:00.72 & 2020/12/30 & 4$\times$10 & 94   \\ 
40 & HD 249879 & 05:58:55.78 & +16:39:57.33 & 2018/11/17 & 4$\times$275 & 63   \\ 
41 & HD 250550 & 06:01:59.99 & +16:30:56.72 & 2017/11/16 & 4$\times$40 & 56   \\ 
42 & V791 Mon & 06:02:14.89 & -10:00:59.51 & 2020/12/29 & 6$\times$10 & 69   \\ 
43 & AE Lep & 06:03:37.07 & -14:53:03.17 & 2020/12/31 & 8$\times$90 & 76   \\ 
44 & LkHa 208 & 06:07:49.53 & +18:39:26.49 & 2018/11/16 & 4$\times$300 & 60   \\ 
45 & IRAS 06071+2925 & 06:10:17.33 & +29:25:16.61 & 2018/11/15 & 8$\times$400 & 44   \\ 
46 & LkHa 338 & 06:10:47.13 & -06:12:50.61 & 2018/11/18 & 4$\times$250 & 66   \\ 
47 & LkHa 339 & 06:10:57.84 & -06:14:39.66 & 2018/11/18 & 4$\times$450 & 48   \\ 
48 & MWC 137 & 06:18:45.52 & +15:16:52.24 & 2018/11/13 & 4$\times$175 & 102   \\ 
49 & HD 45677 & 06:28:17.42 & -13:03:11.13 & 2020/12/29 & 4$\times$2 & 79   \\ 
50 & HD 46060 & 06:30:49.81 & -09:39:14.90 & 2018/11/18 & 4$\times$220 & 53   \\ 
51 & LkHa 341 & 06:30:50.14 & +10:33:09.81 & 2018/11/14 & 2$\times$220 & 48   \\ 
52 & VY Mon & 06:31:06.92 & +10:26:04.98 & 2018/11/13 & 2$\times$90 & 73   \\ 
53 & LkHa 215 & 06:32:41.77 & +10:09:34.21 & 2018/11/15 & 2$\times$220 & 83   \\ 
54 & HD 259431 & 06:33:05.19 & +10:19:19.99 & 2018/11/13 & 2$\times$100 & 74   \\ 
55 & R Mon & 06:39:09.95 & +08:44:09.56 & 2018/11/14 & 4$\times$85 & 89   \\ 
56 & V590 Mon & 06:40:44.64 & +09:48:02.15 & 2018/11/14 & 8$\times$320 & 50   \\ 
57 & HBC 222 & 06:40:51.19 & +09:44:46.11 & 2018/11/16 & 4$\times$350 & 52   \\ 
58 & HD 50138 & 06:51:33.40 & -06:57:59.45 & 2020/12/29 & 4$\times$2 & 84   \\ 
59 & HD 52721 & 07:01:49.51 & -11:18:03.32 & 2018/11/19 & 2$\times$35 & 55   \\ 
60 & LkHa 218 & 07:02:42.53 & -11:26:11.83 & 2018/11/19 & 2$\times$325 & 54   \\ 
61 & LkHa 220 & 07:04:06.70 & -11:26:08.50 & 2018/11/19 & 4$\times$450 & 70   \\ 
62 & HD 53367 & 07:04:25.52 & -10:27:15.61 & 2018/11/19 & 2$\times$50 & 72   \\ 
63 & RAFGL 5223 & 07:08:38.79 & -04:19:04.84 & 2020/12/28 & 6$\times$80 & 67   \\ 
64 & NX Pup & 07:19:28.29 & -44:35:11.24 & 2020/12/28 & 4$\times$6 & 81   \\ 
65 & DW Cma & 07:19:35.94 & -17:39:17.93 & 2018/11/19 & 4$\times$70 & 88   \\ 
66 & HD 58647 & 07:25:56.10 & -14:10:43.55 & 2018/11/19 & 2$\times$25 & 69   \\ 
67 & V388 Vel & 08:42:16.59 & -40:44:09.96 & 2020/02/07 & 4$\times$280 & 41   \\ 
68 & HD 76534 & 08:55:08.71 & -43:27:59.89 & 2020/12/29 & 6$\times$15 & 62   \\ 
69 & HD 85567 & 09:50:28.54 & -60:58:02.95 & 2020/02/13 & 4$\times$3 & 57   \\ 
70 & HD 94509 & 10:53:27.26 & -58:25:24.52 & 2020/02/07 & 4$\times$100 & 75   \\ 
71 & HD 95881 & 11:01:57.62 & -71:30:48.31 & 2020/02/07 & 4$\times$6 & 77   \\ 
72 & HD 97048 & 11:08:03.31 & -77:39:17.49 & 2020/02/07 & 4$\times$6 & 66   \\ 
73 & HD 98922 & 11:22:31.67 & -53:22:11.46 & 2020/02/07 & 4$\times$2 & 76   \\ 
74 & HD 100453 & 11:33:05.58 & -54:19:28.54 & 2020/02/07 & 4$\times$6 & 75   \\ 
75 & HD 100546 & 11:33:25.44 & -70:11:41.24 & 2020/02/07 & 4$\times$2 & 53   \\ 
76 & HD 101412 & 11:39:44.46 & -60:10:27.72 & 2020/02/07 & 4$\times$15 & 65   \\ 
77 & HD 104237 & 12:00:05.09 & -78:11:34.56 & 2020/02/13 & 4$\times$2 & 38   \\ 
78 & Hen 3-847 & 13:01:17.80 & -48:53:18.78 & 2020/02/06 & 6$\times$15 & 59   \\ 
79 & CQ Uma & 13:40:21.38 & +57:12:27.38 & 2018/12/20 & 4$\times$90 & 84   \\ 
80 & PX Vul & 19:26:40.25 & +23:53:50.78 & 2018/09/27 & 4$\times$80 & 50   \\ 
81 & IRAS 19343+2926 & 19:36:18.93 & +29:32:49.84 & 2018/09/29 & 4$\times$30 & 77   \\ 
82 & MWC 342 & 20:23:03.62 & +39:29:49.92 & 2018/09/29 & 4$\times$25 & 100   \\ 
83 & BD +41 3731 & 20:24:15.72 & +42:18:01.38 & 2018/09/27 & 4$\times$350 & 41   \\ 
84 & LkHa 131 & 20:46:36.84 & +43:44:35.29 & 2018/09/29 & 4$\times$650 & 47   \\ 
85 & V2018 Cyg & 20:46:45.65 & +43:45:11.50 & 2018/09/27 & 4$\times$400 & 43   \\ 
86 & V517 Cyg & 20:47:23.60 & +43:44:39.74 & 2018/09/28 & 4$\times$270 & 45   \\ 
87 & V1977 Cyg & 20:47:37.47 & +43:47:24.97 & 2018/09/29 & 4$\times$30 & 66   \\ 
88 & LkHa 134 & 20:48:04.78 & +43:47:25.85 & 2018/09/29 & 4$\times$100 & 70   \\ 
89 & LkHa 135 & 20:48:20.34 & +43:39:48.29 & 2018/09/29 & 4$\times$30 & 63   \\ 
90 & LkHa 147 & 20:51:02.72 & +43:49:31.90 & 2018/09/29 & 4$\times$350 & 52   \\ 
91 & LkHa 167 & 20:52:04.64 & +44:37:30.43 & 2018/09/30 & 4$\times$100 & 60   \\ 
92 & LkHa 168 & 20:52:06.05 & +44:17:16.03 & 2018/09/30 & 4$\times$120 & 53   \\ 
93 & LkHa 169 & 20:52:07.65 & +44:03:44.43 & 2018/09/30 & 4$\times$150 & 48   \\ 
94 & LkHa 176 & 20:52:58.83 & +44:15:03.79 & 2018/09/29 & 4$\times$350 & 47   \\ 
95 & LkHa 183 & 20:55:10.30 & +45:03:02.63 & 2018/09/29 & 4$\times$350 & 50   \\ 
96 & LkHa 192 & 20:59:17.22 & +44:17:46.40 & 2018/09/30 & 4$\times$200 & 32   \\ 
97 & LkHa 324 & 21:03:54.23 & +50:15:09.99 & 2018/09/30 & 4$\times$300 & 40   \\ 
98 & V1578 Cyg & 21:52:34.10 & +47:13:43.60 & 2018/09/26 & 4$\times$30 & 61   \\ 
99 & LkHa 257 & 21:54:18.78 & +47:12:09.67 & 2018/09/26 & 4$\times$350 & 41   \\ 
100 & BO Cep & 22:16:54.06 & +70:03:44.98 & 2018/09/26 & 4$\times$400 & 52   \\ 
101 & LkHa 350 & 22:49:07.21 & +62:11:08.37 & 2018/09/26 & 4$\times$160 & 60   \\ 
102 & V374 Cep & 23:05:07.46 & +62:15:36.50 & 2018/09/26 & 4$\times$50 & 50   \\ 
\enddata
\end{deluxetable*}

\section{Analysis}\label{sec: analysis}

\subsection{Br$\gamma$ Analysis}\label{subsect: brg analysis}

We aim to study if and how the accretion paradigm changes for systems with different stellar and disk properties. Therefore, we derive the \brgamma\ equivalent width, EW(\brgamma); the line flux, \Fbrg; the line luminosity, \Lbrg; and the accretion luminosity based on \brgamma, \Lacc, for the 102 targets in the IGRINS Sample. We correct the EW(\brgamma) for photospheric absorption that can occur for A and B spectral types and for veiling of lines due to the hot disk material in order to determine the \brgamma\ emission coming from accretion only. We follow the methods of \citet{garcialopez06}, \citet{donehew-brittain11}, and \citet{fairlamb17}, which we outline below. For objects with no known \Av\ values in the literature, we calculate \Av\ using the methods of \citet{cardelli89}, 
\begin{equation}
    A_V = \frac{E(B-V)}{(-0.0018+1.0495/R_V)},
\end{equation}
where $E(B-V)$ is the color excess relative to the colors in \citet{k&h95} and $R_{V}$ is the ratio of total to selective extinction. Based on the findings of \citet{hernandez04}, we use $R_V=5.0$, which yields the best agreement for extinction calculated from different colors for HAeBe objects. This process leads to an unphysical negative \Av\ for V451 Ori and CQ Uma. We list these values in Table~\ref{tab: stellar properties} and adopt an extinction of zero for the analysis. The literature value for HD 245465 is also negative as listed in Table~\ref{tab: stellar properties}. We also make that zero for the analysis. In the Auxiliary Sample, one object (CQ Tau) has a negative \Av\ from the literature, that is listed in Table~\ref{tab: Aux Sample} and made zero for the analysis. All target photometry is collected from Vizier \citep{vizier}.

The process is as follows.

1. We flux-calibrate the IGRINS spectra by scaling the continuum, convolved with the K-band bandpass, to the literature flux at K-band, after correcting for interstellar reddening using the \citet{mathis90} and \citet{mcclure09} extinction curves. Objects with $A_{Ks}<0.3$ are dereddened using the \citet{mathis90} extinction law with $R_V=5.0$, and those with $A_{Ks}\geq0.3$ are dereddened using the \citet{mcclure09} law.

2. To remove the photospheric component from EW(\brgamma), we use the closest corresponding effective temperature, \Teff, and surface gravity, $\log(g)$, in the BT Settl photosphere model grids \citep{allard12}. We determine $\log(g)$ from the masses and radii collected from the literature. If an object lacks mass and/or radius measurements, we follow \citet{donehew-brittain11} and adopt $\log(g)=4.0$, which is typical for T Tauri and HAeBe stars. We scale the appropriate model to the dereddened V-band magnitude for each object. If literature V-band photometry is not available, we scale the model to G-band. We use the V-band, or G-band, filter transmission curve to determine the appropriate magnitude for the photospheric model. 

3. We calculate the observed \brgamma\ equivalent width, EW$_{obs}$, of the target; the intrinsic equivalent width of the corresponding photosphere model, EW$_{phot}$; and the difference between the dereddened K-band magnitude of the target and the magnitude of the photosphere, $\Delta m_K$, to account for veiling due to hot disk material. To determine $\Delta m_K$, after we scale the photospheric model as described in step 2, then we determine $\Delta m_K$ as the difference between the dereddened K-band magnitude of the target and the K-band magnitude of the scaled photospheric model. As in step 2, we use the filter transmission curve to get the appropriate K-band magnitude for the model. We then find EW(\brgamma) after correcting for both the disk veiling and the photospheric absorption using the following relationship, 
\begin{equation}
    {\rm EW}_{corr} = {\rm EW}_{obs} - {\rm EW}_{phot}10^{-\Delta m_{K}/2.5}.
\end{equation}

4. We find F$_{Br\gamma}$, the flux of the line, by multiplying EW$_{corr}$ with the continuum flux at the center of the line. The continuum flux is estimated by extrapolating continuum regions near the line on either side through the line region, following \citet{fairlamb17}. F$_{Br\gamma}$ can then be converted into the line luminosity, \Lbrg, using the distances listed in Table~\ref{tab: stellar properties}. From \Lbrg, we use the empirical relation for HAeBe stars from \citet{fairlamb17} to determine \Lacc, 
\begin{equation}
\log_{10}(\frac{{\rm L}_{acc}}{L_{\odot}}) = (1.30 \pm 0.09)\log_{10}(\frac{{\rm L}_{Br\gamma}}{L_{\odot}}) + (4.46 \pm 0.23).
\end{equation}
This can then be converted to the mass accretion rate (\Mdot), for objects with known masses and radii, using the relationship, 
\begin{equation}
    \dot{M} = \frac{L_{acc}R_{*}}{GM_{*}}.
\end{equation}


We propagate the error through each step of this process, largely following the error analysis steps of \citet{fairlamb17}. The uncertainty on EW$_{obs}$ is $\sigma_{EW_{obs}}=\Delta\lambda$/SNR, where $\Delta\lambda$ is the line region, 2.1638 to 2.1684 \mic, in our case, and SNR is the signal-to-noise of the red continuum region (2.172 to 2.177 \mic). We propagate error through the EW$_{corr}$ calculation, adopting a 10\% error on EW$_{phot}$ to account for uncertainties in the stellar parameters used to determine the corresponding photosphere model and the coarseness of the BT Settl model grid. We assume that the uncertainty on $\Delta m_{K}$ is negligible due to the precision of K-band photometry. The uncertainty in L$_{Br\gamma}$ takes into account the uncertainty on F$_{Br\gamma}$ and the distance uncertainty. We propagate error through the \Lacc\ and \Mdot\ calculations. We present EW$_{obs}$, $\Delta m_{K}$, EW$_{corr}$, F$_{Br\gamma}$, L$_{Br\gamma}$, \Lacc, and \Mdot\ for each object in our IGRINS sample in Table~\ref{tab: results}.

We also classify the observed \brgamma\ line shapes by eye into four categories: 1) double-peaked, 2) single-peaked, 3) absorption, and 4) other. There are 48 double-peaked ($\sim$47\%), 28 singled-peaked ($\sim$27\%), 23 absorption/non-detection ($\sim$23\%), and 3 ``other'' ($\sim$3\%) profiles. The line shape classification for each object is included in Table~\ref{tab: results}.

\subsection{Group Determination}\label{subsec: group determination}
The spectral energy distribution (SED) of Herbig Ae/Be stars can be used to determine the geometry of the star/disk system. \citet{meeus01} classified their sample such that the SEDs of Group I objects could be reproduced by a black body and a power-law component while the SEDs of Group II objects required only a power-law to be reproduced. Further classification depends on the presence of solid state emission bands at 10 and 20 \mic. This classification suggested a differing geometry with Group I objects having flared outer disks that have a large surface area to be irradiated by the central star and Group II objects having geometrically thin disks. Many Group I objects have been imaged to have gaps in their dust distribution \citep{maaskant13}, making them (pre-)transitional disks \citep{espaillat14}. Much work has been done in recent years to determine whether one group evolves into the other or whether they are distinct groups from an early age. 

Several methods have been used in the literature to determine group classification, mainly focusing on the shape of the mid-infrared (MIR) emission relative to the NIR. \citet{vanboekel05} finds \LNIR/\LIR\ relative to the m$_{12}$-m$_{60}$ color effective in separating the two groups. \citet{garufi17} uses the flux ratio of 30 \mic\ to 13.5 \mic, with Group II sources having F$_{30}$/F$_{13.5}$ less than 2.1 and Group I sources being above 2.1. To determine group classification for our sample consistently, we follow the methods of \citet{vanboekel05} using photometry compiled for all objects in both the IGRINS Sample and the Auxiliary Sample. The stellar properties for the Auxiliary Sample are given in Table~\ref{tab: Aux Sample}. The process is as follows:

1.  We deredden the photometry collected from the Vizier database \citep{vizier} using \Av\ from the literature or, if unavailable, from (B-V) colors as described in Section~\ref{subsect: brg analysis}. We utilize data in the following bands: Johnson:J, Johnson:H, Johnson:K$_{S}$, Johnson:L, Johnson:M, \textit{Spitzer}/IRAC 3.6 and 4.5 \mic, WISE 3.4, 4.5, and 12 \mic,  and IRAS 12, 25, and 60 \mic. 

2. We then determine \LNIR\ and \LIR. \LNIR\ is determined by using a cubic spline interpolation from 1 to 5 \mic\ using the dereddened photometry (we only use objects who have both J- and M-band photometry to avoid extrapolation). This is done using the \texttt{CubicSpline} function from the SciPy library \citep{scipy}. \LIR\ is similarly determined from 12 to 60 \mic\ using WISE and IRAS photometry. Once interpolated, we integrate over the wavelength range in each case using Simpson's rule and the SciPy \texttt{integrate.simps} function \citep{scipy}. This is converted to a luminosity using the distance to each source. 

3. We calculate m$_{12}$-m$_{60}$ using the dereddened photometry as m$_{12}$-m$_{60}$ = -2.5$log_{10}$(F$_{12}$/F$_{60}$). We note that although we use dereddened fluxes here, the extinction is negligible at these long wavelengths. 

4. We follow the criteria of \citet{vanboekel05} that Group I sources have \LNIR/\LIR $\leq$(m$_{12}$ - m$_{60}$)+1.5 and Group II sources have \LNIR/\LIR $>$ (m$_{12}$ - m$_{60}$)+1.5. 

\begin{figure}
    \centering
    \includegraphics[scale=0.55]{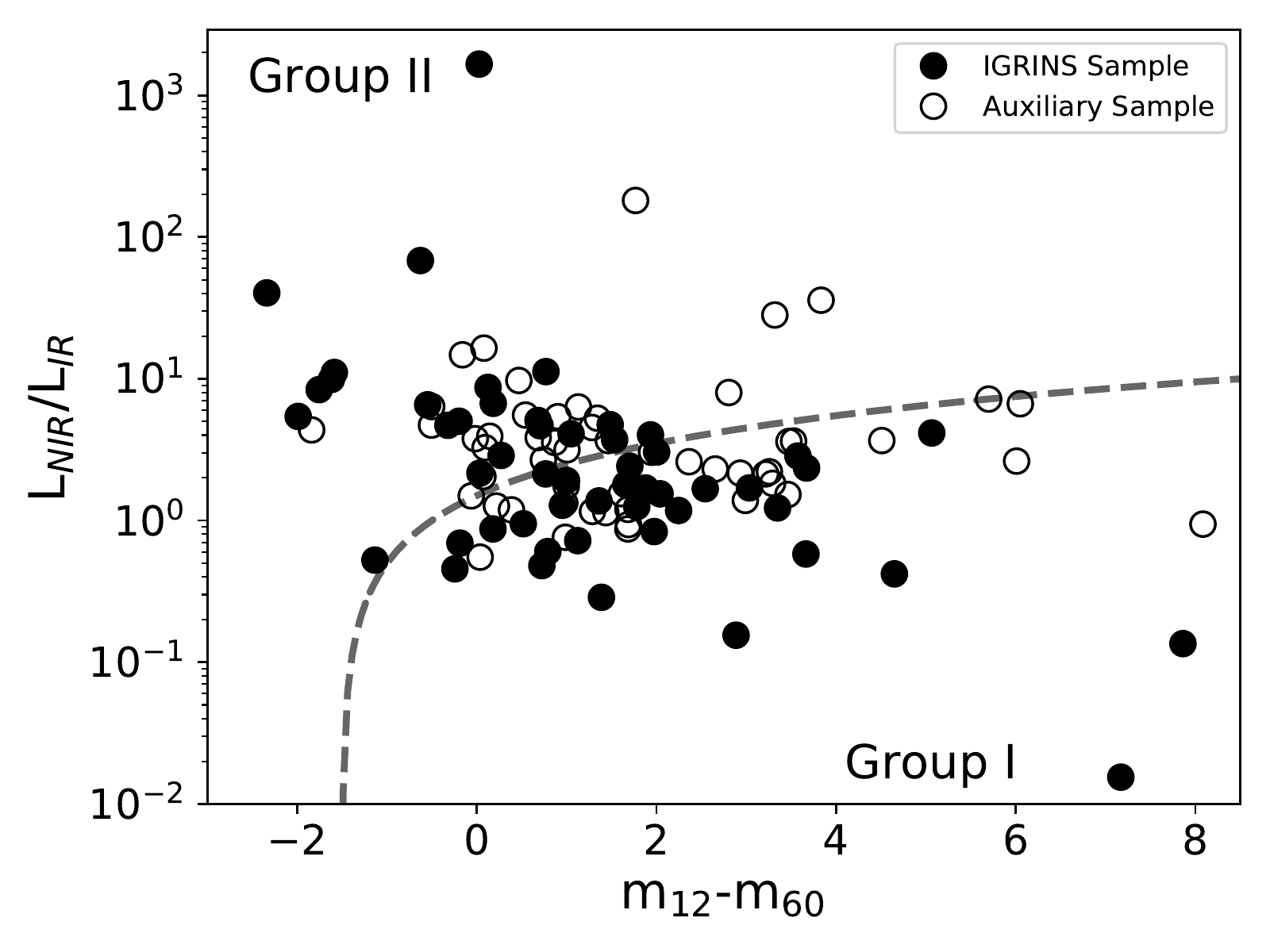}
    \caption{The group determination for the IGRINS Sample (filled) and the Auxiliary Sample (\citealt{fairlamb15} and \citet{donehew-brittain11}; open). The group criteria is shown as the gray dashed line with Group I sources located in the bottom right of the plot. } 
    \label{fig: Group determination}
\end{figure}

We show the criteria plot in Figure~\ref{fig: Group determination} and present the group determination for the IGRINS Sample in Table~\ref{tab: stellar properties} and for the Auxiliary Sample in Table~\ref{tab: Aux Sample}. The values of $m_{12}-m_{60}$, $L_{NIR}$, and $L_{IR}$ are available in Table~\ref{tab: results}. We determine the group classification this way for 110 objects (and for analysis we use an additional four from the literature for the IGRINS Sample, including for MWC 137 which is a group III object in \citealt{baines12} and an additional one from the literature for the Auxiliary Sample). Of these 110 determinations, 69 are from the IGRINS Sample (34 Group I and 35 Group II) and 41 are from the Auxiliary Sample (25 Group I and 16 Group II). Of the 69 from the IGRINS Sample, 30 are new determinations. Of the 41 from the Auxiliary Sample, 20 are new (we note that this is 22 counting the reclassification of two that were ``embedded'' sources in \citealt{acke-anker06}). For the IGRINS Sample, all but two of our determinations agree with those that were previously derived in the literature. The exceptions are V380 Ori, which we find to be a Group I and \citet{acke-anker06} list as ''II?'' indicating that the determination is uncertain, and V599 Ori, which we find to be a Group I and \citet{acke-anker06} classify as an embedded source. For the Auxiliary Sample we agree with all but three in the literature. We find HD 132947 to be a Group I object and \citet{acke-anker06} find it to be Group II. MWC 297 we find to be Group I, while \citet{acke-anker06} classify this as an embedded source. Lastly, we find CQ Tau to be a Group I source, while \citet{meeus12} finds it to be a Group II.

\FloatBarrier 
\startlongtable
\begin{deluxetable*}{ccccccccccc}
\tablewidth{0pt}
\tabletypesize{\scriptsize}
\tablecaption{Stellar Properties of IGRINS Sample\label{tab: stellar properties}}
\tablehead{
\colhead{Num.} & \colhead{Target} & \colhead{T$_{eff}$}  & \colhead{A$_V$}  & \colhead{Distance}  & \colhead{log$_{10}$(L$_{*}$)} 
& \colhead{M$_{*}$} &  \colhead{Age} & \colhead{Group} & \colhead{References} \\ \colhead{} &
\colhead{} & \colhead{(K)}  & \colhead{(mag)} & \colhead{(pc)}  & \colhead{(L$_{\odot}$)} & \colhead{(M$_{\odot}$)}  & \colhead{(Myr)} & \colhead{} & \colhead{}
}
\startdata
1   &   UX Ori   &   8500$_{-200}^{+200}$   &   0.99$_{-0.03}^{+0.04}$   &   325$_{-8}^{+9}$   &   1.18$_{-0.09}^{+0.09}$   &   1.82$_{-0.08}^{+0.07}$   &   7.1$_{-0.5}^{+0.5}$   &   II   &   a,a,a,a,a,a  \\ 
2   &   IRAS 05044-0325   &   17000$_{-2000}^{+2000}$   &   3.4$_{-0.4}^{+0.3}$   &   400$_{-9}^{+10}$   &   2.0$_{-0.4}^{+0.4}$   &   3.2$_{-0.9}^{+0.6}$   &   3$_{-1}^{+3}$   &      &   a,a,a,a,a,a  \\ 
3   &   V1012 Ori   &   8500$_{-200}^{+200}$   &   1.15$_{-0.06}^{+0.05}$   &   390$_{-10}^{+10}$   &   0.77$_{-0.05}^{+0.05}$   &   1.30$_{-0.06}^{+0.06}$   &   8$_{-0.9}^{+1}$   &   I   &   a,a,b,b,b,b  \\ 
4   &   HD 35187   &   9800$_{-300}^{+900}$   &   0.8$_{-0.5}^{+0.5}$   &   163$_{-4}^{+5}$   &   1.4$_{-0.2}^{+0.3}$   &   2.1$_{-0.2}^{+0.3}$   &   0$_{-2}^{+20}$   &   II   &   b,b,b,b,b,b  \\ 
5   &   HD 287823   &   8400$_{-100}^{+100}$   &   0.39$_{-0.03}^{+0.03}$   &   360$_{-10}^{+10}$   &   1.27$_{-0.07}^{+0.07}$   &   1.80$_{-0.04}^{+0.07}$   &   6.6$_{-0.6}^{+0.6}$   &   I   &   a,a,a,a,a,a  \\ 
6   &   HD 290409   &   9800$_{-500}^{+500}$   &   0.41$_{-0.03}^{+0.03}$   &   460$_{-20}^{+30}$   &   1.4$_{-0.2}^{+0.2}$   &   2.2$_{-0.2}^{+0.01}$   &   5$_{-0.4}^{+3}$   &   I   &   a,a,a,a,a,a  \\ 
7   &   HD 35929   &   7000$_{-200}^{+200}$   &   0.35$_{-0.04}^{+0.05}$   &   390$_{-10}^{+10}$   &   1.9$_{-0.1}^{+0.1}$   &   3.4$_{-0.3}^{+0.3}$   &   1.0$_{-0.2}^{+0.4}$   &   II   &   a,a,a,a,a,a  \\ 
8   &   HD 290500   &   9500$_{-500}^{+500}$   &   1.03$_{-0.07}^{+0.06}$   &   440$_{-20}^{+20}$   &   1.2$_{-0.2}^{+0.2}$   &   1.9$_{-0.1}^{+0.2}$   &   8$_{-3}^{+5}$   &   I   &   a,a,a,a,a,a  \\ 
9   &   HD 244314   &   8500$_{-200}^{+200}$   &   0.51$_{-0.06}^{+0.06}$   &   430$_{-20}^{+20}$   &   1.3$_{-0.1}^{+0.1}$   &   1.9$_{-0.07}^{+0.2}$   &   6$_{-1}^{+0.9}$   &   II   &   a,a,a,a,a,a  \\ 
10   &   HD 36112   &   7600$_{-300}^{+200}$   &   0.2$_{-0.2}^{+0.2}$   &   160$_{-3}^{+3}$   &   1.0$_{-0.08}^{+0.1}$   &   1.6$_{-0.08}^{+0.1}$   &   8$_{-1}^{+0.4}$   &   I   &   b,b,b,b,b,b  \\ 
11   &   V451 Ori   &   10700   &   -1.04   &   800$_{-700}^{+800}$   &      &      &      &      &   c,This Work,d,-,-,-  \\ 
12   &   HK Ori   &   8500$_{-500}^{+500}$   &   1.3$_{-0.2}^{+0.2}$   &   440$_{-90}^{+90}$   &   1.1$_{-0.3}^{+0.2}$   &   1.7$_{-0.2}^{+0.2}$   &   9$_{-3}^{+3}$   &   II   &   a,a,e,e,e,e  \\ 
13   &   HD 244604   &   9000$_{-200}^{+200}$   &   0.59$_{-0.05}^{+0.04}$   &   420$_{-20}^{+20}$   &   1.6$_{-0.1}^{+0.1}$   &   2.3$_{-0.2}^{+0.2}$   &   3.8$_{-0.8}^{+0.8}$   &   II   &   a,a,a,a,a,a  \\ 
14   &   UY Ori   &   9800$_{-200}^{+200}$   &   1.4$_{-0.1}^{+0.1}$   &   360$_{-20}^{+20}$   &   0.39$_{-0.06}^{+0.07}$   &   2.0$_{-0.1}^{+0.1}$   &   6$_{-2}^{+2}$   &   I   &   a,a,b,b,e,e  \\ 
15   &   RY Ori   &   6200$_{-80}^{+200}$   &   1.0$_{-0.2}^{+0.2}$   &   368$_{-8}^{+9}$   &   0.9$_{-0.09}^{+0.1}$   &   1.5$_{-0.08}^{+0.2}$   &   7$_{-2}^{+0.9}$   &   II   &   b,b,b,b,b,b  \\ 
16   &   HD 36408   &   11900$_{-800}^{+800}$   &   0.4$_{-0.2}^{+0.1}$   &   440$_{-30}^{+30}$   &   3.1$_{-0.2}^{+0.2}$   &   6$_{-1}^{+1}$   &   0.2$_{-0.09}^{+0.2}$   &   II$^{f}$   &   b,b,b,b,b,b  \\ 
17   &   Brun 216   &   6200$_{-200}^{+80}$   &   0.1$_{-0.07}^{+0.2}$   &   390$_{-9}^{+10}$   &   0.98$_{-0.05}^{+0.09}$   &   1.8$_{-0.09}^{+0.2}$   &   4$_{-1}^{+0.2}$   &   I   &   b,b,b,b,b,b  \\ 
18   &   HD 245185   &   10000$_{-500}^{+500}$   &   0.40$_{-0.02}^{+0.03}$   &   430$_{-30}^{+40}$   &   1.5$_{-0.2}^{+0.2}$   &   2.2$_{-0.2}^{+0.04}$   &   5$_{-0.8}^{+3}$   &   I   &   a,a,a,a,a,a  \\ 
19   &   MX Ori   &   6150   &   0.76   &   400$_{-300}^{+500}$   &      &      &      &      &   g,g,d,-,-,-  \\ 
20   &   NV Ori   &   6800$_{-200}^{+200}$   &   0.1$_{-0.1}^{+0.3}$   &   390$_{-10}^{+20}$   &   1.2$_{-0.08}^{+0.2}$   &   1.8$_{-0.1}^{+0.2}$   &   5$_{-1}^{+1.0}$   &   II$^{h}$   &   b,b,b,b,b,b  \\ 
21   &   T Ori   &   9000$_{-500}^{+500}$   &   1.8$_{-0.1}^{+0.1}$   &   410$_{-10}^{+10}$   &   1.7$_{-0.2}^{+0.2}$   &   2.3$_{-0.3}^{+0.4}$   &   3$_{-1}^{+1}$   &   II$^{f}$   &   a,a,a,a,a,a  \\ 
22   &   V380 Ori   &   9800$_{-800}^{+800}$   &   2.0$_{-0.4}^{+0.3}$   &   480$_{-50}^{+80}$   &   2.0$_{-0.2}^{+0.2}$   &   2.8$_{-0.4}^{+0.6}$   &   2$_{-0.8}^{+1}$   &   I   &   a,a,b,b,b,b  \\ 
23   &   HD 245465   &   6640   &   -0.11   &   300$_{-300}^{+300}$   &   1.58   &   2.7   &      &      &   i,j,d,i,i,-  \\ 
24   &   HD 37258   &   9800$_{-500}^{+500}$   &   0.54$_{-0.04}^{+0.03}$   &   360$_{-20}^{+20}$   &   1.2$_{-0.1}^{+0.1}$   &   1.9$_{-0.1}^{+0.1}$   &   10$_{-2}^{+10}$   &   II   &   a,a,b,b,b,b  \\ 
25   &   HD 290770   &   10500$_{-200}^{+200}$   &   0.34$_{-0.05}^{+0.05}$   &   400$_{-20}^{+20}$   &   1.7$_{-0.1}^{+0.1}$   &   2.4$_{-0.1}^{+0.05}$   &   4$_{-0.2}^{+1}$   &   II   &   a,a,a,a,a,a  \\ 
26   &   BF Ori   &   9000$_{-200}^{+200}$   &   0.74$_{-0.07}^{+0.06}$   &   390$_{-10}^{+10}$   &   1.5$_{-0.1}^{+0.1}$   &   2.0$_{-0.08}^{+0.2}$   &   5.1$_{-0.9}^{+0.8}$   &   II   &   a,a,a,a,a,a  \\ 
27   &   HD 37357   &   9500$_{-200}^{+200}$   &   0.40$_{-0.03}^{+0.03}$   &   600$_{-100}^{+300}$   &   2.0$_{-0.2}^{+0.4}$   &   3.0$_{-0.4}^{+1.0}$   &   1.7$_{-0.9}^{+0.9}$   &   II   &   a,a,b,b,b,b  \\ 
28   &   HD 290764   &   7900$_{-400}^{+400}$   &   0.58$_{-0.08}^{+0.07}$   &   400$_{-20}^{+20}$   &   1.4$_{-0.2}^{+0.2}$   &   1.9$_{-0.2}^{+0.2}$   &   5$_{-1}^{+2}$   &   I   &   a,a,a,a,a,a  \\ 
29   &   V1787 Ori   &   8200$_{-200}^{+200}$   &   4.1$_{-0.2}^{+0.2}$   &   390$_{-10}^{+10}$   &   1.2$_{-0.09}^{+0.1}$   &   1.66$_{-0.08}^{+0.09}$   &   7$_{-1}^{+0.6}$   &   II   &   b,b,b,b,b,b  \\ 
30   &   HD 37411   &   9800$_{-200}^{+200}$   &   0.59$_{-0.07}^{+0.07}$   &   400$_{-100}^{+100}$   &   1.3$_{-0.4}^{+0.3}$   &   1.9$_{-0.4}^{+0.4}$   &   9$_{-4}^{+4}$   &   II   &   a,a,e,e,e,e  \\ 
31   &   V599 Ori   &   8000$_{-200}^{+200}$   &   4.9$_{-0.4}^{+0.2}$   &   410$_{-10}^{+10}$   &   1.6$_{-0.2}^{+0.2}$   &   2.3$_{-0.4}^{+0.4}$   &   3$_{-1}^{+2}$   &   I   &   a,a,a,a,a,a  \\ 
32   &   HD 245906   &   8000$_{-200}^{+200}$   &   0.9$_{-0.2}^{+0.2}$   &   700$_{-200}^{+400}$   &   1.7$_{-0.3}^{+0.5}$   &   2$_{-0.5}^{+1}$   &   3$_{-2}^{+3}$   &   II   &   b,b,b,b,b,b  \\ 
33   &   RR Tau   &   10000$_{-200}^{+200}$   &   1.6$_{-0.1}^{+0.3}$   &   770$_{-40}^{+50}$   &   2.0$_{-0.1}^{+0.2}$   &   2.8$_{-0.2}^{+0.5}$   &   2.0$_{-0.7}^{+0.4}$   &   II   &   b,b,b,b,b,b  \\ 
34   &   V350 Ori   &   9000$_{-200}^{+200}$   &   1.06$_{-0.08}^{+0.08}$   &   390$_{-30}^{+30}$   &   1.0$_{-0.09}^{+0.1}$   &   1.71$_{-0.08}^{+0.09}$   &   12$_{-5}^{+8}$   &   II   &   a,a,b,b,b,b  \\ 
35   &   HD 37806   &   10000$_{-700}^{+1000}$   &   0.1$_{-0.1}^{+0.2}$   &   430$_{-20}^{+20}$   &   2.2$_{-0.1}^{+0.2}$   &   3.1$_{-0.3}^{+0.5}$   &   1.6$_{-0.6}^{+0.6}$   &   II   &   b,b,b,b,b,b  \\ 
36   &   HD 38120   &   10700$_{-900}^{+800}$   &   0.2$_{-0.2}^{+0.5}$   &   400$_{-20}^{+20}$   &   1.7$_{-0.2}^{+0.3}$   &   2.4$_{-0.2}^{+0.4}$   &   0$_{-1}^{+10}$   &   I   &   b,b,b,b,b,b  \\ 
37   &   HD 38238   &   7800$_{-200}^{+200}$   &   0.0$_{-0.03}^{+0.2}$   &   342$_{-8}^{+9}$   &   1.4$_{-0.03}^{+0.1}$   &   2.0$_{-0.10}^{+0.2}$   &   5$_{-1}^{+0.2}$   &   I   &   b,b,b,b,b,b  \\ 
38   &   MWC 778   &   15700   &   2.1   &   2000$_{-900}^{+4000}$   &      &      &      &      &   k,l,d,-,-,-  \\ 
39   &   V1818 Ori   &   13000$_{-2000}^{+1000}$   &   3.7$_{-0.2}^{+0.2}$   &   700$_{-60}^{+80}$   &   3.0$_{-0.3}^{+0.2}$   &   5$_{-1}^{+1}$   &   0.4$_{-0.2}^{+0.4}$   &   II   &   b,b,b,b,b,b  \\ 
40   &   HD 249879   &   12000$_{-800}^{+2000}$   &   0.2$_{-0.1}^{+0.2}$   &   670$_{-50}^{+70}$   &   1.6$_{-0.2}^{+0.3}$   &   2.2$_{-0.2}^{+0.5}$   &   10$_{-2}^{+10}$   &   II   &   b,b,b,b,b,b  \\ 
41   &   HD 250550   &   11000$_{-500}^{+500}$   &   1.6$_{-0.1}^{+0.1}$   &   700$_{-60}^{+90}$   &   3.5$_{-0.2}^{+0.2}$   &   9$_{-2}^{+2}$   &   0.08$_{-0.03}^{+0.06}$   &   I   &   a,a,a,a,a,a  \\ 
42   &   V791 Mon   &   15000$_{-2000}^{+2000}$   &   1.5$_{-0.1}^{+0.1}$   &   890$_{-40}^{+50}$   &   2.7$_{-0.3}^{+0.3}$   &   4.3$_{-0.4}^{+0.9}$   &   0.8$_{-0.4}^{+0.4}$   &   II   &   a,a,a,a,a,a  \\ 
43   &   AE Lep   &   9800$_{-300}^{+900}$   &   0.2$_{-0.1}^{+0.3}$   &   900$_{-90}^{+100}$   &   1.8$_{-0.2}^{+0.3}$   &   2.5$_{-0.2}^{+0.7}$   &   3$_{-1}^{+0.9}$   &   I   &   b,b,b,b,b,b  \\ 
44   &   LkHa 208   &   7800$_{-600}^{+500}$   &   0.4$_{-0.4}^{+0.3}$   &   600$_{-90}^{+200}$   &   1.0$_{-0.3}^{+0.4}$   &   1.6$_{-0.1}^{+0.5}$   &   10$_{-5}^{+10}$   &   I   &   b,b,b,b,b,b  \\ 
45   &   IRAS 06071+2925   &   10700$_{-900}^{+800}$   &   3.0$_{-0.3}^{+0.1}$   &   1070$_{-50}^{+60}$   &   1.8$_{-0.2}^{+0.2}$   &   2.4$_{-0.3}^{+0.3}$   &   0$_{-1}^{+10}$   &   I   &   b,b,b,b,b,b  \\ 
46   &   LkHa 338   &   10700$_{-900}^{+800}$   &   2.6$_{-0.3}^{+0.1}$   &   880$_{-50}^{+60}$   &   1.1$_{-0.3}^{+0.2}$   &   1.88$_{-0.09}^{+0.09}$   &   10$_{-2}^{+10}$   &      &   b,b,b,b,b,b  \\ 
47   &   LkHa 339   &   10500$_{-200}^{+200}$   &   3.1$_{-0.5}^{+0.4}$   &   860$_{-30}^{+30}$   &   1.8$_{-0.3}^{+0.2}$   &   2.5$_{-0.3}^{+0.5}$   &   3$_{-1}^{+2}$   &   I   &   a,a,a,a,a,a  \\ 
48   &   MWC 137   &   29000$_{-4000}^{+4000}$   &   4.6$_{-0.3}^{+0.2}$   &   2900$_{-400}^{+600}$   &   4.9$_{-0.4}^{+0.4}$   &   20$_{-7}^{+10}$   &   0.02$_{-0.008}^{+0.02}$   &   III$^{f}$   &   b,b,b,b,b,b  \\ 
49   &   HD 45677   &   16000$_{-800}^{+3000}$   &   0.6$_{-0.2}^{+0.2}$   &   620$_{-30}^{+40}$   &   2.9$_{-0.2}^{+0.3}$   &   5$_{-0.4}^{+1}$   &   1$_{-0.3}^{+4}$   &   II   &   b,b,b,b,b,b  \\ 
50   &   HD 46060   &   21000$_{-4000}^{+3000}$   &   1.8$_{-0.2}^{+0.3}$   &   930$_{-70}^{+100}$   &   3.9$_{-0.4}^{+0.3}$   &   10$_{-2}^{+3}$   &   0.1$_{-0.05}^{+0.1}$   &   II   &   b,b,b,b,b,b  \\ 
51   &   LkHa 341   &   6720   &   3.0   &   1000$_{-1000}^{+1000}$   &   1.58   &   2.6   &   3.13   &      &   i,i,d,i,i,i  \\ 
52   &   VY Mon   &   12000$_{-4000}^{+4000}$   &   5.6$_{-0.5}^{+0.3}$   &   1500$_{-300}^{+600}$   &   3.6$_{-0.7}^{+0.6}$   &   9$_{-4}^{+7}$   &   0.1$_{-0.07}^{+0.5}$   &   I   &   a,a,b,b,b,b  \\ 
53   &   LkHa 215   &   14000$_{-1000}^{+1000}$   &   2.0$_{-0.2}^{+0.2}$   &   710$_{-30}^{+40}$   &   2.6$_{-0.2}^{+0.2}$   &   3.8$_{-0.4}^{+0.6}$   &   1.0$_{-0.4}^{+0.3}$   &   I   &   b,b,b,b,b,b  \\ 
54   &   HD 259431   &   14000$_{-3000}^{+2000}$   &   1.1$_{-0.3}^{+0.2}$   &   720$_{-40}^{+40}$   &   3.0$_{-0.4}^{+0.3}$   &   5$_{-1}^{+2}$   &   0.4$_{-0.3}^{+0.5}$   &   I   &   b,b,b,b,b,b  \\ 
55   &   R Mon   &   12000$_{-2000}^{+2000}$   &   2.5$_{-0.3}^{+0.2}$   &   800$_{-200}^{+200}$   &   2.2$_{-0.4}^{+0.3}$   &   3.1$_{-0.6}^{+0.8}$   &   2$_{-0.9}^{+1}$   &   I   &   a,a,e,e,e,e  \\ 
56   &   V590 Mon   &   12000$_{-1000}^{+1000}$   &   1.0$_{-0.2}^{+0.2}$   &   800$_{-100}^{+200}$   &   1.4$_{-0.2}^{+0.3}$   &   2.3$_{-0.1}^{+0.1}$   &   10$_{-1}^{+10}$   &   I   &   a,a,b,b,b,b  \\ 
57   &   HBC 222   &   6200$_{-80}^{+200}$   &   0.1$_{-0.1}^{+0.2}$   &   710$_{-30}^{+30}$   &   0.8$_{-0.08}^{+0.1}$   &   1.5$_{-0.08}^{+0.1}$   &   7$_{-2}^{+0.9}$   &      &   b,b,b,b,b,b  \\ 
58   &   HD 50138   &   9400$_{-400}^{+400}$   &   0.0$_{-0.03}^{+0.1}$   &   380$_{-10}^{+20}$   &   2.5$_{-0.09}^{+0.1}$   &   4.2$_{-0.3}^{+0.5}$   &   0.6$_{-0.2}^{+0.2}$   &   II   &   b,b,b,b,b,b  \\ 
59   &   HD 52721   &   22000$_{-2000}^{+2000}$   &   0.8$_{-0.1}^{+0.1}$   &   500$_{-200}^{+200}$   &   3.9$_{-0.6}^{+0.6}$   &   10$_{-7}^{+7}$   &   0.1$_{-0.1}^{+0.1}$   &   II   &   a,a,e,e,e,e  \\ 
60   &   LkHa 218   &   10500$_{-500}^{+500}$   &   1.61$_{-0.09}^{+0.09}$   &   1120$_{-70}^{+80}$   &   2.0$_{-0.2}^{+0.2}$   &   2.9$_{-0.4}^{+0.4}$   &   2.1$_{-0.7}^{+1.0}$   &   I   &   a,a,a,a,a,a  \\ 
61   &   LkHa 220   &   13000$_{-200}^{+200}$   &   1.29$_{-0.07}^{+0.07}$   &   1200$_{-80}^{+100}$   &   2.3$_{-0.1}^{+0.1}$   &   3.3$_{-0.02}^{+0.3}$   &   1.7$_{-0.4}^{+0.1}$   &   I   &   a,a,a,a,a,a  \\ 
62   &   HD 53367   &   30000$_{-1000}^{+1000}$   &   2.1$_{-0.1}^{+0.1}$   &   130$_{-20}^{+30}$   &   3.1$_{-0.2}^{+0.2}$   &   12$_{-4}^{+4}$   &   0.08$_{-0.08}^{+0.08}$   &   I   &   a,a,b,b,e,e  \\ 
63   &   RAFGL 5223   &   26000$_{-2000}^{+2000}$   &   2.9$_{-0.4}^{+0.3}$   &   2900$_{-400}^{+600}$   &   4.2$_{-0.4}^{+0.4}$   &   13$_{-3}^{+5}$   &   0.06$_{-0.03}^{+0.06}$   &      &   a,a,a,a,a,a  \\ 
64   &   NX Pup   &   7000$_{-200}^{+200}$   &   0.3$_{-0.2}^{+0.1}$   &   1700$_{-400}^{+600}$   &   2.5$_{-0.2}^{+0.3}$   &   5$_{-0.8}^{+1}$   &   0.3$_{-0.2}^{+0.2}$   &   II   &   a,a,b,b,b,b  \\ 
65   &   DW Cma   &   18000$_{-4000}^{+4000}$   &   5.0$_{-0.5}^{+0.3}$   &   2600$_{-300}^{+500}$   &   4.2$_{-0.7}^{+0.6}$   &   13$_{-6}^{+10}$   &   0.0$_{-0.03}^{+0.2}$   &   I   &   a,a,a,a,a,a  \\ 
66   &   HD 58647   &   10500$_{-200}^{+200}$   &   0.4$_{-0.1}^{+0.2}$   &   318$_{-7}^{+7}$   &   2.4$_{-0.09}^{+0.1}$   &   3.9$_{-0.2}^{+0.3}$   &   0.8$_{-0.2}^{+0.1}$   &   II   &   b,b,b,b,b,b  \\ 
67   &   V388 Vel   &   9500$_{-600}^{+300}$   &   4.0$_{-0.4}^{+0.1}$   &   2500$_{-300}^{+500}$   &   2.4$_{-0.3}^{+0.2}$   &   4$_{-0.9}^{+1}$   &   0.7$_{-0.3}^{+0.7}$   &      &   b,b,b,b,b,b  \\ 
68   &   HD 76534   &   19000$_{-500}^{+500}$   &   0.95$_{-0.08}^{+0.07}$   &   910$_{-50}^{+60}$   &   3.7$_{-0.1}^{+0.1}$   &   8.5$_{-0.9}^{+0.9}$   &   0.13$_{-0.03}^{+0.05}$   &   I   &   a,a,a,a,a,a  \\ 
69   &   HD 85567   &   13000$_{-500}^{+500}$   &   1.0$_{-0.2}^{+0.2}$   &   1020$_{-40}^{+50}$   &   3.3$_{-0.2}^{+0.2}$   &   7$_{-1}^{+1}$   &   0.2$_{-0.07}^{+0.1}$   &   II   &   a,a,a,a,a,a  \\ 
70   &   HD 94509   &   12000$_{-1000}^{+1000}$   &   0.35$_{-0.02}^{+0.01}$   &   1800$_{-200}^{+200}$   &   3.1$_{-0.2}^{+0.2}$   &   6$_{-1}^{+1}$   &   0.2$_{-0.09}^{+0.2}$   &      &   a,a,a,a,a,a  \\ 
71   &   HD 95881   &   10000$_{-200}^{+200}$   &   0.72$_{-0.04}^{+0.03}$   &   1170$_{-70}^{+80}$   &   3.1$_{-0.1}^{+0.1}$   &   7.0$_{-0.7}^{+0.6}$   &   0.14$_{-0.03}^{+0.05}$   &   II   &   a,a,a,a,a,a  \\ 
72   &   HD 97048   &   10500$_{-500}^{+500}$   &   1.37$_{-0.06}^{+0.05}$   &   185$_{-2}^{+2}$   &   1.8$_{-0.1}^{+0.1}$   &   2.4$_{-0.03}^{+0.2}$   &   3.6$_{-0.6}^{+0.2}$   &   I   &   a,a,a,a,a,a  \\ 
73   &   HD 98922   &   10500$_{-200}^{+200}$   &   0.45$_{-0.07}^{+0.07}$   &   690$_{-20}^{+30}$   &   3.2$_{-0.1}^{+0.1}$   &   7.2$_{-0.7}^{+0.7}$   &   0.14$_{-0.03}^{+0.05}$   &   II   &   a,a,a,a,a,a  \\ 
74   &   HD 100453   &   7200$_{-200}^{+200}$   &   0.12$_{-0.04}^{+0.04}$   &   104.2$_{-0.7}^{+0.7}$   &   0.85$_{-0.09}^{+0.08}$   &   1.48$_{-0.02}^{+0.03}$   &   11$_{-1}^{+0.8}$   &   I   &   a,a,a,a,a,a  \\ 
75   &   HD 100546   &   9800$_{-500}^{+500}$   &   0.17$_{-0.03}^{+0.02}$   &   110$_{-1}^{+1}$   &   1.4$_{-0.1}^{+0.1}$   &   2.2$_{-0.2}^{+0.02}$   &   5$_{-0.2}^{+2}$   &   I   &   a,a,a,a,a,a  \\ 
76   &   HD 101412   &   9800$_{-200}^{+200}$   &   0.69$_{-0.03}^{+0.03}$   &   411$_{-8}^{+8}$   &   1.79$_{-0.08}^{+0.08}$   &   2.4$_{-0.1}^{+0.2}$   &   3.2$_{-0.6}^{+0.5}$   &   II   &   a,a,a,a,a,a  \\ 
77   &   HD 104237   &   8000$_{-200}^{+200}$   &   0.29$_{-0.06}^{+0.05}$   &   108$_{-1}^{+1}$   &   1.33$_{-0.01}^{+0.04}$   &   1.85$_{-0.09}^{+0.09}$   &   5.5$_{-0.4}^{+0.3}$   &   II   &   a,a,b,b,b,b  \\ 
78   &   Hen 3-847   &   14000$_{-500}^{+500}$   &   0.8$_{-0.1}^{+0.1}$   &   800$_{-100}^{+300}$   &   2.1$_{-0.2}^{+0.3}$   &   3.0$_{-0.2}^{+0.6}$   &   2$_{-1}^{+9}$   &   II   &   a,a,b,b,b,b  \\ 
79   &   CQ Uma   &   8840   &   -1.78   &   90$_{-90}^{+90}$   &      &      &      &      &   m,This Work,d,-,-,-  \\ 
80   &   PX Vul   &   6800$_{-200}^{+300}$   &   1.2$_{-0.2}^{+0.3}$   &   630$_{-40}^{+50}$   &   1.4$_{-0.1}^{+0.2}$   &   2.1$_{-0.2}^{+0.3}$   &   3$_{-1}^{+1}$   &   II   &   b,b,b,b,b,b  \\ 
81   &   IRAS 19343+2926   &   24000$_{-5000}^{+4000}$   &   2.6$_{-0.3}^{+0.3}$   &   700$_{-200}^{+400}$   &   2.9$_{-0.6}^{+0.6}$   &   5$_{-0.3}^{+3}$   &   1$_{-0.4}^{+1}$   &   I   &   b,b,b,b,b,b  \\ 
82   &   MWC 342   &   26000$_{-4000}^{+6000}$   &   4.3$_{-0.3}^{+0.2}$   &   1800$_{-100}^{+100}$   &   4.9$_{-0.3}^{+0.4}$   &   22$_{-6}^{+9}$   &   0.02$_{-0.006}^{+0.02}$   &   II   &   b,b,b,b,b,b  \\ 
83   &   BD +41 3731   &   17000$_{-1000}^{+1000}$   &   1.1$_{-0.5}^{+0.5}$   &   1000$_{-50}^{+60}$   &   3.0$_{-0.3}^{+0.3}$   &   5$_{-0.7}^{+1}$   &   1$_{-0.3}^{+3}$   &      &   b,b,b,b,b,b  \\ 
84   &   LkHa 131   &   15700   &   3.32   &   1000$_{-800}^{+1000}$   &      &      &      &      &   n,o,d,-,-,-  \\ 
85   &   V2018 Cyg   &   10700   &   1.29   &   700$_{-700}^{+700}$   &      &      &      &      &   p,This Work,d,-,-,-  \\ 
86   &   V517 Cyg   &   9700   &   2.0   &   800$_{-700}^{+800}$   &      &      &      &      &   q,q,d,-,-,-  \\ 
87   &   V1977 Cyg   &   11000$_{-200}^{+200}$   &   2.3$_{-0.1}^{+0.1}$   &   860$_{-30}^{+40}$   &   2.5$_{-0.1}^{+0.1}$   &   3.9$_{-0.3}^{+0.4}$   &   0.8$_{-0.2}^{+0.2}$   &      &   b,b,b,b,b,b  \\ 
88   &   LkHa 134   &   11000$_{-200}^{+200}$   &   2.4$_{-0.2}^{+0.2}$   &   840$_{-30}^{+40}$   &   2.4$_{-0.2}^{+0.2}$   &   3.8$_{-0.6}^{+0.6}$   &   1.0$_{-0.3}^{+0.6}$   &      &   a,a,a,a,a,a  \\ 
89   &   LkHa 135   &   14000   &   1.98   &   900$_{-800}^{+900}$   &      &      &      &      &   r,o,d,-,-,-  \\ 
90   &   LkHa 147   &   20000$_{-3000}^{+5000}$   &   5.5$_{-0.2}^{+0.3}$   &   2100$_{-200}^{+200}$   &   3.7$_{-0.3}^{+0.4}$   &   8$_{-2}^{+4}$   &   0.1$_{-0.08}^{+0.2}$   &      &   b,b,b,b,b,b  \\ 
91   &   LkHa 167   &   8900$_{-400}^{+600}$   &   5.8$_{-0.2}^{+0.4}$   &   1100$_{-100}^{+200}$   &   2.2$_{-0.2}^{+0.3}$   &   3$_{-0.5}^{+1}$   &   1.2$_{-0.7}^{+0.7}$   &      &   b,b,b,b,b,b  \\ 
92   &   LkHa 168   &   6400$_{-200}^{+200}$   &   2.8$_{-0.2}^{+0.3}$   &   1400$_{-200}^{+400}$   &   1.9$_{-0.2}^{+0.3}$   &   3$_{-0.5}^{+1}$   &   0.9$_{-0.5}^{+0.5}$   &      &   b,b,b,b,b,b  \\ 
93   &   LkHa 169   &   17000   &   2.85   &   800$_{-800}^{+900}$   &      &      &      &      &   s,This Work,d,-,-,-  \\ 
94   &   LkHa 176   &   15700   &   3.3   &   800$_{-800}^{+800}$   &      &      &      &      &   n,This Work,d,-,-,-  \\ 
95   &   LkHa 183   &   29000   &   2.05   &   3000$_{-3000}^{+4000}$   &      &      &      &      &   t,This Work,d,-,-,-  \\ 
96   &   LkHa 192   &   15700   &   5.92   &   2000$_{-2000}^{+2000}$   &      &      &      &      &   r,This Work,d,-,-,-  \\ 
97   &   LkHa 324   &   12500$_{-500}^{+500}$   &   3.9$_{-0.2}^{+0.1}$   &   600$_{-10}^{+20}$   &   2.3$_{-0.2}^{+0.2}$   &   3.4$_{-0.3}^{+0.4}$   &   1.5$_{-0.4}^{+0.5}$   &      &   a,a,a,a,a,a  \\ 
98   &   V1578 Cyg   &   10500$_{-500}^{+500}$   &   1.46$_{-0.08}^{+0.08}$   &   770$_{-30}^{+30}$   &   2.4$_{-0.2}^{+0.2}$   &   3.7$_{-0.4}^{+0.4}$   &   1.0$_{-0.3}^{+0.4}$   &   II   &   a,a,a,a,a,a  \\ 
99   &   LkHa 257   &   9200$_{-200}^{+200}$   &   2.28$_{-0.08}^{+0.07}$   &   790$_{-20}^{+20}$   &   1.4$_{-0.1}^{+0.1}$   &   1.98$_{-0.04}^{+0.06}$   &   6$_{-0.5}^{+1}$   &   I   &   a,a,a,a,a,a  \\ 
100   &   BO Cep   &   6600$_{-100}^{+100}$   &   0.1$_{-0.1}^{+0.2}$   &   374$_{-4}^{+5}$   &   0.47$_{-0.06}^{+0.09}$   &   1.22$_{-0.06}^{+0.06}$   &   17$_{-2}^{+0.9}$   &   I   &   b,b,b,b,b,b  \\ 
101   &   LkHa 350   &   12500   &   8.46   &   2000$_{-2000}^{+3000}$   &      &      &      &      &   i,This Work,d,-,-,-  \\ 
102   &   V374 Cep   &   16000$_{-1000}^{+1000}$   &   3.2$_{-0.2}^{+0.2}$   &   870$_{-40}^{+40}$   &   3.5$_{-0.2}^{+0.2}$   &   8$_{-1}^{+1}$   &   0.2$_{-0.06}^{+0.1}$   &      &   a,a,a,a,a,a  \\ 
\enddata
\tablecomments{
References (In order of appearance in the table): 
$^a$\cite{wichittanakom20}, $^b$\cite{vioque18}, $^c$\cite{joy49}, $^d$\cite{bailer-jones18}, $^e$\cite{fairlamb15}, $^f$\cite{baines12}, $^g$\cite{kim16}, $^h$\cite{mendigutia12}, $^i$\cite{hernandez04}, $^j$\cite{valenti00}, $^k$\cite{suarez06}, $^l$\cite{sartori10}, $^m$\cite{abt-morell95}, $^n$\cite{herbig58}, $^o$\cite{cohen-kuhi79}, $^p$\cite{stephenson-sanduleak77}, $^q$\cite{melnikov01}, $^r$\cite{welin73}, $^s$\cite{straizys99}, $^t$\cite{straizys08}.
We do not list references for the Group determinations, as these are primarily from this work. Group determinations from the literature are denoted with a superscript corresponding to the reference, as listed above. The spectral types for MWC 778, LkHa 192, and LkHa 131 are given as B and with no subclass in the literature. We adopt \Teff\ values of 15,700 K, corresponding to a B5 star, for the analysis. V451 Ori, HD 245465, and CQ Uma have negative \Av\ values, either from the literature or our determinations. We list the negative values here but make the \Av\ zero for these objects in the analysis.  }
 
\end{deluxetable*}   
\FloatBarrier
\clearpage

\begin{longrotatetable}
\FloatBarrier 
\startlongtable
\begin{deluxetable}{cccccccccccc}
\rotate
\tablewidth{0pt}
\tabletypesize{\scriptsize}
\tablecaption{Stellar Properties of the Auxiliary Sample\label{tab: Aux Sample}}
\tablehead{
\colhead{Lit. Source} & \colhead{Object Name} & \colhead{RA} & \colhead{Dec.} & \colhead{T$_{eff}$}  & \colhead{A$_V$} & \colhead{Distance}  & \colhead{log$_{10}$(L$_{*}$)} & \colhead{M$_{*}$} & \colhead{Age} & \colhead{Group}  & \colhead{Ref.} \\
\colhead{} & \colhead{} & \colhead{(J2000)} & \colhead{(J2000)} & \colhead{(K)} &  \colhead{(mag)} & \colhead{(pc)} &  \colhead{(L$_{\odot}$)} & \colhead{(M$_{\odot}$)}  &  \colhead{(Myr)} &  \colhead{} & \colhead{}
}
\startdata
DB11	&	HD 278937	&	03:40:47.0	&	+31:32:53.7	&	8000$_{-400}^{+300}$	&	0.5$_{-0.4}^{+0.2}$	&	310$_{-10}^{+10}$	&	0.9$_{-0.2}^{+0.1}$	&	1.6$_{-0.1}^{+0.1}$	&	12$_{-3}^{+8}$	&	I	&	b,b,b,b,b,b  \\
DB11	&	XY Per	&	03:49:36.3	&	+38:58:55.6	&	9800$_{-200}^{+200}$	&	1.5$_{-0.1}^{+0.1}$	&	460$_{-30}^{+30}$	&	2.0$_{-0.1}^{+0.1}$	&	2.8$_{-0.2}^{+0.3}$	&	2.0$_{-0.4}^{+0.4}$	&	II	&	b,b,b,b,b,b  \\
DB11	&	AB Aur	&	04:55:45.9	&	+30:33:04.2	&	9500$_{-800}^{+800}$	&	0.4$_{-0.4}^{+0.3}$	&	163$_{-2}^{+3}$	&	1.6$_{-0.2}^{+0.2}$	&	2.2$_{-0.2}^{+0.4}$	&	4$_{-1}^{+1}$	&	I	&	b,b,b,b,b,b  \\
DB11	&	MWC 480	&	04:58:46.3	&	+29:50:36.9	&	8200$_{-200}^{+200}$	&	0.1$_{-0.06}^{+0.3}$	&	162$_{-3}^{+3}$	&	1.3$_{-0.05}^{+0.1}$	&	1.8$_{-0.09}^{+0.1}$	&	6$_{-1}^{+0.3}$	&	II	&	b,b,b,b,b,b  \\
F15	&	HD 34282	&	05:16:00.4	&	-09:48:35.3	&	9500$_{-200}^{+200}$	&	0.55$_{-0.02}^{+0.03}$	&	312$_{-7}^{+8}$	&	1.21$_{-0.08}^{+0.08}$	&	1.9$_{-0.04}^{+0.1}$	&	9$_{-2}^{+3}$	&	I	&	a,a,a,a,a,a  \\
F15	&	HD 287841	&	05:24:42.8	&	+01:43:48.2	&	7800$_{-200}^{+200}$	&	0.33$_{-0.03}^{+0.03}$	&	370$_{-9}^{+10}$	&	1.04$_{-0.09}^{+0.09}$	&	1.60$_{-0.02}^{+0.07}$	&	9$_{-1}^{+0.6}$	&	I	&	a,a,a,a,a,a  \\
DB11	&	CQ Tau	&	05:35:58.5	&	+24:44:54.1	&	6800$_{-200}^{+400}$	&	0.4$_{-0.3}^{+0.4}$	&	163$_{-4}^{+4}$	&	0.9$_{-0.1}^{+0.2}$	&	1.5$_{-0.1}^{+0.2}$	&	9$_{-3}^{+3}$	&	I	&	b,b,b,b,b,b  \\
F15	&	PDS 124	&	06:06:58.4	&	-05:55:06.7	&	10200$_{-200}^{+200}$	&	1.79$_{-0.01}^{+0.02}$	&	850$_{-50}^{+60}$	&	1.6$_{-0.1}^{+0.1}$	&	2.33$_{-0.04}^{+0.02}$	&	4.0$_{-0.3}^{+0.6}$	&	I	&	a,a,a,a,a,a  \\
F15	&	PDS 24	&	06:48:41.6	&	-16:48:05.6	&	10500$_{-500}^{+500}$	&	1.4$_{-0.1}^{+0.1}$	&	1130$_{-40}^{+40}$	&	1.4$_{-0.2}^{+0.2}$	&	2.1$_{-0.1}^{+0.3}$	&	6$_{-2}^{+3}$	&	I	&	a,a,a,a,a,a  \\
F15	&	PDS 130	&	06:49:58.5	&	-07:38:52.2	&	10500$_{-200}^{+200}$	&	2.4$_{-0.2}^{+0.1}$	&	1320$_{-50}^{+60}$	&	1.9$_{-0.2}^{+0.1}$	&	2.6$_{-0.2}^{+0.3}$	&	3.0$_{-0.8}^{+0.7}$	&	I	&	a,a,a,a,a,a  \\
F15	&	PDS 229N	&	06:55:40.0	&	-03:09:50.5	&	12500$_{-200}^{+200}$	&	2.2$_{-0.2}^{+0.2}$	&	900$_{-200}^{+500}$	&	1.7$_{-0.3}^{+0.4}$	&	2.5$_{-0.1}^{+0.5}$	&	0$_{-3}^{+10}$	&		&	a,a,b,b,b,b  \\
Both	&	Z CMa	&	07:03:43.1	&	-11:33:06.2	&	8500$_{-500}^{+500}$	&	3.1$_{-0.3}^{+0.2}$	&	200$_{-50}^{+100}$	&	2.2$_{-0.3}^{+0.5}$	&	4$_{-0.8}^{+2}$	&	0.8$_{-0.6}^{+0.8}$	&	I	&	a,a,b,b,b,b  \\
F15	&	PDS 133	&	07:25:04.9	&	-25:45:49.6	&	14000$_{-2000}^{+2000}$	&	1.9$_{-0.1}^{+0.09}$	&	1480$_{-80}^{+90}$	&	2.2$_{-0.3}^{+0.2}$	&	3.3$_{-0.4}^{+0.4}$	&	2$_{-0.6}^{+1}$	&	I	&	a,a,a,a,a,a  \\
F15	&	HD 59319	&	07:28:36.7	&	-21:57:49.2	&	12500$_{-500}^{+500}$	&	0.06$_{-0.01}^{+0.02}$	&	670$_{-30}^{+40}$	&	2.5$_{-0.1}^{+0.1}$	&	3.9$_{-0.4}^{+0.4}$	&	1.0$_{-0.3}^{+0.3}$	&	I	&	a,a,a,a,a,a  \\
F15	&	PDS 134	&	07:32:26.6	&	-21:55:35.7	&	14000$_{-500}^{+500}$	&	1.70$_{-0.07}^{+0.06}$	&	2600$_{-300}^{+400}$	&	2.9$_{-0.2}^{+0.2}$	&	5.0$_{-0.7}^{+0.9}$	&	0.5$_{-0.2}^{+0.2}$	&		&	a,a,a,a,a,a  \\
F15	&	HD 68695	&	08:11:44.5	&	-44:05:08.7	&	9200$_{-200}^{+200}$	&	0.37$_{-0.05}^{+0.05}$	&	400$_{-9}^{+10}$	&	1.35$_{-0.09}^{+0.09}$	&	2.0$_{-0.03}^{+0.1}$	&	5.8$_{-0.4}^{+0.7}$	&	I	&	a,a,a,a,a,a  \\
F15	&	HD 72106	&	08:29:34.8	&	-38:36:21.1	&	8800$_{-200}^{+200}$	&	0.0$_{-0.0}^{+0.03}$	&	600$_{-200}^{+400}$	&	1.8$_{-0.4}^{+0.5}$	&	3$_{-0.7}^{+1}$	&	2$_{-2}^{+3}$	&	II	&	a,a,b,b,b,b  \\
F15	&	TYC 8581-2002-1	&	08:44:23.6	&	-59:56:57.8	&	9800$_{-200}^{+200}$	&	1.54$_{-0.07}^{+0.05}$	&	560$_{-10}^{+10}$	&	1.50$_{-0.09}^{+0.09}$	&	2.1$_{-0.03}^{+0.1}$	&	4.8$_{-0.3}^{+0.2}$	&		&	a,a,a,a,a,a  \\
F15	&	PDS 33	&	08:48:45.6	&	-40:48:21.0	&	9800$_{-200}^{+200}$	&	1.03$_{-0.02}^{+0.03}$	&	950$_{-40}^{+40}$	&	1.4$_{-0.09}^{+0.1}$	&	2.2$_{-0.2}^{+0.04}$	&	5$_{-0.3}^{+2}$	&	I	&	a,a,a,a,a,a  \\
F15	&	PDS 281	&	08:55:45.9	&	-44:25:14.1	&	16000$_{-2000}^{+2000}$	&	2.32$_{-0.09}^{+0.09}$	&	930$_{-40}^{+50}$	&	3.8$_{-0.2}^{+0.2}$	&	9$_{-2}^{+2}$	&	0.09$_{-0.04}^{+0.07}$	&	I	&	a,a,a,a,a,a  \\
F15	&	PDS 286	&	09:06:00.0	&	-47:18:58.1	&	30000$_{-3000}^{+3000}$	&	6.4$_{-0.6}^{+0.4}$	&	1800$_{-200}^{+200}$	&	5.4$_{-0.5}^{+0.4}$	&	40$_{-10}^{+20}$	&	0.01$_{-0.0}^{+0.01}$	&	II	&	a,a,a,a,a,a  \\
F15	&	PDS 297	&	09:42:40.3	&	-56:15:34.1	&	10800$_{-200}^{+200}$	&	1.23$_{-0.06}^{+0.06}$	&	1600$_{-100}^{+100}$	&	2.2$_{-0.1}^{+0.1}$	&	3.1$_{-0.3}^{+0.4}$	&	1.8$_{-0.5}^{+0.6}$	&		&	a,a,a,a,a,a  \\
F15	&	HD 87403	&	10:02:51.4	&	-59:16:54.6	&	10000$_{-200}^{+200}$	&	0.34$_{-0.05}^{+0.04}$	&	1900$_{-200}^{+300}$	&	3.0$_{-0.2}^{+0.2}$	&	6.2$_{-0.8}^{+1.0}$	&	0.2$_{-0.08}^{+0.1}$	&		&	a,a,a,a,a,a  \\
F15	&	PDS 37	&	10:10:00.3	&	-57:02:07.3	&	18000$_{-4000}^{+4000}$	&	5.6$_{-0.6}^{+0.4}$	&	1900$_{-200}^{+400}$	&	4.0$_{-0.8}^{+0.6}$	&	11$_{-5}^{+8}$	&	0.1$_{--0.1}^{+0.3}$	&	I	&	a,a,a,a,a,a  \\
F15	&	HD 305298	&	10:33:04.9	&	-60:19:51.3	&	34000$_{-1000}^{+1000}$	&	1.5$_{-0.1}^{+0.1}$	&	4000$_{-400}^{+600}$	&	4.7$_{-0.2}^{+0.2}$	&	19$_{-3}^{+3}$	&	0.03$_{-0.01}^{+0.03}$	&		&	a,a,a,a,a,a  \\
F15	&	HD 96042	&	11:03:40.5	&	-59:25:59.0	&	26000$_{-2000}^{+2000}$	&	1.13$_{-0.08}^{+0.08}$	&	3100$_{-400}^{+500}$	&	5.0$_{-0.2}^{+0.3}$	&	24$_{-5}^{+8}$	&	0.01$_{-0.01}^{+0.01}$	&	I	&	a,a,a,a,a,a  \\
F15	&	PDS 344	&	11:40:32.8	&	-64:32:05.7	&	15200$_{-500}^{+500}$	&	1.27$_{-0.07}^{+0.07}$	&	2400$_{-100}^{+200}$	&	2.4$_{-0.1}^{+0.2}$	&	3.8$_{-0.2}^{+0.3}$	&	1.3$_{-0.3}^{+0.5}$	&		&	a,a,a,a,a,a  \\
F15	&	PDS 361S	&	13:03:21.4	&	-62:13:26.2	&	18000$_{-1000}^{+1000}$	&	2.2$_{-0.2}^{+0.1}$	&	3000$_{-300}^{+400}$	&	3.3$_{-0.3}^{+0.3}$	&	6$_{-0.6}^{+1}$	&	0.3$_{-0.1}^{+0.1}$	&		&	a,a,a,a,a,a  \\
F15	&	HD 114981	&	13:14:40.6	&	-38:39:05.6	&	16000$_{-500}^{+500}$	&	0.15$_{-0.03}^{+0.02}$	&	700$_{-40}^{+60}$	&	3.3$_{-0.1}^{+0.1}$	&	6.5$_{-0.6}^{+0.7}$	&	0.25$_{-0.07}^{+0.09}$	&	II	&	a,a,a,a,a,a  \\
F15	&	PDS 364	&	13:20:03.5	&	-62:23:54.0	&	12000$_{-1000}^{+1000}$	&	2.0$_{-0.2}^{+0.2}$	&	2400$_{-400}^{+700}$	&	2.3$_{-0.3}^{+0.3}$	&	3.3$_{-0.5}^{+0.9}$	&	1.5$_{-0.7}^{+0.7}$	&		&	a,a,b,b,b,b  \\
F15	&	PDS 69	&	13:57:44.1	&	-39:58:44.2	&	15000$_{-2000}^{+2000}$	&	1.6$_{-0.2}^{+0.2}$	&	640$_{-30}^{+30}$	&	2.8$_{-0.4}^{+0.4}$	&	4$_{-0.7}^{+1}$	&	0.8$_{-0.4}^{+0.7}$	&	I	&	a,a,a,a,a,a  \\
F15	&	DG Cir	&	15:03:23.7	&	-63:22:58.8	&	11000$_{-3000}^{+3000}$	&	4.0$_{-0.3}^{+0.2}$	&	830$_{-40}^{+50}$	&	1.7$_{-0.8}^{+0.6}$	&	2.4$_{-0.8}^{+0.8}$	&	0$_{-2}^{+10}$	&	I	&	a,a,a,a,a,a  \\
F15	&	HD 132947	&	15:04:56.0	&	-63:07:52.6	&	10200$_{-200}^{+200}$	&	0.29$_{-0.02}^{+0.03}$	&	380$_{-10}^{+20}$	&	1.73$_{-0.08}^{+0.09}$	&	2.4$_{-0.04}^{+0.1}$	&	3.7$_{-0.5}^{+0.3}$	&	I	&	a,a,a,a,a,a  \\
F15	&	HD 135344B	&	15:15:48.4	&	-37:09:16.0	&	6400$_{-100}^{+100}$	&	0.5$_{-0.2}^{+0.2}$	&	136$_{-2}^{+2}$	&	0.9$_{-0.1}^{+0.1}$	&	1.7$_{-0.2}^{+0.2}$	&	6$_{-2}^{+2}$	&	I	&	a,a,a,a,a,a  \\
F15	&	HD 139614	&	15:40:46.3	&	-42:29:53.5	&	7800$_{-200}^{+200}$	&	0.21$_{-0.04}^{+0.04}$	&	135$_{-2}^{+2}$	&	0.86$_{-0.09}^{+0.08}$	&	1.6$_{-0.1}^{+0.02}$	&	10$_{-0.6}^{+10}$	&	I	&	a,a,a,a,a,a  \\
F15	&	HD 141926	&	15:54:21.7	&	-55:19:44.3	&	28000$_{-2000}^{+2000}$	&	2.6$_{-0.2}^{+0.2}$	&	1300$_{-100}^{+200}$	&	4.9$_{-0.3}^{+0.3}$	&	22$_{-5}^{+6}$	&	0.02$_{-0.01}^{+0.01}$	&	II	&	a,a,a,a,a,a  \\
F15	&	HD 142527	&	15:56:41.8	&	-42:19:23.2	&	6500$_{-200}^{+200}$	&	1.20$_{-0.05}^{+0.04}$	&	157$_{-2}^{+2}$	&	1.4$_{-0.1}^{+0.09}$	&	2.3$_{-0.1}^{+0.2}$	&	3.1$_{-0.6}^{+0.5}$	&	I	&	a,a,a,a,a,a  \\
F15	&	PDS 144S	&	15:49:15.3	&	-26:00:54.7	&	7800$_{-500}^{+500}$	&	1.02$_{-0.07}^{+0.07}$	&	150$_{-4}^{+5}$	&	-0.67$_{-0.06}^{+0.06}$	&		&		&		&	a,a,b,b,-,-  \\
F15	&	HD 141569	&	15:49:57.7	&	-03:55:16.3	&	9500$_{-200}^{+200}$	&	0.38$_{-0.03}^{+0.02}$	&	110.6$_{-0.9}^{+0.9}$	&	1.34$_{-0.07}^{+0.06}$	&	2.1$_{-0.2}^{+0.02}$	&	6$_{-0.6}^{+2}$	&	II	&	a,a,a,a,a,a  \\
Both	&	HD 142666	&	15:56:40.0	&	-22:01:40.0	&	7200$_{-200}^{+200}$	&	0.82$_{-0.08}^{+0.07}$	&	148$_{-2}^{+2}$	&	1.1$_{-0.1}^{+0.1}$	&	1.6$_{-0.1}^{+0.1}$	&	8$_{-1}^{+2}$	&	II	&	a,a,a,a,a,a  \\
Both	&	HD 144432	&	16:06:57.9	&	-27:43:09.7	&	7500$_{-200}^{+200}$	&	0.48$_{-0.05}^{+0.04}$	&	155$_{-2}^{+2}$	&	1.18$_{-0.09}^{+0.09}$	&	1.7$_{-0.1}^{+0.1}$	&	7$_{-1}^{+1}$	&	II	&	a,a,a,a,a,a  \\
F15	&	HD 144668	&	16:08:34.2	&	-39:06:18.3	&	8500$_{-200}^{+200}$	&	0.87$_{-0.03}^{+0.03}$	&	161$_{-3}^{+3}$	&	1.95$_{-0.08}^{+0.08}$	&	3.0$_{-0.2}^{+0.2}$	&	1.7$_{-0.2}^{+0.4}$	&	II	&	a,a,a,a,a,a  \\
F15	&	PDS 415N	&	16:18:37.2	&	-24:05:18.1	&	6200$_{-200}^{+200}$	&	1.5$_{-0.1}^{+0.1}$	&	144$_{-4}^{+5}$	&	0.4$_{-0.1}^{+0.2}$	&	1.2$_{-0.09}^{+0.2}$	&	13$_{-4}^{+5}$	&	I	&	a,a,b,b,b,b  \\
F15	&	AK Sco	&	16:54:44.8	&	-36:53:18.5	&	6200$_{-200}^{+200}$	&	0.51$_{-0.02}^{+0.03}$	&	141$_{-2}^{+2}$	&	0.82$_{-0.09}^{+0.09}$	&	1.6$_{-0.1}^{+0.1}$	&	7$_{-1}^{+2}$	&	II	&	a,a,a,a,a,a  \\
F15	&	PDS 431	&	16:54:59.1	&	-43:21:49.7	&	10500$_{-500}^{+500}$	&	2.1$_{-0.1}^{+0.1}$	&	1800$_{-100}^{+200}$	&	2.0$_{-0.2}^{+0.2}$	&	2.9$_{-0.4}^{+0.5}$	&	2$_{-0.8}^{+1}$	&		&	a,a,a,a,a,a  \\
F15	&	HD 145718	&	16:13:11.5	&	-22:29:06.6	&	7800$_{-200}^{+200}$	&	1.10$_{-0.06}^{+0.06}$	&	152$_{-3}^{+3}$	&	1.0$_{-0.1}^{+0.1}$	&	1.62$_{-0.03}^{+0.07}$	&	9$_{-1}^{+0.8}$	&	II	&	a,a,a,a,a,a  \\
F15	&	HD 150193	&	16:40:17.9	&	-23:53:45.1	&	9200$_{-200}^{+200}$	&	1.9$_{-0.2}^{+0.2}$	&	151$_{-2}^{+3}$	&	1.6$_{-0.2}^{+0.1}$	&	2.1$_{-0.1}^{+0.2}$	&	5$_{-1}^{+0.9}$	&	II	&	a,a,a,a,a,a  \\
F15	&	KK Oph	&	17:10:08.1	&	-27:15:18.8	&	8500$_{-500}^{+500}$	&	1.6$_{-0.1}^{+0.1}$	&	220$_{-10}^{+10}$	&	0.7$_{-0.1}^{+0.1}$	&	1.51$_{-0.08}^{+0.08}$	&	18$_{-1}^{+2}$	&	II	&	a,a,b,b,b,b  \\
Both	&	HD 163296	&	17:56:21.2	&	-21:57:21.8	&	9000$_{-200}^{+200}$	&	0.29$_{-0.02}^{+0.01}$	&	102$_{-2}^{+2}$	&	1.31$_{-0.07}^{+0.07}$	&	1.95$_{-0.07}^{+0.07}$	&	6.0$_{-0.3}^{+0.3}$	&	II	&	a,a,a,a,a,a  \\
F15	&	MWC 297	&	18:27:39.5	&	-03:49:52.1	&	24000$_{-2000}^{+2000}$	&	7.9$_{-0.6}^{+0.4}$	&	380$_{-20}^{+20}$	&	4.4$_{-0.5}^{+0.4}$	&	15$_{-5}^{+6}$	&	0.04$_{-0.02}^{+0.07}$	&	I	&	a,a,a,a,a,a  \\
DB11	&	VV Ser	&	18:28:47.8	&	+00:08:39.9	&	13800$_{-200}^{+200}$	&	2.9$_{-0.1}^{+0.1}$	&	420$_{-10}^{+10}$	&	2.0$_{-0.08}^{+0.1}$	&	2.9$_{-0.1}^{+0.1}$	&	3$_{-0.2}^{+8}$	&	II	&	b,b,b,b,b,b  \\
DB11	&	HD 179218	&	19:11:11.2	&	+15:47:15.6	&	9500$_{-200}^{+200}$	&	0.5$_{-0.3}^{+0.1}$	&	266$_{-5}^{+6}$	&	2.0$_{-0.1}^{+0.09}$	&	3.0$_{-0.3}^{+0.2}$	&	1.7$_{-0.3}^{+0.5}$	&	I	&	b,b,b,b,b,b  \\
DB11	&	V1686 Cyg	&	20:20:29.4	&	+21:41:28.4	&	6000$_{-100}^{+200}$	&	1.9$_{-0.2}^{+0.2}$	&	1100$_{-200}^{+300}$	&	1.5$_{-0.2}^{+0.3}$	&	2.9$_{-0.6}^{+0.7}$	&	1$_{-0.6}^{+1}$	&	Ia$^{c}$	&	b,b,b,b,b,b  \\ 
\enddata
\tablecomments{Objects in \citet{fairlamb15,fairlamb17} and \citet{donehew-brittain11} are denoted with ``F15'' and ``DB11'' in the first column, respectively. References (In order of appearance in the table): 
$^a$\cite{wichittanakom20}, $^b$\cite{vioque18}, $^c$\cite{juhasz10}.
We do not list references for the Group determinations, as these are primarily from this work. Group determinations from the literature are denoted with a superscript corresponding to the reference, as listed above.  
}
\end{deluxetable}   
\FloatBarrier
\end{longrotatetable}
\clearpage

\begin{longrotatetable}
\FloatBarrier 
\startlongtable
\begin{deluxetable*}{ccccccccccccccc}
\tablewidth{0pt}
\rotate
\tablecaption{\brgamma\ Line Properties of the IGRINS Sample\label{tab: results}}
\vspace{0.5cm}
\tablehead{
\colhead{Num.} & \colhead{Object Name} & \colhead{{\rm EW}$_{obs}$} &  \colhead{$\Delta m_{K}$} & \colhead{{\rm EW}$_{corr}$} & \colhead{{\rm F}$_{Br\gamma}$} & \colhead{{\rm L}$_{Br\gamma}$} & \colhead{{\rm L}$_{acc}$} & \colhead{$\dot{M}$} & \colhead{Br$\gamma$ Line} & \colhead{$m_{12}-m_{60}$} & \colhead{$L_{NIR}$} & \colhead{$L_{IR}$} & \colhead{Variability} \\ \colhead{} & 
\colhead{} & \colhead{(\angstrom)} & \colhead{(mag)} & \colhead{(\angstrom)} & \colhead{($10^{-14}$ erg/s/cm$^2$)} & \colhead{($10^{-4}$ \Lsun)} & \colhead{(\Lsun)} & \colhead{($10^{-6}$ $M_{\odot}/yr$)} & \colhead{Shape} & \colhead{(mag)} & \colhead{($L_{\odot}$)} & \colhead{($L_{\odot}$)} & \colhead{Source}
}
\startdata
1 & UX Ori & -2.8$\pm$0.3 & 2.34 & -3.6$\pm$0.3 & 24$\pm$2 & 8.0$\pm$0.8 & 3$\pm$2 & 0.09$\pm$0.07 & Double & 0.27 & 2.56 & 2.1 & Both   \\ 
2 & IRAS 05044-0325 & 0.6$\pm$0.4 & 0.94 & -1.5$\pm$0.5 & 4$\pm$1 & 1.8$\pm$0.6 & 0.4$\pm$0.4 & 0.004$\pm$0.005 & Single &   &   &   & F17   \\ 
3 & V1012 Ori & -0.4$\pm$0.4 & 2.47 & -1.2$\pm$0.4 & 1.6$\pm$0.5 & 0.7$\pm$0.3 & 0.1$\pm$0.1 & 0.004$\pm$0.005 & Single & 2.04 & 2.05 & 1.86 & F17   \\ 
4 & HD 35187 & -0.1$\pm$0.7 & 1.18 & -2.6$\pm$0.7 & 50$\pm$10 & 4$\pm$1 & 1$\pm$1 & 0.03$\pm$0.03 & Single & 0.68 & 2.48 & 1.78 & DB11   \\ 
5 & HD 287823 & 1.2$\pm$0.4 & 1.13 & -1.6$\pm$0.5 & 5$\pm$1 & 1.9$\pm$0.6 & 0.4$\pm$0.4 & 0.02$\pm$0.02 & Single & 1.7 & 2.39 & 2.01 & F17   \\ 
6 & HD 290409 & -1.8$\pm$0.6 & 1.01 & -4.6$\pm$0.7 & 7$\pm$1 & 4.7$\pm$1.0 & 1$\pm$1 & 0.04$\pm$0.03 & Absorption & 1.12 & 2.21 & 2.35 & F17   \\ 
7 & HD 35929 & 2.0$\pm$0.6 & 0.41 & -3.0$\pm$0.7 & 28$\pm$7 & 13$\pm$3 & 5$\pm$4 & 0.3$\pm$0.3 & Absorption & -0.63 & 2.95 & 1.12 & F17   \\ 
8 & HD 290500 & 0.2$\pm$0.6 & 0.57 & -4.0$\pm$0.7 & 2.8$\pm$0.5 & 1.7$\pm$0.4 & 0.4$\pm$0.4 & 0.010$\pm$0.009 & Double & 3.35 & 2.02 & 1.93 & F17   \\ 
9 & HD 244314 & -5.3$\pm$0.4 & 1.53 & -7.2$\pm$0.4 & 20$\pm$1 & 11$\pm$1 & 4$\pm$4 & 0.2$\pm$0.1 & Absorption & 0.7 & 2.47 & 1.8 & Both   \\ 
10 & HD 36112 & -1.2$\pm$0.5 & 1.84 & -2.7$\pm$0.5 & 50$\pm$10 & 4.4$\pm$0.9 & 1$\pm$1 & 0.05$\pm$0.04 & Double & 2.0 & 2.52 & 2.04 & DB11   \\ 
11 & V451 Ori & -7.8$\pm$0.6 & 0.13 & -13.7$\pm$0.9 & 7.4$\pm$0.5 & 10$\pm$30 & 10$\pm$10 &   & Double &   &   &   &     \\ 
12 & HK Ori & -4.2$\pm$0.4 & 2.98 & -4.6$\pm$0.4 & 26$\pm$3 & 16$\pm$6 & 7$\pm$6 & 0.2$\pm$0.2 & Double & -0.63 & 2.83 & 2.33 & Both   \\ 
13 & HD 244604 & -2.0$\pm$0.6 & 1.71 & -3.5$\pm$0.7 & 22$\pm$4 & 12$\pm$3 & 5$\pm$4 & 0.2$\pm$0.2 & Absorption & -0.2 & 2.73 & 2.03 & Both   \\ 
14 & UY Ori & -2.2$\pm$0.7 & 1.57 & -3.9$\pm$0.7 & 3.7$\pm$0.7 & 1.4$\pm$0.3 & 0.3$\pm$0.3 & 0.008$\pm$0.008 & Double & -0.19 & 1.88 & 2.04 & F17   \\ 
15 & RY Ori & 0.7$\pm$0.5 & 1.5 & -0.7$\pm$0.5 & 2$\pm$1 & 0.7$\pm$0.5 & 0.1$\pm$0.2 & 0.005$\pm$0.007 & Absorption & 1.54 & 2.23 & 1.66 &     \\ 
16 & HD 36408 & 5.0$\pm$0.3 & 0.61 & 1.3$\pm$0.5 & 23$\pm$9 & 14$\pm$6 & 5$\pm$5 & 0.2$\pm$0.2 & Double &   &   &   & DB11   \\ 
17 & Brun 216 & 1.3$\pm$0.4 & 0.89 & -1.0$\pm$0.5 & 3$\pm$1 & 1.3$\pm$0.6 & 0.2$\pm$0.3 & 0.01$\pm$0.01 & Double & 1.88 & 2.27 & 2.04 &     \\ 
18 & HD 245185 & -7.4$\pm$0.5 & 1.63 & -9.0$\pm$0.6 & 24$\pm$1 & 14$\pm$3 & 5$\pm$5 & 0.1$\pm$0.1 & Single & 0.52 & 2.49 & 2.51 & Both   \\ 
19 & MX Ori & -1.4$\pm$0.7 & 1.03 & -3.4$\pm$0.7 & 19$\pm$4 & 10$\pm$20 & 0$\pm$10 &   & Double & 7.86 &   &   &     \\ 
20 & NV Ori & 0.2$\pm$0.7 & 1.36 & -1.6$\pm$0.7 & 6$\pm$3 & 3$\pm$1 & 0.7$\pm$0.8 & 0.04$\pm$0.04 & Double &   &   &   &     \\ 
21 & T Ori & -1.7$\pm$0.6 & 2.15 & -2.7$\pm$0.7 & 34$\pm$8 & 18$\pm$4 & 8$\pm$6 & 0.3$\pm$0.3 & Double &   &   &   & F17   \\ 
22 & V380 Ori & -9.3$\pm$0.7 & 2.68 & -10.0$\pm$0.7 & 220$\pm$10 & 160$\pm$50 & 100$\pm$100 & 4$\pm$3 & Double & 2.37 & 3.46 & 3.06 & Both   \\ 
23 & HD 245465 & 3.5$\pm$0.6 & 0.39 & -1.0$\pm$0.8 & 2$\pm$2 & 1$\pm$2 & 0.1$\pm$0.4 & 0.01 & Single &   &   &   &     \\ 
24 & HD 37258 & -4.6$\pm$0.8 & 1.66 & -6.1$\pm$0.8 & 23$\pm$3 & 10$\pm$2 & 3$\pm$3 & 0.10$\pm$0.09 & Single & 1.5 & 2.46 & 1.88 & F17   \\ 
25 & HD 290770 & -8.3$\pm$0.4 & 2.03 & -9.4$\pm$0.4 & 59$\pm$3 & 29$\pm$3 & 10$\pm$10 & 0.4$\pm$0.3 & Absorption & 0.68 & 2.73 & 2.14 & F17   \\ 
26 & BF Ori & -1.6$\pm$0.4 & 1.4 & -3.7$\pm$0.4 & 11$\pm$1 & 5.0$\pm$0.7 & 1$\pm$1 & 0.05$\pm$0.05 & Double & 0.89 & 2.47 & 1.77 & Both   \\ 
27 & HD 37357 & -3.7$\pm$0.5 & 1.12 & -6.3$\pm$0.5 & 30$\pm$3 & 40$\pm$30 & 20$\pm$30 & 0.5$\pm$0.7 & Double & 0.9 & 3.08 & 2.5 & F17   \\ 
28 & HD 290764 & -0.9$\pm$0.5 & 1.42 & -3.0$\pm$0.5 & 14$\pm$3 & 7$\pm$1 & 2$\pm$2 & 0.10$\pm$0.09 & Absorption & 3.04 & 2.65 & 2.42 & F17   \\ 
29 & V1787 Ori & -1.8$\pm$0.4 & 1.94 & -3.1$\pm$0.4 & 13$\pm$2 & 6.0$\pm$0.9 & 2$\pm$2 & 0.07$\pm$0.06 & Double & 1.05 & 2.67 & 2.18 &     \\ 
30 & HD 37411 & -2.7$\pm$0.4 & 1.78 & -4.0$\pm$0.4 & 17$\pm$2 & 7$\pm$4 & 2$\pm$2 & 0.05$\pm$0.06 & Double & 1.04 & 2.5 & 1.98 & F17   \\ 
31 & V599 Ori & -1.2$\pm$0.4 & 1.46 & -3.2$\pm$0.4 & 22$\pm$3 & 11$\pm$2 & 4$\pm$4 & 0.2$\pm$0.2 & Double & 2.92 & 2.8 & 2.45 & F17   \\ 
32 & HD 245906 & -1.0$\pm$0.4 & 1.6 & -2.8$\pm$0.4 & 8$\pm$1 & 10$\pm$10 & 5$\pm$8 & 0.2$\pm$0.4 & Double & 1.93 & 2.88 & 2.28 &     \\ 
33 & RR Tau & -2.3$\pm$0.4 & 2.11 & -3.3$\pm$0.5 & 18$\pm$2 & 34$\pm$6 & 20$\pm$10 & 0.7$\pm$0.5 & Single & 1.49 & 3.3 & 2.62 &     \\ 
34 & V350 Ori & -1.8$\pm$0.4 & 2.56 & -2.5$\pm$0.4 & 5.2$\pm$0.9 & 2.5$\pm$0.6 & 0.6$\pm$0.6 & 0.02$\pm$0.02 & Double & 0.73 & 2.24 & 1.8 & Both   \\ 
35 & HD 37806 & -6.8$\pm$0.8 & 2.39 & -7.5$\pm$0.8 & 200$\pm$20 & 110$\pm$20 & 90$\pm$60 & 3$\pm$2 & Other & -0.55 & 3.44 & 2.62 & DB11   \\ 
36 & HD 38120 & -9.5$\pm$0.8 & 1.85 & -10.7$\pm$0.8 & 63$\pm$5 & 32$\pm$5 & 20$\pm$10 & 0.5$\pm$0.4 & Double & 0.79 & 2.69 & 2.91 & DB11   \\ 
37 & HD 38238 & 0.8$\pm$0.5 & 1.53 & -1.2$\pm$0.6 & 9$\pm$4 & 3$\pm$2 & 0.8$\pm$0.9 & 0.04$\pm$0.04 & Absorption & 3.67 & 2.76 & 2.39 &     \\ 
38 & MWC 778 & -6.4$\pm$0.5 & 3.34 & -6.7$\pm$0.5 & 29$\pm$2 & 0$\pm$1000 & 0$\pm$1000 &   & Single & 2.89 &   &   &     \\ 
39 & V1818 Ori & -1.5$\pm$0.3 & 2.17 & -2.3$\pm$0.3 & 61$\pm$9 & 90$\pm$30 & 60$\pm$50 & 2$\pm$2 & Double & 0.34 & 4.01 & 3.6 &     \\ 
40 & HD 249879 & -7.2$\pm$0.5 & 1.51 & -8.8$\pm$0.5 & 8.7$\pm$0.5 & 12$\pm$2 & 5$\pm$4 & 0.10$\pm$0.09 & Double & 0.03 & 2.58 & 2.24 & DB11   \\ 
41 & HD 250550 & -6.0$\pm$0.5 & 1.51 & -7.7$\pm$0.6 & 76$\pm$6 & 120$\pm$30 & 90$\pm$70 & 5$\pm$4 & Single & 0.77 & 3.47 & 3.14 & Both   \\ 
42 & V791 Mon & -12.8$\pm$0.4 & 2.03 & -13.7$\pm$0.4 & 70$\pm$2 & 170$\pm$20 & 150$\pm$100 & 4$\pm$3 & Single & -0.07 & 3.33 & 3.14 & F17   \\ 
43 & AE Lep & -6.8$\pm$0.4 & 1.61 & -8.5$\pm$0.4 & 7.4$\pm$0.4 & 17$\pm$6 & 7$\pm$6 & 0.3$\pm$0.3 & Absorption & 0.91 & 2.66 & 3.01 &     \\ 
44 & LkHa 208 & -1.8$\pm$0.5 & 4.92 & -1.8$\pm$0.5 & 2.2$\pm$0.6 & 3$\pm$2 & 0.7$\pm$0.8 & 0.02$\pm$0.03 & Absorption & 0.72 & 2.43 & 2.75 &     \\ 
45 & IRAS 06071+2925 & -17.8$\pm$0.7 & 1.5 & -19.5$\pm$0.7 & 13.3$\pm$0.5 & 48$\pm$6 & 30$\pm$20 & 0.8$\pm$0.6 & Double & 0.18 & 2.67 & 2.72 &     \\ 
46 & LkHa 338 & -9.2$\pm$0.4 & 3.89 & -9.4$\pm$0.4 & 18.1$\pm$0.9 & 44$\pm$7 & 30$\pm$20 & 0.5$\pm$0.3 & Double &   &   &   &     \\ 
47 & LkHa 339 & -4.8$\pm$0.6 & 1.28 & -6.8$\pm$0.7 & 5.0$\pm$0.5 & 11$\pm$1 & 4$\pm$4 & 0.1$\pm$0.1 & Double & 7.17 & 2.64 & 4.45 & F17   \\ 
48 & MWC 137 & -32.0$\pm$0.3 & 2.18 & -32.6$\pm$0.3 & 628$\pm$6 & 17000$\pm$7000 & 60000$\pm$40000 & 900$\pm$800 & Single &   &   &   &     \\ 
49 & HD 45677 & -21.2$\pm$0.4 & 3.36 & -21.5$\pm$0.4 & 1310$\pm$20 & 1600$\pm$200 & 3000$\pm$2000 & 60$\pm$40 & Double & -1.94 & 4.61 & 4.12 &     \\ 
50 & HD 46060 & 3.8$\pm$0.6 & 0.26 & 2.6$\pm$0.6 & 8$\pm$2 & 21$\pm$6 & 10$\pm$8 & 0.2$\pm$0.2 & Absorption & 1.05 & 3.34 & 2.73 &     \\ 
51 & LkHa 341 & -3.8$\pm$0.6 & 1.92 & -5.0$\pm$0.6 & 18$\pm$2 & 100$\pm$200 & 100$\pm$200 & 2.96 & Absorption &   &   &   &     \\ 
52 & VY Mon & -3.5$\pm$0.4 & 2.82 & -3.9$\pm$0.4 & 200$\pm$20 & 1000$\pm$1000 & 2000$\pm$3000 & 30$\pm$90 & Absorption & 1.0 & 5.04 & 4.75 & F17   \\ 
53 & LkHa 215 & -5.5$\pm$0.4 & 2.28 & -6.2$\pm$0.4 & 54$\pm$3 & 80$\pm$10 & 60$\pm$40 & 2$\pm$1 & Single & 3.66 & 3.42 & 3.66 & DB11   \\ 
54 & HD 259431 & -6.4$\pm$0.4 & 2.46 & -7.0$\pm$0.4 & 190$\pm$10 & 300$\pm$40 & 300$\pm$200 & 10$\pm$8 & Double & 2.54 & 3.92 & 3.69 & DB11   \\ 
55 & R Mon & -1.7$\pm$0.3 & 2.09 & -2.6$\pm$0.3 & 39$\pm$5 & 80$\pm$30 & 50$\pm$50 & 2$\pm$2 & Single & 1.08 & 4.46 & 4.4 & F17   \\ 
56 & V590 Mon & -8.7$\pm$0.6 & 3.33 & -9.0$\pm$0.6 & 13.0$\pm$0.9 & 30$\pm$10 & 10$\pm$10 & 0.4$\pm$0.4 & Absorption & -0.24 & 2.59 & 2.94 & F17   \\ 
57 & HBC 222 & 1.5$\pm$0.6 & 0.97 & -0.7$\pm$0.6 & 0.4$\pm$0.4 & 0.7$\pm$0.6 & 0.1$\pm$0.2 & 0.005$\pm$0.008 & Double &   &   &   &     \\ 
58 & HD 50138 & -11.7$\pm$0.3 & 2.36 & -12.6$\pm$0.4 & 1080$\pm$30 & 490$\pm$40 & 600$\pm$300 & 30$\pm$20 & Absorption & -1.66 & 3.91 & 3.34 & DB11   \\ 
59 & HD 52721 & -6.6$\pm$0.5 & 0.39 & -7.6$\pm$0.6 & 140$\pm$10 & 120$\pm$70 & 90$\pm$100 & 2$\pm$2 & Double & 5.06 & 3.54 & 2.64 & F17   \\ 
60 & LkHa 218 & -7.5$\pm$0.6 & 1.97 & -8.6$\pm$0.6 & 15$\pm$1 & 61$\pm$10 & 40$\pm$30 & 1$\pm$1 & Double & 4.65 & 3.17 & 3.55 & F17   \\ 
61 & LkHa 220 & -13.9$\pm$0.4 & 1.86 & -15.0$\pm$0.4 & 16.4$\pm$0.5 & 70$\pm$10 & 50$\pm$30 & 1.2$\pm$0.9 & Double & 1.36 & 2.97 & 2.83 & F17   \\ 
62 & HD 53367 & -7.2$\pm$0.4 & 0.79 & -8.9$\pm$0.5 & 400$\pm$20 & 21$\pm$10 & 9$\pm$9 & 0.1$\pm$0.1 & Double & 5.07 & 2.48 & 1.86 & Both   \\ 
63 & RAFGL 5223 & -2.1$\pm$0.5 & 0.06 & -3.4$\pm$0.5 & 2.4$\pm$0.3 & 60$\pm$30 & 40$\pm$40 & 0.6$\pm$0.6 & Single &   &   &   & F17   \\ 
64 & NX Pup & -3.0$\pm$0.4 & 3.06 & -3.4$\pm$0.4 & 54$\pm$6 & 500$\pm$400 & 500$\pm$600 & 10$\pm$10 & Single & 1.14 & 4.49 & 3.7 & F17   \\ 
65 & DW Cma & -5.2$\pm$0.3 & 3.32 & -5.5$\pm$0.3 & 106$\pm$7 & 2100$\pm$800 & 4000$\pm$3000 & 100$\pm$100 & Absorption & 0.95 & 4.98 & 4.87 & F17   \\ 
66 & HD 58647 & -3.9$\pm$0.4 & 1.16 & -6.3$\pm$0.5 & 180$\pm$10 & 57$\pm$5 & 40$\pm$30 & 1$\pm$1 & Double & -2.34 & 3.32 & 1.71 &     \\ 
67 & V388 Vel & -13.5$\pm$0.7 & 0.88 & -16.8$\pm$0.8 & 4.8$\pm$0.2 & 90$\pm$40 & 60$\pm$60 & 3$\pm$3 & Single &   &   &   &     \\ 
68 & HD 76534 & -12.4$\pm$0.5 & 0.14 & -17.3$\pm$0.7 & 58$\pm$2 & 150$\pm$20 & 120$\pm$80 & 3$\pm$2 & Double & 6.04 & 3.39 & 2.94 & F17   \\ 
69 & HD 85567 & -4.9$\pm$0.5 & 2.2 & -5.8$\pm$0.5 & 130$\pm$10 & 430$\pm$60 & 500$\pm$300 & 20$\pm$10 & Double & -1.62 & 4.17 & 3.17 & F17   \\ 
70 & HD 94509 & -6.1$\pm$0.4 & 0.04 & -12.7$\pm$0.8 & 14.9$\pm$0.9 & 160$\pm$40 & 130$\pm$90 & 6$\pm$5 & Double &   &   &   & F17   \\ 
71 & HD 95881 & -2.5$\pm$0.4 & 1.91 & -3.8$\pm$0.4 & 86$\pm$9 & 370$\pm$60 & 400$\pm$300 & 20$\pm$10 & Other & -1.59 & 4.3 & 3.26 & F17   \\ 
72 & HD 97048 & -8.8$\pm$0.4 & 1.41 & -10.6$\pm$0.5 & 215$\pm$10 & 23$\pm$1 & 11$\pm$8 & 0.3$\pm$0.3 & Double & 2.25 & 2.69 & 2.62 & F17   \\ 
73 & HD 98922 & -4.3$\pm$0.4 & 2.2 & -5.2$\pm$0.4 & 450$\pm$30 & 660$\pm$70 & 900$\pm$500 & 50$\pm$30 & Double & -1.75 & 4.4 & 3.47 & F17   \\ 
74 & HD 100453 & 1.1$\pm$0.4 & 1.35 & -1.0$\pm$0.4 & 20$\pm$10 & 0.8$\pm$0.4 & 0.1$\pm$0.2 & 0.005$\pm$0.006 & Double & 1.78 & 2.18 & 2.08 & F17   \\ 
75 & HD 100546 & -7.7$\pm$0.5 & 1.14 & -10.3$\pm$0.6 & 290$\pm$20 & 11.0$\pm$0.6 & 4$\pm$3 & 0.11$\pm$0.09 & Absorption & 1.39 & 2.39 & 2.93 & F17   \\ 
76 & HD 101412 & -1.1$\pm$0.5 & 1.23 & -3.4$\pm$0.5 & 16$\pm$2 & 8$\pm$1 & 3$\pm$2 & 0.10$\pm$0.09 & Single & -0.32 & 2.77 & 2.09 & F17   \\ 
77 & HD 104237 & -5.9$\pm$0.7 & 1.34 & -8.2$\pm$0.7 & 510$\pm$50 & 19$\pm$2 & 8$\pm$6 & 0.4$\pm$0.3 & Absorption & 0.18 & 2.61 & 1.79 & F17   \\ 
78 & Hen 3-847 & -12.1$\pm$0.5 & 2.37 & -12.8$\pm$0.5 & 42$\pm$2 & 80$\pm$60 & 50$\pm$60 & 3$\pm$3 & Absorption & -1.13 & 3.63 & 3.91 & F17   \\ 
79 & CQ Uma & 5.8$\pm$0.4 & 0.08 & -1.0$\pm$0.8 & 10$\pm$10 & 0.4$\pm$0.8 & 0.0$\pm$0.1 &   & Single & 0.03 &   &   &     \\ 
80 & PX Vul & -0.8$\pm$0.6 & 1.62 & -2.2$\pm$0.6 & 7$\pm$2 & 9$\pm$3 & 3$\pm$3 & 0.2$\pm$0.2 & Single & 0.13 & 2.88 & 1.94 & DB11   \\ 
81 & IRAS 19343+2926 & -2.2$\pm$0.4 & 3.76 & -2.2$\pm$0.4 & 39$\pm$7 & 60$\pm$60 & 40$\pm$50 & 0.3$\pm$0.5 & Absorption & 1.98 & 3.93 & 4.01 &     \\ 
82 & MWC 342 & -17.4$\pm$0.3 & 2.61 & -17.7$\pm$0.3 & 1310$\pm$20 & 13000$\pm$2000 & 40000$\pm$20000 & 800$\pm$700 & Double & -1.99 & 5.36 & 4.62 &     \\ 
83 & BD +41 3731 & 3.8$\pm$0.7 & 0.24 & -0.3$\pm$0.8 & 0.2$\pm$0.5 & 1$\pm$2 & 0.1$\pm$0.4 & 0.003$\pm$0.008 & Absorption &   &   &   &     \\ 
84 & LkHa 131 & -10.2$\pm$0.6 & 1.21 & -12.0$\pm$0.6 & 5.3$\pm$0.3 & 20$\pm$40 & 10$\pm$20 &   & Single &   &   &   &     \\ 
85 & V2018 Cyg & -19.8$\pm$0.7 & 1.33 & -21.7$\pm$0.7 & 14.6$\pm$0.5 & 20$\pm$50 & 10$\pm$30 &   & Other &   &   &   &     \\ 
86 & V517 Cyg & -0.7$\pm$0.6 & 0.9 & -3.9$\pm$0.7 & 3.5$\pm$0.6 & 10$\pm$10 & 2$\pm$6 &   & Absorption &   &   &   &     \\ 
87 & V1977 Cyg & -3.2$\pm$0.4 & 2.55 & -3.8$\pm$0.4 & 52$\pm$6 & 120$\pm$20 & 90$\pm$60 & 4$\pm$2 & Double &   &   &   &     \\ 
88 & LkHa 134 & -12.5$\pm$0.4 & 1.42 & -14.4$\pm$0.5 & 51$\pm$2 & 110$\pm$10 & 80$\pm$60 & 3$\pm$2 & Absorption &   &   &   &     \\ 
89 & LkHa 135 & -5.6$\pm$0.5 & 3.1 & -6.0$\pm$0.5 & 72$\pm$6 & 200$\pm$300 & 100$\pm$400 &   & Single &   &   &   &     \\ 
90 & LkHa 147 & -9.4$\pm$0.6 & 0.74 & -10.1$\pm$0.6 & 13.0$\pm$0.7 & 170$\pm$40 & 100$\pm$100 & 4$\pm$3 & Double &   &   &   &     \\ 
91 & LkHa 167 & -2.1$\pm$0.5 & 1.93 & -3.4$\pm$0.5 & 15$\pm$2 & 60$\pm$20 & 40$\pm$30 & 2$\pm$2 & Single &   &   &   &     \\ 
92 & LkHa 168 & -1.5$\pm$0.5 & 1.53 & -2.9$\pm$0.6 & 8$\pm$1 & 50$\pm$30 & 30$\pm$30 & 2$\pm$2 & Double &   &   &   &     \\ 
93 & LkHa 169 & 4.4$\pm$0.6 & 0.34 & 0.7$\pm$0.7 & 1$\pm$1 & 3$\pm$7 & 1$\pm$2 &   & Double &   &   &   &     \\ 
94 & LkHa 176 & -42.1$\pm$0.6 & 0.07 & -47.3$\pm$0.8 & 38.3$\pm$0.7 & 100$\pm$200 & 100$\pm$100 &   & Double &   &   &   &     \\ 
95 & LkHa 183 & -17.5$\pm$0.6 & 1.49 & -18.6$\pm$0.6 & 12.3$\pm$0.4 & 400$\pm$900 & 0$\pm$1000 &   & Single &   &   &   &     \\ 
96 & LkHa 192 & -20.2$\pm$0.9 & 1.31 & -21.8$\pm$0.9 & 44$\pm$2 & 1000$\pm$2000 & 1000$\pm$3000 &   & Single &   &   &   &     \\ 
97 & LkHa 324 & -2.8$\pm$0.7 & 0.12 & -8.5$\pm$0.9 & 8.3$\pm$0.9 & 9$\pm$1 & 3$\pm$3 & 0.10$\pm$0.09 & Single &   &   &   &     \\ 
98 & V1578 Cyg & -3.4$\pm$0.5 & 2.2 & -4.3$\pm$0.5 & 42$\pm$5 & 80$\pm$10 & 50$\pm$40 & 2$\pm$2 & Double & 0.77 & 3.49 & 2.44 &     \\ 
99 & LkHa 257 & -0.4$\pm$0.7 & 1.4 & -2.4$\pm$0.7 & 1.7$\pm$0.5 & 3$\pm$1 & 0.9$\pm$0.9 & 0.03$\pm$0.03 & Double & 1.66 & 2.34 & 2.08 &     \\ 
100 & BO Cep & 1.9$\pm$0.5 & 0.97 & -0.7$\pm$0.6 & 0.4$\pm$0.4 & 0.2$\pm$0.2 & 0.02$\pm$0.03 & 0.001$\pm$0.001 & Single & 3.58 & 1.72 & 1.27 &     \\ 
101 & LkHa 350 & -10.6$\pm$0.5 & 1.92 & -11.7$\pm$0.5 & 61$\pm$3 & 1000$\pm$3000 & 2000$\pm$5000 &   & Double &   &   &   &     \\ 
102 & V374 Cep & -4.0$\pm$0.6 & 1.41 & -5.6$\pm$0.6 & 51$\pm$5 & 120$\pm$20 & 90$\pm$60 & 3$\pm$2 & Single &   &   &   &     \\ 
\enddata
\tablecomments{Objects that over lap with \citet{fairlamb15,fairlamb17} or \citet{donehew-brittain11} are denoted with ``F17'' and ``DB11'' in the Variability Source column, respectively. If the object is in both, it is denoted ``Both''. Objects with unknown uncertainties on \mstar\ or \rstar\ have \Mdot\ values listed without uncertainties.}
\end{deluxetable*}
\FloatBarrier
\end{longrotatetable}
\clearpage

\section{Results}\label{sec: results}

\subsection{Trends with System Properties}\label{subsec: trends}
Once we have determined EW(\brgamma), \Lbrg, and \Lacc, we test for trends between the accretion properties and system properties such as stellar effective temperature, stellar luminosity, stellar mass, SED shape (based on the Meeus group, \citealt{meeus01}), and age.

\begin{figure*}
    \centering
    \includegraphics[scale=0.45]{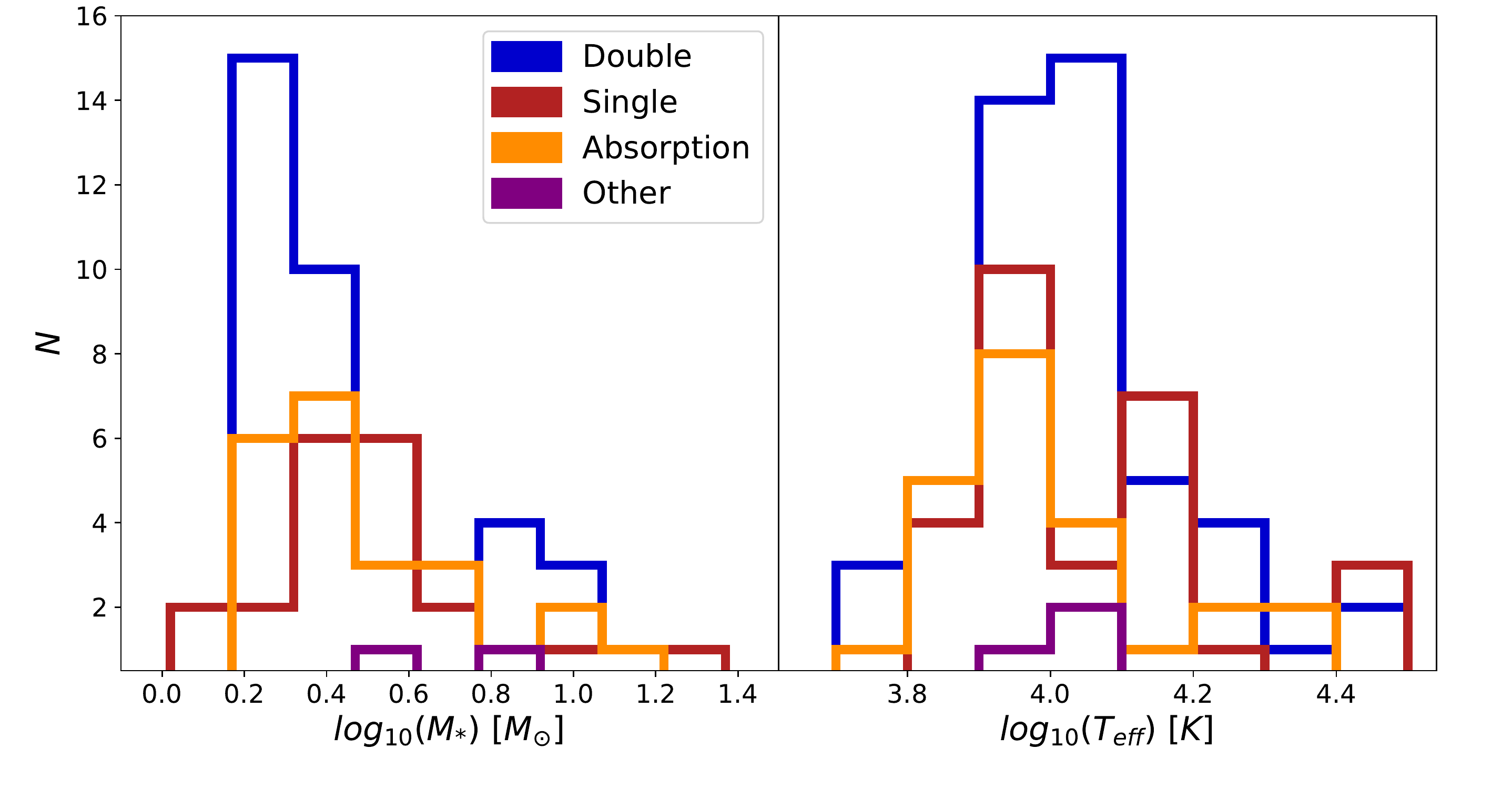}
    \caption{The distribution of stellar masses (left) and effective temperatures (right) for the four categories of the \brgamma\ line profile. We find no dependence on the line profile with stellar mass or effective temperature, and thus stellar spectral type.}
   \label{fig: line shape}
\end{figure*}

Figure~\ref{fig: line shape} shows the four \brgamma\ line shape categories found in the IGRINS data as a function of stellar mass and effective temperature. We find no trend of line profile with the mass or effective temperature, and thus the spectral type. We discuss this result in more detail in Section~\ref{subsect: variability results}. 

\begin{figure}
    \centering
    \includegraphics[scale=0.45]{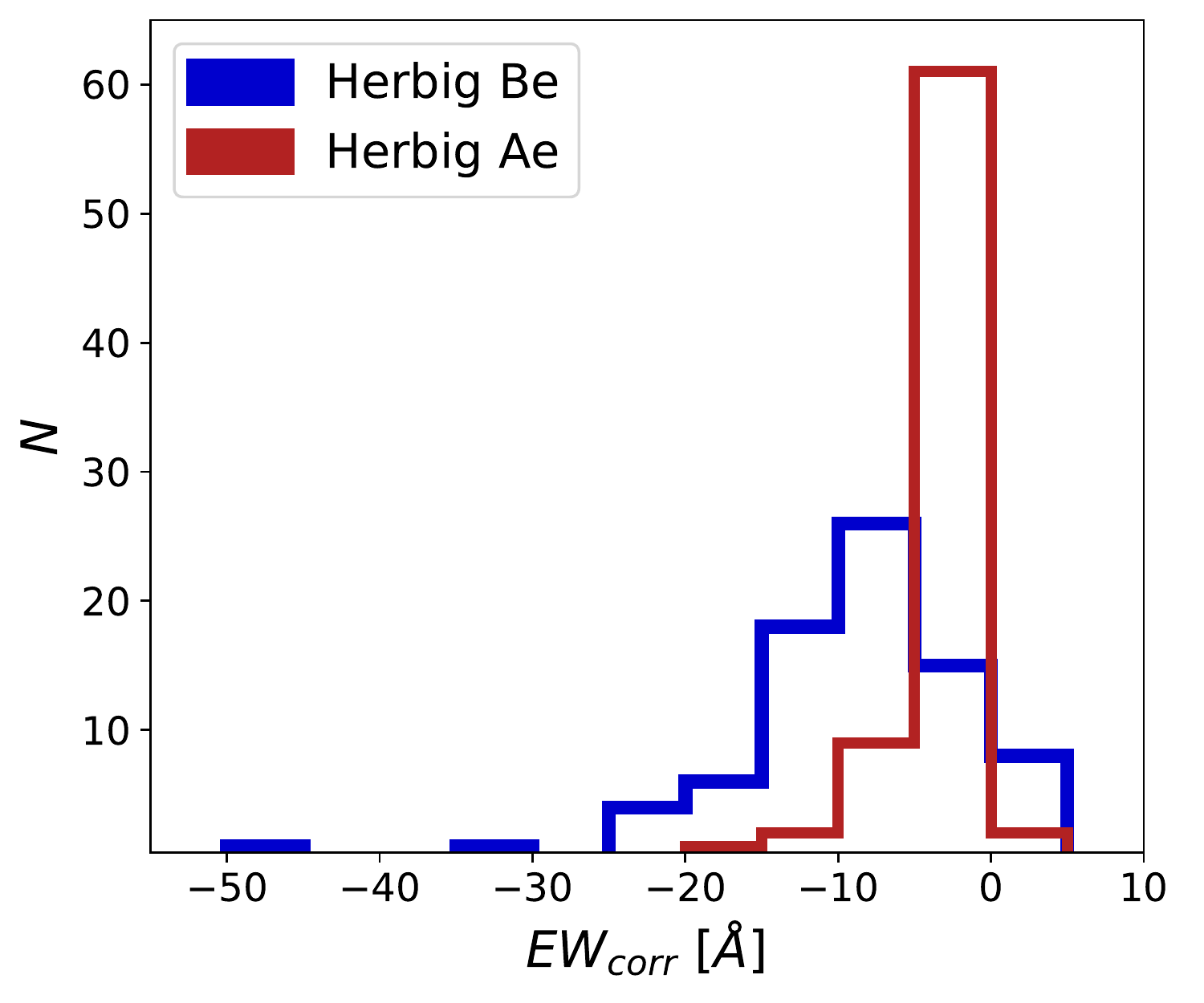}
    \caption{Equivalent width (EW, absorption is positive, emission is negative) of the \brgamma\ line in Herbig Be (blue) and Herbig Ae (red) stars after correcting for photospheric absorption for the IGRINS sample and the Auxiliary sample from \citet{donehew-brittain11} and \citet{fairlamb17}. a two-sample Kolmogorov-Smirnov test results in a p-value of 2.14$\times$10$^{-10}$, indicating that the distributions are statistically different.}
   \label{fig: EW}
\end{figure}

Figure~\ref{fig: EW} shows that the HBe stars peak at more negative corrected EWs (i.e., showing more emission relative to the photosphere) than the HAe stars. We perform a two-sample Kolmogorov-Smirnov (KS) test between the two distributions using the \texttt{scipy.stats.ks\_2samp} function in python \citep{scipy}. This results in a p-value of 2.14$\times$10$^{-10}$, indicating that the distributions are statistically different. They also show a greater spread in the measured EWs. This is also seen for H$\alpha$ EWs by \citet{vioque18}.

\begin{figure}
    \centering
    \includegraphics[scale=0.45]{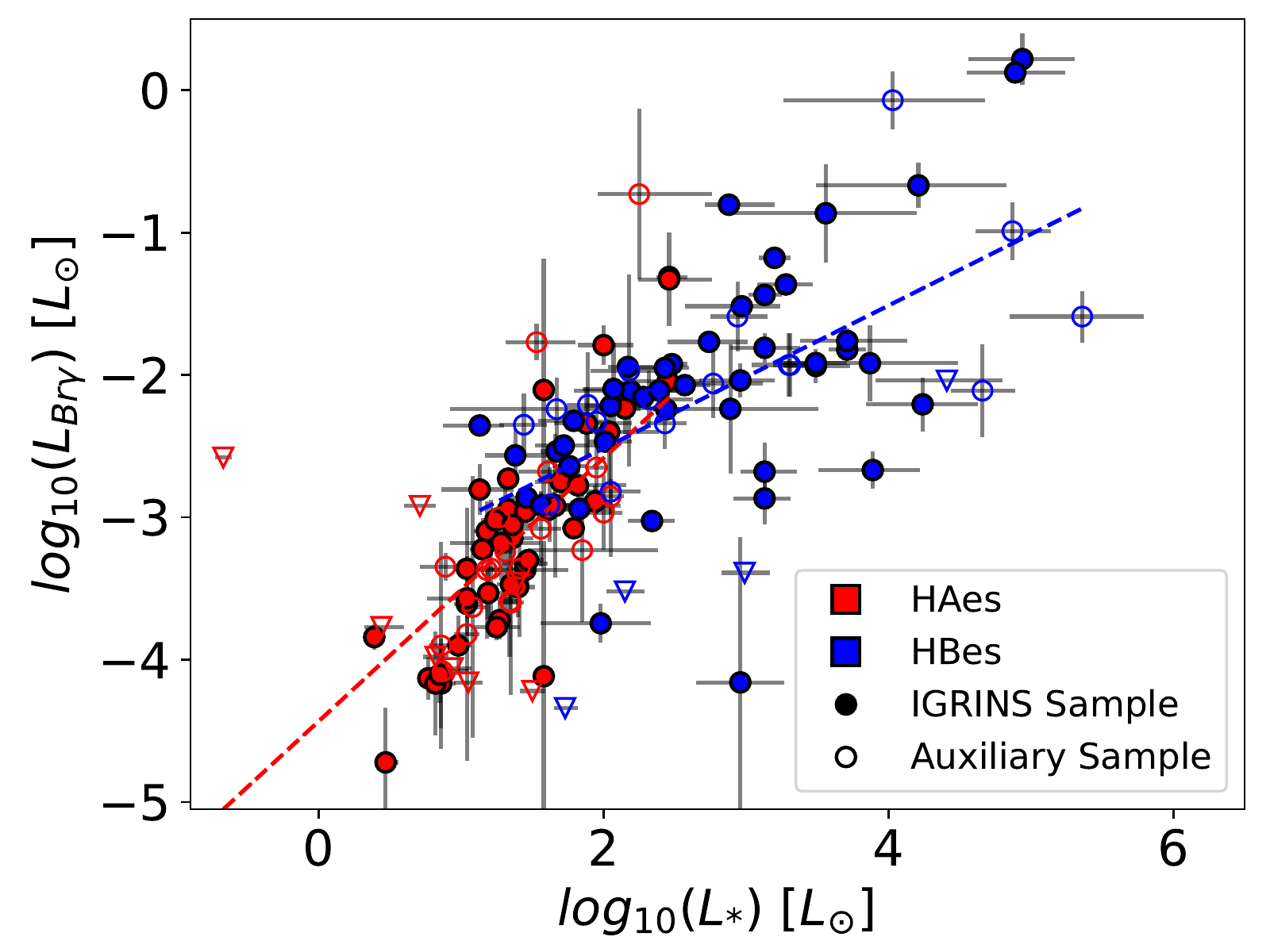}
    \caption{Log$_{10}$(\Lbrg)\ vs.\ log$_{10}$(L$_{*}$) for Herbig Be (blue) and Herbig Ae (red) stars for the IGRINS Sample (filled) and the Auxiliary Sample (\citealt{donehew-brittain11} and \citealt{fairlamb17}; open). Upper limits are shown as downward facing triangles. Uncertainties on \Lstar\ are shown for all objects that have these uncertainties listed in the literature. A linear fit to the HAe data is shown as the red line and a linear fit to the HBe data is shown as the blue line. The \Lbrg-\Lstar\ relationship shallows for the HBes. }
    \label{fig: Lbrg vs L}
\end{figure}

We find a trend between \Lbrg\ and \Lstar\ (Figure~\ref{fig: Lbrg vs L}). We use the \texttt{scikitlearn LinearRegression} function \citep{scikit-learn} to determine best fits to the data in log space. The slope of the IGRINS sample as a whole is 0.662, the fit to the HBes alone (blue line in Figure~\ref{fig: Lbrg vs L}) is more shallow with a slope of 0.502, and the fit to the HAes alone (red line in Figure~\ref{fig: Lbrg vs L}) is steeper with a slope of 0.924. The \Lbrg-\Lstar relationship corresponds to a dependence of \Lacc\ on \Lstar. \citet{mendigutia15} find that \Lbrg\ and \Lacc\ are correlated with L$_{*}$ for all near-UV, optical, and NIR lines. They recommend using \Lbrg/L$_{*}$ or \Lacc/L$_{*}$ to look for trends that will be unbiased by stellar luminosity. Based on this finding, we perform the majority of our analysis using \LaccL\ to mitigate relationships based on the stellar luminosity. However, for comparison with other studies that did not divide \Lacc\ by \Lstar, also include analysis just based on the accretion luminosity. 

\begin{figure}
    \centering
    \includegraphics[scale=0.45]{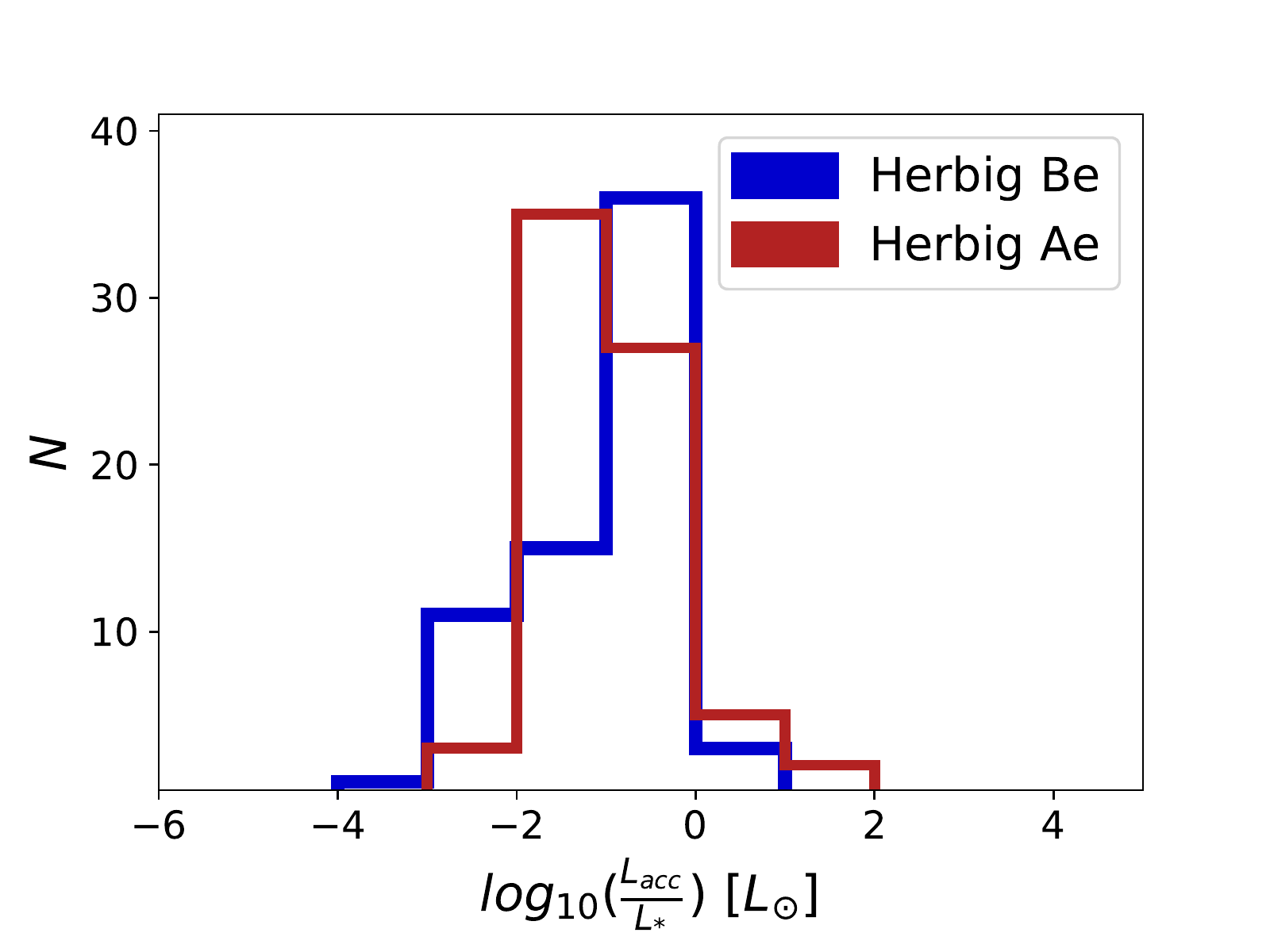}
    \caption{\Lacc/L$_{*}$ for Herbig Be (blue) and Herbig Ae (red) stars based on the accretion-tracing \brgamma\ line for the IGRINS Sample and the Auxiliary Sample (\citealt{donehew-brittain11} and \citealt{fairlamb17}). A two-sample Kolmogorov-Smirnov test returns a p-value of 0.008, indicating that the distributions may be drawn from different distributions.}
    \label{fig: lacc A vs B}
\end{figure}

\begin{figure}
    \centering
    \includegraphics[scale=0.45]{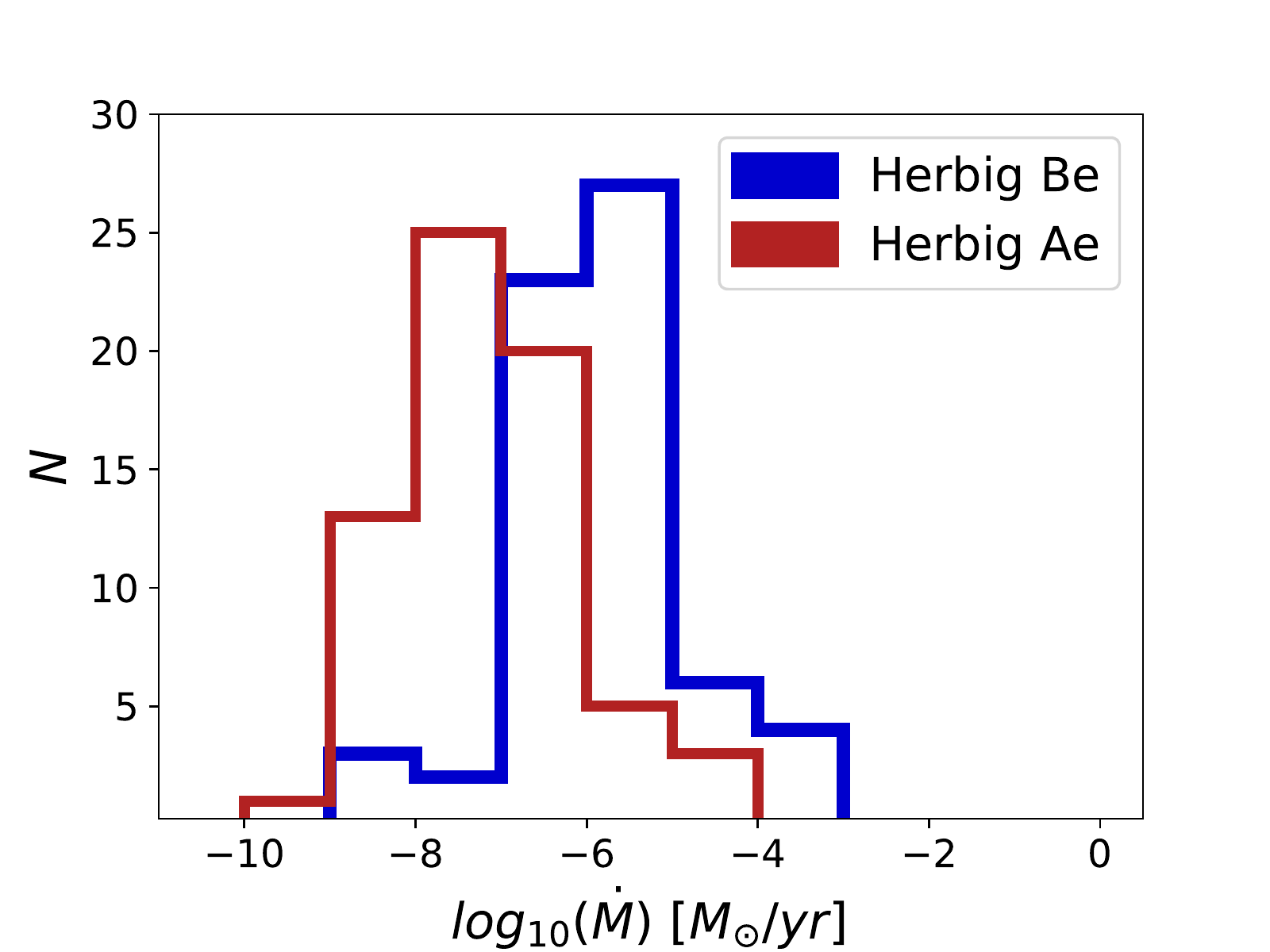}
    \caption{\Mdot\ for Herbig Be (blue) and Herbig Ae (red) stars based on the accretion-tracing \brgamma\ line for the IGRINS Sample and the Auxiliary Sample (\citealt{donehew-brittain11} and \citealt{fairlamb17}). A two-sample Kolmogorov-Smirnov test returns a p-value of 9.446$\times$10$^{-12}$, indicating that the distributions are statistically different.}
    \label{fig: mdot A vs B}
\end{figure}

\begin{figure*}
    \centering
    \includegraphics[scale=0.55]{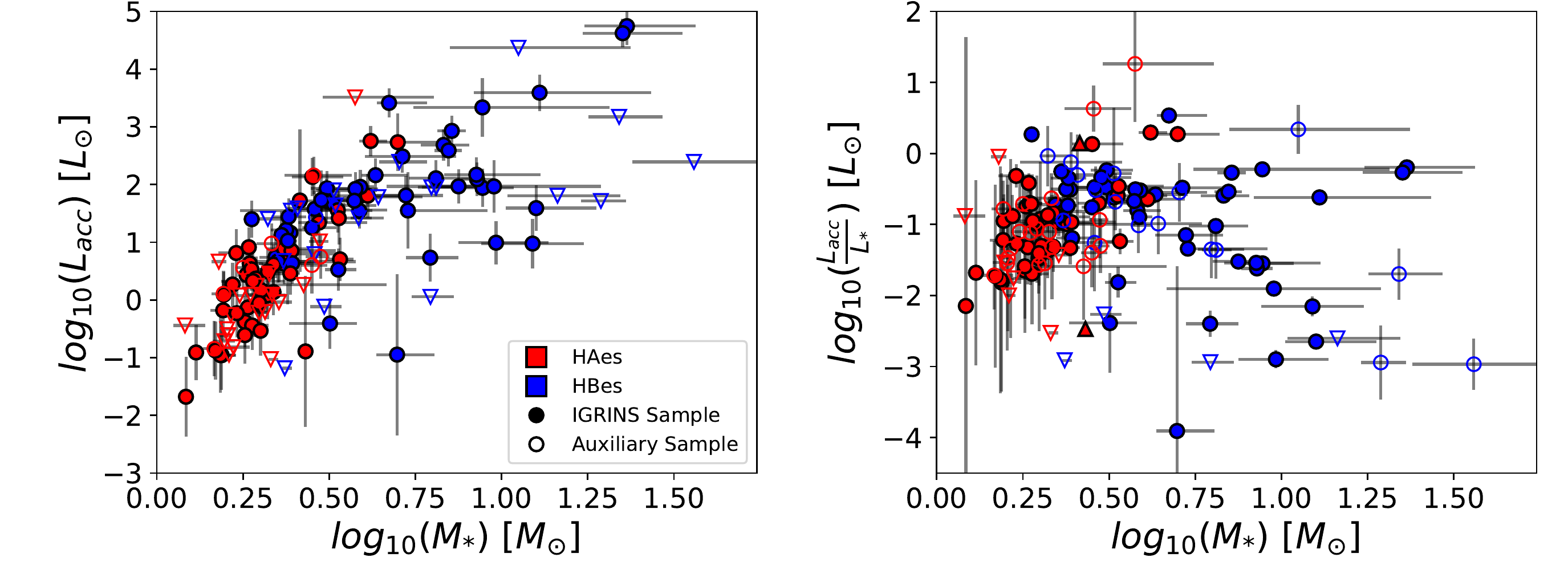}
    \caption{Left: \Lacc\ vs. stellar mass for Herbig Be (blue) and Herbig Ae (red) stars for the IGRINS Sample (filled) and the Auxiliary Sample (\citealt{donehew-brittain11} and \citealt{fairlamb17}; open). The accretion luminosity (not normalized by the stellar luminosity) increases with increasing stellar mass. Right: \Lacc/L$_{*}$ vs. stellar mass for the same samples. When dividing \Lacc\ by the stellar luminosity, the trend with stellar mass disappears.}
    \label{fig: mass lacc}
\end{figure*}

We look for trends between accretion properties and system parameters. Figure~\ref{fig: lacc A vs B} shows that the accretion luminosity distributions are slightly different for the HAe and HBes, when normalized by the stellar luminosity. A two-sample KS test on these populations results in a p-value of 0.008, indicating that the distributions may be drawn from different populations. We show the distributions of \Mdot\ for the HAe and HBe objects in Figure~\ref{fig: mdot A vs B} and a KS test on those distributions results in a p-value of 9.446$\times$10$^{-12}$, indicating that they are significantly different. Since stellar mass and luminosity are related, we show the trend of \Lacc\ and \LaccL\ with stellar mass in Figure~\ref{fig: mass lacc}.

\begin{figure}
    \centering
    \includegraphics[scale=0.55]{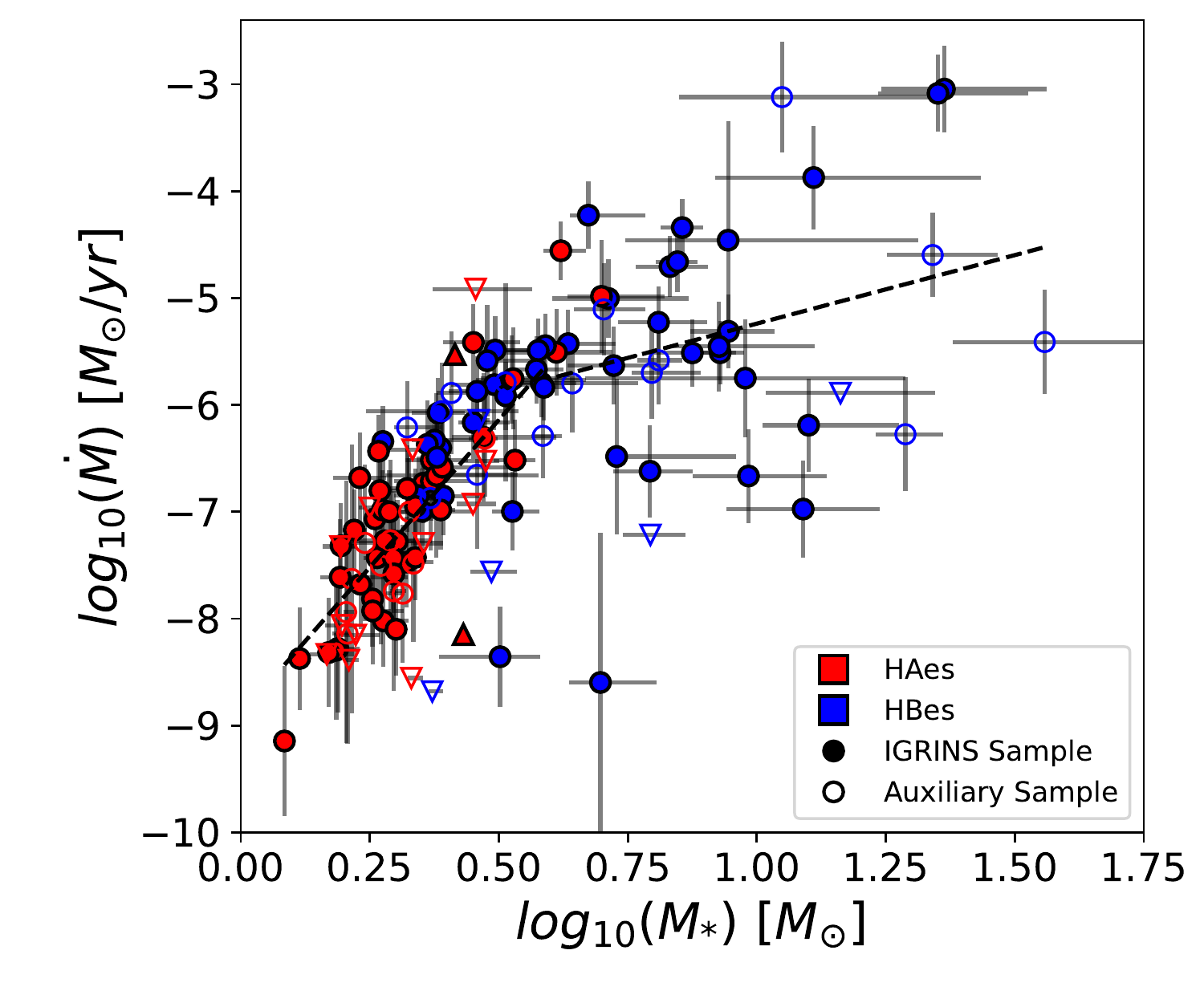}
    \caption{Log$_{10}$(\Mdot) vs. log$_{10}$(\mstar) for Herbig Be (blue) and Herbig Ae (red) stars for the IGRINS Sample (filled) and the Auxiliary Sample (\citealt{donehew-brittain11} and \citealt{fairlamb17}; open). Upward facing triangles are objects for which there is no uncertainty on \Mdot\ due to no literature errors on \mstar\ and/or \rstar. Uncertainties on age are shown for all objects that have these uncertainties listed in the literature. Downward facing triangles are objects with \Lbrg\ upper limits from \citet{fairlamb17}. The mass accretion rate increases with increasing mass. Two fits are shown as the black dashed lines, one for stars with \mstar$<$4 \Msun\ and one for stars with \mstar$\geq$4 \Msun. The slope for the high-mass objects is 1.282, while it is a steeper 5.518 for the low-mass objects.}
    \label{fig: mass mdot}
\end{figure}

\begin{figure*}
    \centering
    \includegraphics[scale=0.55]{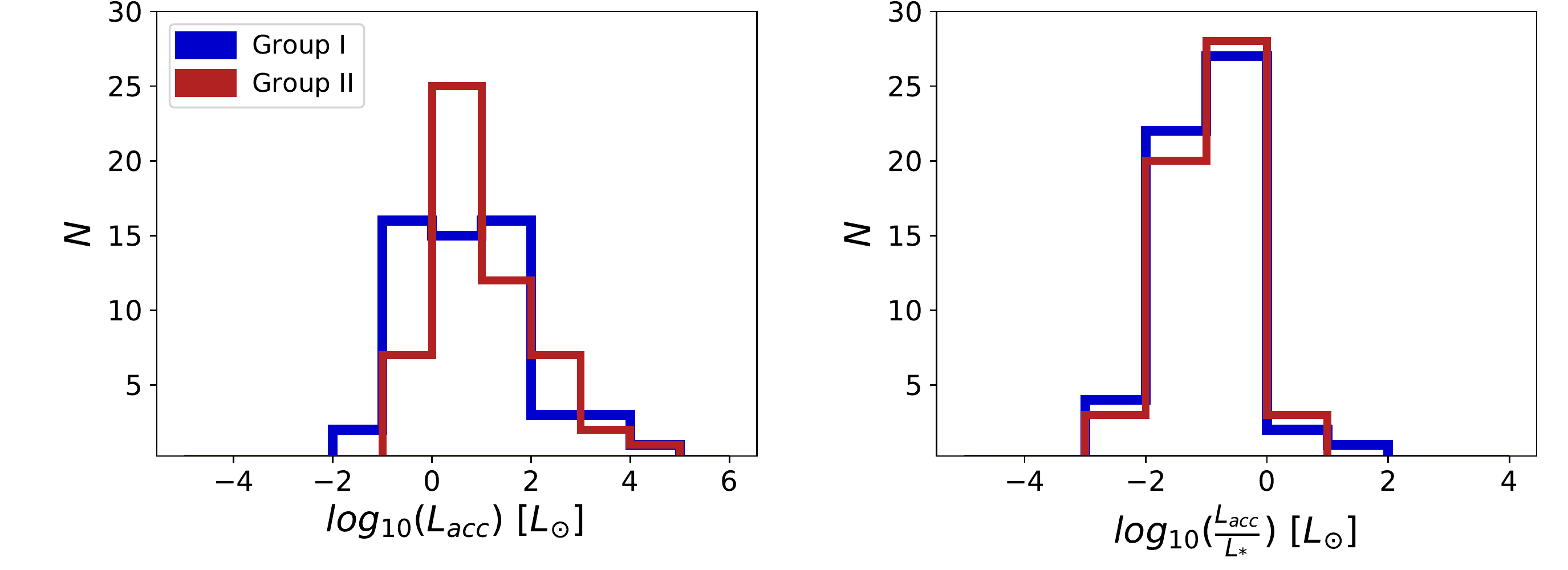}
    \caption{Left: Log$_{10}(L_{acc})$ for Group I (blue) and Group II (red) objects in the IGRINS Sample, and the Auxiliary Sample (\citealt{donehew-brittain11} and \citealt{fairlamb17}). Right: Log$_{10}(L_{acc}/L_{*})$ for the same samples. There is no trend of $\log_{10}(L_{acc})$ or $\log_{10}(L_{acc}/L_{*})$ with group. A Kolmogorov-Smirnov two-sample test returns p-values of 0.060 and 0.471 for the \Lacc\ and \LaccL\ distributions, respectively.}
    \label{fig: lacc group}
\end{figure*}

\begin{figure}
    \centering
    \includegraphics[scale=0.55]{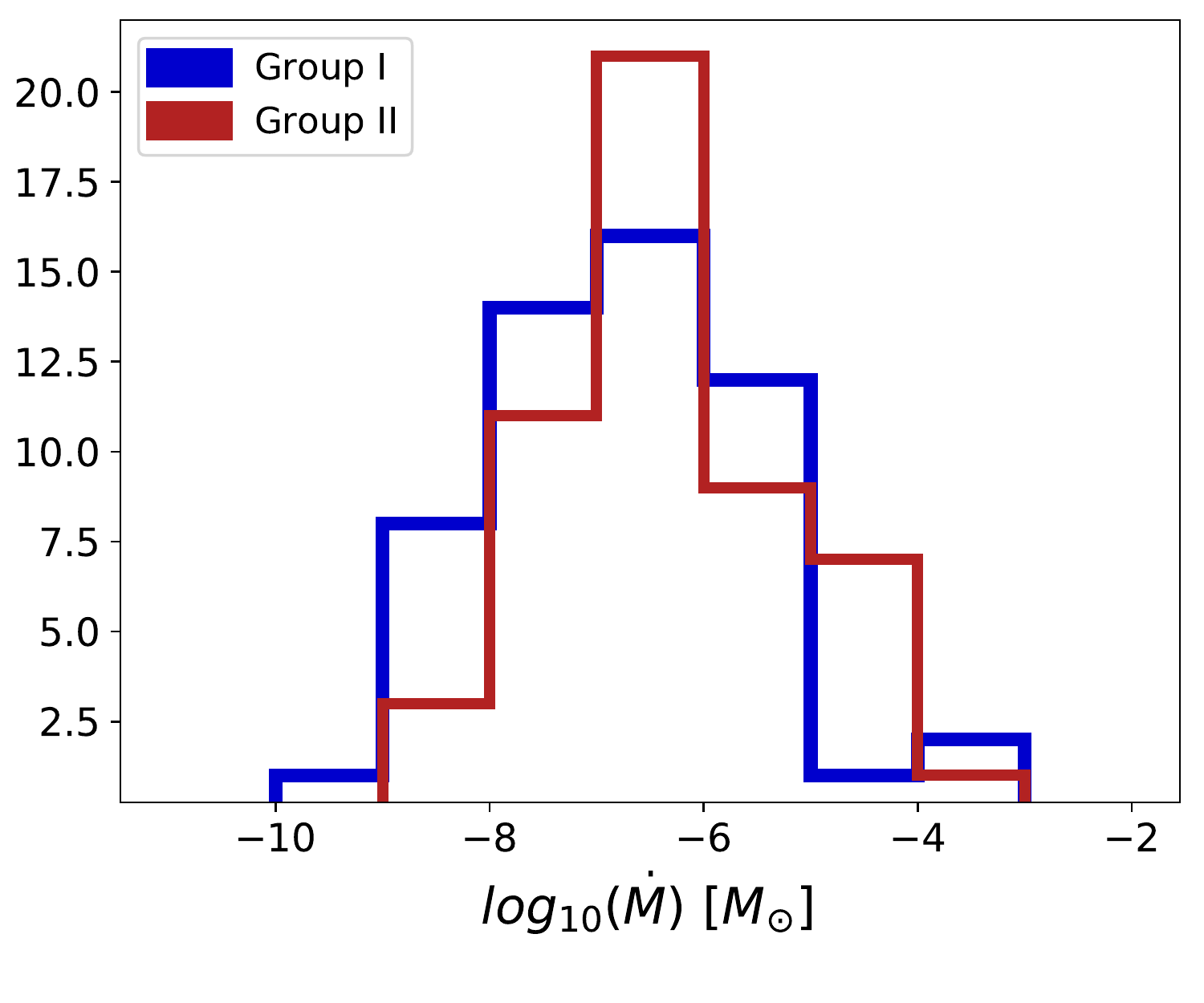}
    \caption{Log$_{10}$(\Mdot) for Group I (blue) and Group II (red) objects in the IGRINS Sample, and the Auxiliary Sample (\citealt{donehew-brittain11} and \citealt{fairlamb17}). There is no trend of $\log_{10}$(\Mdot) with group with a p-value of 0.052 from a two-sample Kolmogorov-Smirnov test. }
    \label{fig: mdot group}
\end{figure}

The relationship between \Mdot\ and stellar mass is shown in Figure~\ref{fig: mass mdot} where we see a positive correlation. We fit the data with a break at 4 \Msun, as is found for the objects in \citet{wichittanakom20}. The slope for the high-mass objects is 1.282, while it is a steeper 5.518 for the low-mass objects. The low-mass slope is steeper than that calculated by \citet{hartmann16}, however that fit was determined for lower mass objects, between 0.1 and 1.0 \Msun. When we look at the Meeus group, as discussed in Section~\ref{subsec: group determination}, we see that there is no distinguishable difference based on group in terms of \Lacc, \LaccL, or \Mdot\ as shown in Figure~\ref{fig: lacc group} and Figure~\ref{fig: mdot group}.  

\begin{figure*}
    \centering
    \includegraphics[scale=0.55]{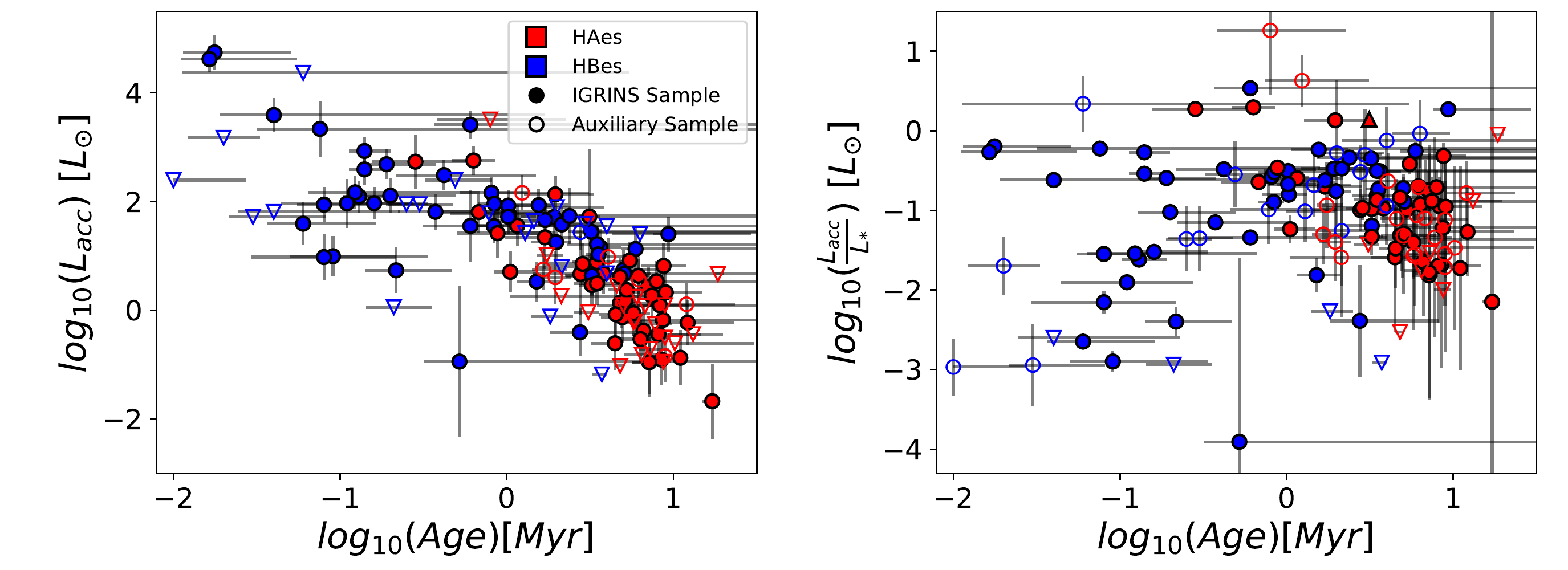}
    \caption{Left: \Lacc\ vs. age for Herbig Be (blue) and Herbig Ae (red) stars for the IGRINS Sample (filled) and the Auxiliary Sample (\citealt{donehew-brittain11} and \citealt{fairlamb17}; open). The accretion luminosity (not normalized by the stellar luminosity) decreases with increasing age. Right: \Lacc/L$_{*}$ vs. age for the same samples. When dividing the accretion luminosity by the stellar luminosity, the trend with age disappears.}
    \label{fig: age lacc}
\end{figure*}

\begin{figure}
    \centering
    \includegraphics[scale=0.55]{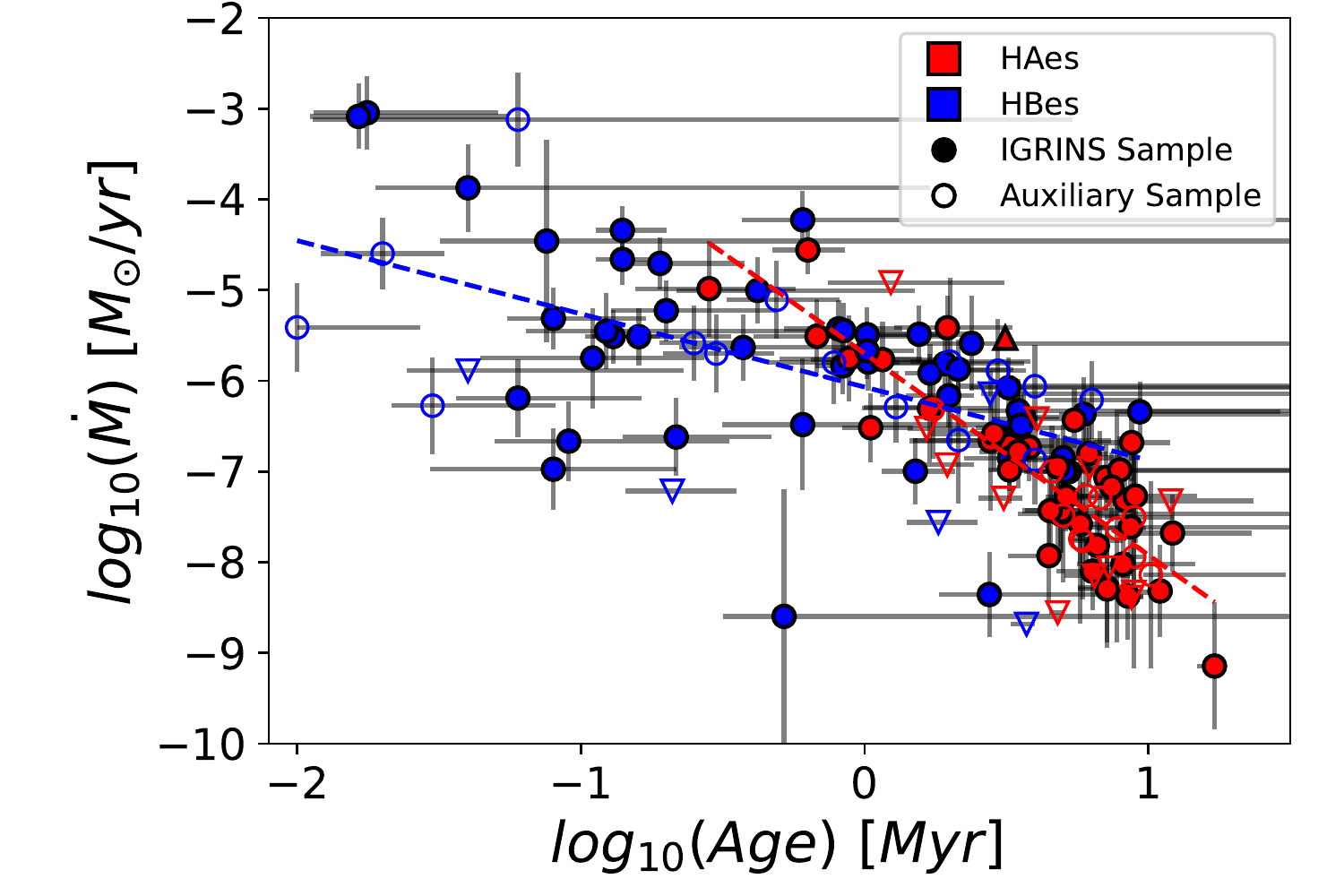}
    \caption{Log$_{10}$(\Mdot) vs. log$_{10}$(Age) for Herbig Be (blue) and Herbig Ae (red) stars for the IGRINS Sample (filled) and the Auxiliary Sample (\citealt{donehew-brittain11} and \citealt{fairlamb17}; open). Upward facing triangles are objects for which there is no uncertainty on \Mdot\ due to no literature errors on \mstar\ and/or \rstar. Uncertainties on age are shown for all objects that have these uncertainties listed in the literature. Downward facing triangles are objects with \Lbrg\ upper limits from \citet{fairlamb17}. A fit to the HBes is shown as the blue line and a find to the HAes is shown as the red line. The mass accretion rate decreases with age.}
    \label{fig: age mdot}
\end{figure}

Ages are available in the literature for 88 of our targets from the IGRINS sample and 53 from the Auxiliary Sample. We use these ages to search for trends with \Lacc, \LaccL, and \Mdot\ in all three samples. We show the relationship between age and \Lacc\ in Figure~\ref{fig: age lacc}, both normalized by \Lstar\ and not. Without normalizing by \Lstar, the \Lacc\ decreases as a function of age. However, normalizing by \Lstar\ makes this trend disappear. The relationship between \Mdot\ and age is shown in Figure~\ref{fig: age mdot} which shows that the mass accretion rate decreases with age. This is discussed in Section~\ref{subsec: age}.

\subsection{Variability}\label{subsect: variability results}
\begin{figure}
    \centering
    \includegraphics[scale=0.48]{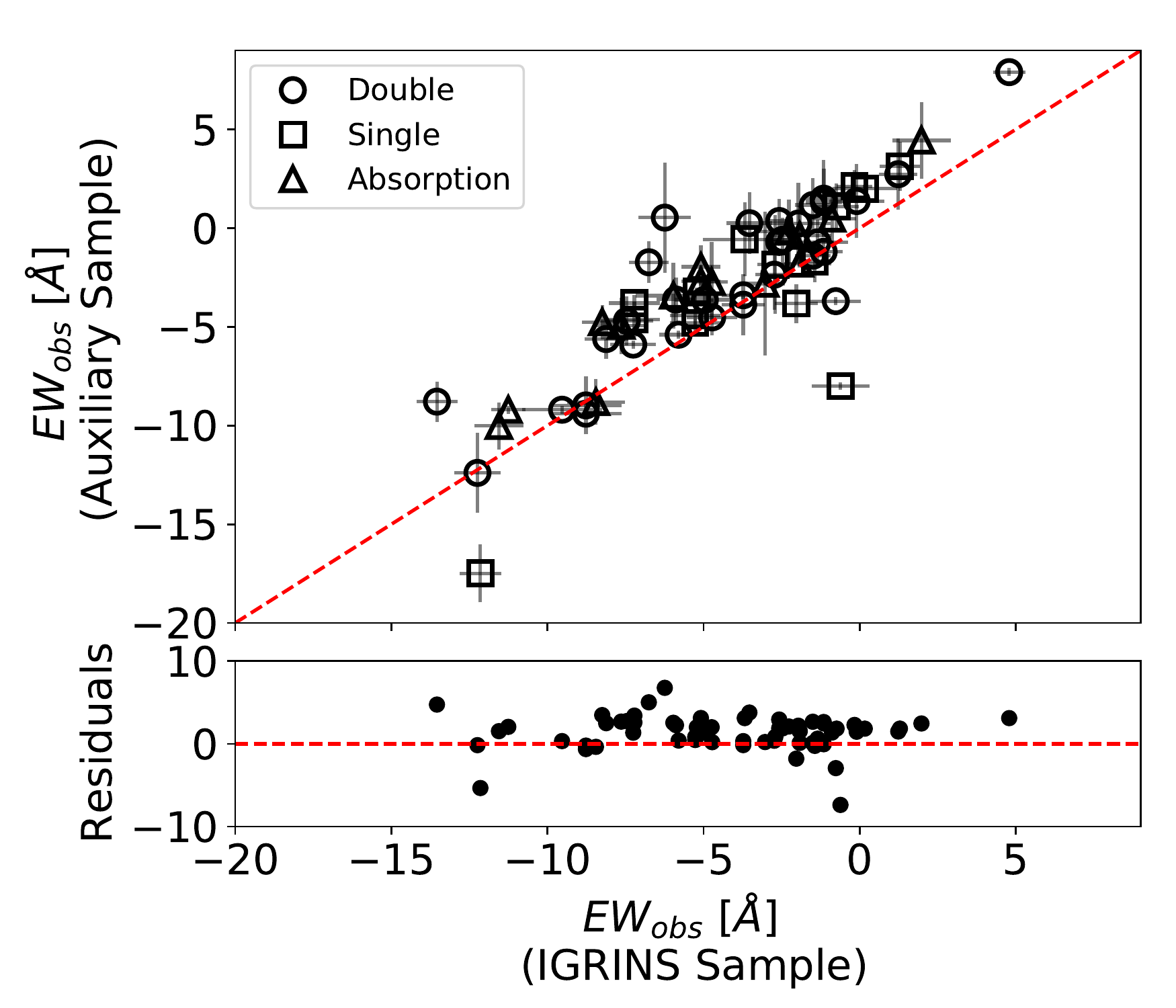}
    \caption{Top: EW$_{obs}$ for the objects that overlap between the IGRINS Sample and the Auxiliary Sample (\citealt{donehew-brittain11} and \citealt{fairlamb15,fairlamb17}). The \brgamma\ line profile shapes in the IGRINS data are indicated by the shape, circle being double-peaked, square being single-peaked, and triangle being absorption. The red, dashed line corresponds to a one-to-one relationship, indicating no variability between observations. Errors are indicted for each point. For objects with no errors in \citet{donehew-brittain11}, we adopt the typical uncertainty in that work of 0.2 \angstrom. Bottom: Residuals between observations.}
    \label{fig: variability}
\end{figure}

To study variability in the objects that overlap between our observations and those taken by \citet{donehew-brittain11} and \citet{fairlamb17}, we compare the \brgamma\ $EW_{obs}$. We choose $EW_{obs}$ to avoid differences caused by different distances, stellar parameters, and models that are incorporated in calculating other parameters like \Lbrg, \Lacc, and \Mdot. Figure~\ref{fig: variability} shows the differences between our own measured $EW_{obs}$ values and those from the previous works. We see that the $EW_{obs}$ values for our sample are lower for most objects than in the previous works. This could be caused by differences in the exact line and continuum regions chosen and differences in the telluric correction (done with telluric standard stars in the case of the IGRINS data and done using \texttt{molecfit}, see \citealt{kausch15} and \citealt{smette15}, in the case of the X-Shooter data). However, there are four objects that have residuals greater than 5 \angstrom. These are (with locations in Figure~\ref{fig: variability} in parenthesis) V791 Mon (-12.15, -17.49), HD 53367 (-6.75, -1.73), HD 52721 (-6.24, 0.53), and PX Vul (-0.62, -8.0). The first three are presented in \citet{fairlamb17} and the last in \citet{donehew-brittain11}.

Here, we discuss the objects that have large residuals between previous observations and those that we present here. V791 Mon was observed on 29 December, 2020 with IGRINS. Our observations have a strong, single-peaked \brgamma\ line with an equivalent width of -12.1$\pm$0.7 \angstrom. In \citet{fairlamb17} \brgamma\ has an observed EW of -17.49$\pm$1.47, therefore a much stronger line was present in the \citet{fairlamb17} observations. HD 53367 was observed with IGRINS on 19 November 2019 with an EW$_{obs}$ of -6.8$\pm$0.6 and a double-peaked profile. The \citet{fairlamb17} observations have \brgamma\ EW$_{obs}$ of -1.73$\pm$1.06, indicating a weaker line. HD 52721 was observed on 19 November, 2018 with IGRINS and has \brgamma\ in emission with EW$_{obs}$ of -6.2$\pm$0.8 \angstrom. In \citet{fairlamb17} the \brgamma\ EW$_{obs}$ is only 0.53$\pm$2.78 \angstrom, with the line then predominantly being in absorption. Our observations show a double-peaked feature. This target has been observed in the optical at H$\alpha$ and He I at 1.0830 \mic\ by \citet{ababakr16} and \citet{reiter18}, respectively. The H$\alpha$ profile was taken on 01 July, 2012 and is single-peaked in emission with an EW of -11.4 \angstrom. The He I profile was observed on 20 September, 2016 and shows no emission or absorption. PX Vul was observed with IGRINS on 27 September 2018 with some emission on either side of the line with a slight absorption feature in the center. We measure EW$_{obs}$ to be -0.6$\pm$0.9 \angstrom. In \citet{donehew-brittain11} PX Vul was observed on 23 September, 2007 with an EW of -8.0 \angstrom. Although the EW values are quite different, the observations look largely consistent between our observations and those in \citet{donehew-brittain11}, with no strong absorption or emission feature.

\section{Discussion}\label{sec: discussion}

\subsection{Herbig Ae vs. Herbig Be Stars}\label{subsec: a vs b}

We find that the HBes show higher \brgamma\ equivalent widths than the HAes such that a two-sample KS test determines that they are drawn from different populations. We also find that the \Lbrg-\Lstar\ relationship is different between these two groups, with the dependence of \Lbrg\ on \Lstar\ being shallower for HBes. This relationship between \Lbrg\ and \Lstar\ (and thus \Lbrg\ and \Lacc) explains many of the subsequent findings. Effective temperature, stellar luminosity, stellar mass, and system age are all related, making the trends correlated. 

Several surveys have looked for differences in accretion mechanism based on stellar mass and spectral type. \citet{donehew-brittain11} and \citet{fairlamb17} find differences in the \brgamma\ emission between HAes and HBes. \citet{donehew-brittain11} find that the \Lbrg-\Lacc\ relationship that holds for T Tauri stars and HAes does not continue to hold for the more massive HBes. The authors calibrate the \Lbrg-\Lacc\ relationship using \Lbrg\ and the veiling of the Balmer discontinuity (accretion can veil the Balmer discontinuity that exists due to the many Balmer absorption lines short-ward of 4000 \angstrom). \citet{fairlamb15} find that seven early HBes in their sample do not fit the magnetospheric accretion paradigm when looking at the Balmer discontinuity. Overall, \citet{fairlamb15} find that the \Lacc-\Lstar \ relationship holds for HAeBes, with the HBes having a shallower dependence than the HAes. \citet{wichittanakom20} finds a break at about 4 \Msun\ for the relationship between \Mdot\ and \mstar. We find that this break explains our data as well. 

If there is a break in accretion mechanism with increasing stellar mass/earlier spectral types, we might expect the \brgamma\ line to reflect that change as it traces the accretion. As stellar magnetic fields grow weaker as they transition from having convective envelopes to radiative ones, the inner edge of the gas disk may extend inward. We might expect objects with stronger magnetic fields to have accretion-tracing line profiles that are wider due to the fast-moving material as it accelerates along the magnetic field lines. Using the line profiles to distinguish between large magnetospheric infall and close-in boundary layer accretion with strong disk winds will require the use of magnetospheric accretion models. The high-resolution data presented here are ideal for future modeling.

\subsection{Group}\label{subsec: group}
We see no trend of \Lacc, \LaccL, or \Mdot\ dependence on group classification, indicating that the disk dust morphology is not playing a role in the accretion process. \citet{mendigutia12} also find that mass accretion rate does not correlate with group which the authors point out is not surprising given that the group classification relies heavily on mid- to far-infrared fluxes, which do not correlate with the accretion rate. \citet{banzatti18} also find no correlation between accretion rate and SED group (broken up into Group II, Group I with high NIR excess, and Group I with low NIR excess). The authors additionally find that the group does not correlate with stellar temperature, mass, luminosity, or age. While group determination is based largely on long wavelength fluxes, \citet{maaskant13} find that Group I objects are gapped such that the disks are transitional or pre-transitional. These gaps indicate that at least the dust is cleared near the star, but given that the accretion rate is similar between groups, it is likely that there is still gas in the gaps that is being accreted onto the star.

\subsection{Age}\label{subsec: age}
We find that the accretion luminosity and mass accretion rate decrease with age. However, when we normalize the accretion luminosity by the stellar luminosity, the trend with age disappears. Several studies have looked at accretion tracers as a function of stellar age. \citet{manoj06} and \citet{vioque18} find that the equivalent width of H$\alpha$ decreases with age. \citet{vioque18} cautions that there may be a bias due to the stellar masses. As in the IGRINS Sample, the most massive stars in the \citet{manoj06} and \citet{vioque18} samples are the youngest but are still close to the ZAMS due to their rapid evolution. Our findings for \brgamma\ thus agree with what \citet{manoj06} and \citet{vioque18} find for H$\alpha$; however, it seems that the trend depends largely on the stellar luminosity, making it hard to determine whether there is a trend with age. We also see a decrease of the mass accretion rate with age, however that also suffers from the correlation between stellar mass and age.

\subsection{Line Shape and Variability}\label{subsec: line shape}

Spectroscopic analysis of accretion (as opposed to U-band photometric excess, for example) allows for kinematic characterization of the line-generating region. \citet{cauley14}, \citet{cauley15}, and \citet{reiter18} are examples of works that use optical spectra to characterize accretion onto HAeBes. \citet{cauley14}, using the He I line at $\lambda\sim1.0830$ \mic, find that the HAeBes have different line morphologies than classical T Tauri stars (CTTSs). Additionally, the authors find that the HBes have different morphologies than the HAes. They find that the HBes do not show redshifted absorption, which would be expected for material accreting onto the star magnetospherically, while they do show blueshifted absorption, indicative of winds. HAes, on the other hand, show both red- and blueshifted absorption. However, as a group, the HAeBes show less of these absorption characteristics than CTTSs. \citet{reiter18} also use the He I $\lambda$ 1.0830 line and target HAeBes that were observed for magnetic field measurements by \citet{alecian13}. Few of their sources had detected magnetic fields, and for the ones that did have detections, the line profiles were not significantly different. In Figure~\ref{fig: line shape} we can see that there is no trend in the \brgamma\ profile shapes based on the stellar mass or effective temperature. If the accretion mechanism does change from magnetospheric to boundary layer in the stellar mass/spectral type region that we explore here, then this indicates that we are not seeing that change manifest in a change in line shape. The line shape is likely heavily influenced by the inclination of the source, which may obscure any influences of the accretion geometry (e.g., \citealt{muzerolle04}). The double- and single-peaked line fractions in the IGRINS sample ($\sim$47\% and $\sim$27\%, respectively) agree well with the line profiles for the 197 objects in \citet{vioque18} (52\% are double-peaked, 31\% are single-peaked, and 17\% are P-Cygni, mostly all regular and not inverse). These fractions are similar to those found by \citet{finkenzeller&mundt84} in H$\alpha$: 50\% double-peaked, 25\% single-peaked, and 20\% P-Cygni. 

\citet{vioque18} use H$\alpha$ EWs and line profiles from the literature, along with data from the Gaia Data Release 2, the 2MASS survey, and the Wide-field Infrared Survey Explorer (WISE, \citealt{cutri13}), to study variability and correlations between H$\alpha$ and IR excess in a large sample of HAeBes. \citet{vioque18} find that high-mass stars ($\gtrsim 7\ M_{\odot}$) have low IR excesses and low variability, perhaps due to their proximity to the main sequence and the tendency of massive stars to clear their disks more quickly. They also find that the most variable objects have double-peaked profiles, which they suggest are caused by edge-on disks.

When we look at our Variability Sample (the objects that overlap between the IGRINS Sample and the Auxiliary Samples of targets from \citealt{donehew-brittain11} and/or \citealt{fairlamb17}), we find that EW$_{obs}$ is largely consistent between the samples. However, our values tend to be lower, perhaps due to differences in the continuum and line regions chosen over which to determine the equivalent width, or potentially due to differences in telluric correction, see Section~\ref{subsect: variability results}. The two objects that show the largest differences are HD 52721 and PX Vul. Although the EW$_{obs}$ determined in this work and that of \citet{donehew-brittain11} are quite different, the line is weak in both cases. The \brgamma\ line of HD 52721 shows a fairly large double-peaked profile while the positive EW$_{obs}$ for this object in \citet{fairlamb17} indicates that the line was in emission when that data was taken. Additionally, H$\alpha$ and He I lines have been observed for this target showing strong, single-peaked emission and no emission, respectively.

While we do not have flux calibrated spectra, variability in the $EW_{obs}$ could reflect a change in the line luminosity or continuum. While UXor objects can vary by as much as four magnitudes, HAeBes typically vary by less than 0.3 magnitudes \citep{davies90,doering_meixner09}. \citet{davies90} report the K-band variability for HD 53367 and HD 52721 as 0.11 and 0.04 mag, respectively. Comparison of the K-band magnitude reported by \citet{miroshnichenko99} with the 2MASS magnitude indicates that the variability of V791 Mon is $\sim$0.14 mag. Similarly comparison of the K-band flux reported by \citet{myers87} with the 2MASS measurements indicates that the variability of PX Vul is $\sim$0.26 mag. All are comparable to the NIR variability of typical HAeBes. Thus we conclude that the variability of the equivalent width due to photometric variability is likely less than 30\%.

\section{Summary and Conclusions}\label{sec: summary and conclusions}

We present high-resolution, NIR spectra for 102 Herbig Ae/Be stars. We analyze the \brgamma\ line at 2.166 \mic\ to explore accretion in these intermediate-mass objects and we expand our sample to 155 objects in total by including targets analyzed in \citet{donehew-brittain11} and \citet{fairlamb17}. We find the following: 

\begin{itemize}

\item The \brgamma\ line profiles are varied in profile shape with 48 double-peaked ($\sim$47\%), 28 single-peaked ($\sim$27\%), 23 absorption/non-detection ($\sim$23\%), and 3 ``other'' ($\sim$3\%) profiles. There is no trend of line profile with stellar effective temperature. This indicates that the line shape, if it is influenced by the kinematics of the accretion flow, is not indicating a change in the accretion mechanism. However, it is likely that inclination effects are impacting the \brgamma\ line shape, potentially obscuring line shape changes based on the accretion geometry.

\item We find a dependence of \Lacc\ and \Mdot\ with stellar mass, in line with relationships found in the literature (e.g., \citealt{garcialopez06,donehew-brittain11, fairlamb15, wichittanakom20}). \LaccL\ shows no dependence with stellar mass due to the strong correlation between stellar mass and luminosity. Similarly, \Lacc\ and age are correlated (as is seen in e.g., \citealt{wichittanakom20}), however, we find that the trend disappears for \LaccL.

\item The \Mdot--\mstar\ relationship slope is smaller for stars with \mstar$\geq$4\Msun\ (1.282) than for \mstar$<$4\Msun\ (5.518), which may indicate a change in accretion mechanism. These slopes, determined using \brgamma\ as the accretion tracer, are similar to those found by \citealt{wichittanakom20} using H$\alpha$. 

\item We find no dependence on the group classification, which was derived using the IR SED, indicating that gas accretion onto the star is not affected by the disk's dust structure. This has been seen in other surveys as well \citep{mendigutia12,banzatti18} and is confirmed with this large sample size.

\item For objects that overlap between our observations and previous \brgamma\ line measurements, we compare the results and find them to be largely consistent, with the exception of V791 Mon, HD 53367, HD 52721, and PX Vul. For the objects with large variations, we find it unlikely that this is due to overall K-band variability and may reflect a change in the accretion rates for these objects.

\end{itemize}

This work presents and analyses a large sample of HAeBe objects with high-resolution, NIR spectroscopy. This data can be used in future analysis, including being paired with magnetospheric accretion models to determine precise accretion properties and to determine in what systems the disk wind contribution to the \brgamma\ line is important. 

\begin{acknowledgements}
We thank the referee for their thoughtful and constructive comments that improved the manuscript. We thank the support staff and telescope operators at the Lowell Discovery Telescope. Additionally, we thank Greg Mace, IGRINS Instrument Manager at the University of Texas at Austin and Hwihyun Kim, Assistant Scientist at Gemini South, for their assistance with our observations at Gemini South. 

This material is based upon work supported by the National Science Foundation under Grant Number AST-1455042.This work used the Immersion Grating Infrared Spectrometer (IGRINS) that was developed under a collaboration between the University of Texas at Austin and the Korea Astronomy and Space Science Institute (KASI) with the financial support of the US National Science Foundation under grants AST-1229522 and AST-1702267, of the University of Texas at Austin, and of the Korean GMT Project of KASI. These results made use of the Lowell Discovery Telescope, supported by Lowell Observatory, Boston University, the University of Maryland, the University of Toledo, and Northern Arizona University. This work is also based on observations obtained at the international Gemini Observatory (program GS-2020A-Q-318), a program of NSF’s OIR Lab, which is managed by the Association of Universities for Research in Astronomy (AURA) under a cooperative agreement with the National Science Foundation. on behalf of the Gemini Observatory partnership: the National Science Foundation (United States), National Research Council (Canada), Agencia Nacional de Investigaci\'{o}n y Desarrollo (Chile), Ministerio de Ciencia, Tecnolog\'{i}a e Innovaci\'{o}n (Argentina), Minist\'{e}rio da Ci\^{e}ncia, Tecnologia, Inova\c{c}\~{o}es e Comunica\c{c}\~{o}es (Brazil), and Korea Astronomy and Space Science Institute (Republic of Korea). 
\end{acknowledgements}

\bibliographystyle{aasjournal}
\bibliography{main}

\begin{thebibliography}{}
\expandafter\ifx\csname natexlab\endcsname\relax\def\natexlab#1{#1}\fi

\bibitem[{{Ababakr} {et~al.}(2016){Ababakr}, {Oudmaijer}, \&
  {Vink}}]{ababakr16}
{Ababakr}, K.~M., {Oudmaijer}, R.~D., \& {Vink}, J.~S. 2016, \mnras, 461, 3089

\bibitem[{{Ababakr} {et~al.}(2017){Ababakr}, {Oudmaijer}, \&
  {Vink}}]{ababakr17}
---. 2017, \mnras, 472, 854

\bibitem[{{Abt} \& {Morrell}(1995)}]{abt-morell95}
{Abt}, H.~A., \& {Morrell}, N.~I. 1995, \apjs, 99, 135

\bibitem[{{Acke} {et~al.}(2009){Acke}, {Min}, {van den Ancker}, {Bouwman},
  {Ochsendorf}, {Juhasz}, \& {Waters}}]{acke09}
{Acke}, B., {Min}, M., {van den Ancker}, M.~E., {et~al.} 2009, \aap, 502, L17

\bibitem[{{Acke} \& {van den Ancker}(2006)}]{acke-anker06}
{Acke}, B., \& {van den Ancker}, M.~E. 2006, \aap, 457, 171

\bibitem[{{Alecian} {et~al.}(2013){Alecian}, {Wade}, {Catala}, {Grunhut},
  {Landstreet}, {Bagnulo}, {B{\"o}hm}, {Folsom}, {Marsden}, \&
  {Waite}}]{alecian13}
{Alecian}, E., {Wade}, G.~A., {Catala}, C., {et~al.} 2013, \mnras, 429, 1001

\bibitem[{{Allard} {et~al.}(2012){Allard}, {Homeier}, \& {Freytag}}]{allard12}
{Allard}, F., {Homeier}, D., \& {Freytag}, B. 2012, Philosophical Transactions
  of the Royal Society of London Series A, 370, 2765

\bibitem[{{Alonso-Albi} {et~al.}(2009){Alonso-Albi}, {Fuente}, {Bachiller},
  {Neri}, {Planesas}, {Testi}, {Bern{\'e}}, \& {Joblin}}]{alonso-albi09}
{Alonso-Albi}, T., {Fuente}, A., {Bachiller}, R., {et~al.} 2009, \aap, 497, 117

\bibitem[{{Aret} {et~al.}(2016){Aret}, {Kraus}, \& {{\v{S}}lechta}}]{aret16}
{Aret}, A., {Kraus}, M., \& {{\v{S}}lechta}, M. 2016, \mnras, 456, 1424

\bibitem[{{Arun} {et~al.}(2019){Arun}, {Mathew}, {Manoj}, {Ujjwal}, {Kartha},
  {Viswanath}, {Narang}, \& {Paul}}]{arun19}
{Arun}, R., {Mathew}, B., {Manoj}, P., {et~al.} 2019, \aj, 157, 159

\bibitem[{{Bailer-Jones} {et~al.}(2018){Bailer-Jones}, {Rybizki}, {Fouesneau},
  {Mantelet}, \& {Andrae}}]{bailer-jones18}
{Bailer-Jones}, C.~A.~L., {Rybizki}, J., {Fouesneau}, M., {Mantelet}, G., \&
  {Andrae}, R. 2018, \aj, 156, 58

\bibitem[{{Baines} {et~al.}(2006){Baines}, {Oudmaijer}, {Porter}, \&
  {Pozzo}}]{baines06}
{Baines}, D., {Oudmaijer}, R.~D., {Porter}, J.~M., \& {Pozzo}, M. 2006, \mnras,
  367, 737

\bibitem[{{Baines} {et~al.}(2012){Baines}, {Gonzalez}, {Arviset}, {Barbarisi},
  {Laruelo}, {Leon}, {Ortiz de Landaluce}, {Osuna}, {Rios}, \&
  {Salgado}}]{baines12}
{Baines}, D., {Gonzalez}, J., {Arviset}, C., {et~al.} 2012, Baltic Astronomy,
  21, 379

\bibitem[{{Banzatti} {et~al.}(2018){Banzatti}, {Garufi}, {Kama}, {Benisty},
  {Brittain}, {Pontoppidan}, \& {Rayner}}]{banzatti18}
{Banzatti}, A., {Garufi}, A., {Kama}, M., {et~al.} 2018, \aap, 609, L2

\bibitem[{{Calvet} \& {Gullbring}(1998)}]{calvet&gullbring98}
{Calvet}, N., \& {Gullbring}, E. 1998, \apj, 509, 802

\bibitem[{{Calvet} {et~al.}(2004){Calvet}, {Muzerolle}, {Brice{\~n}o},
  {Hern{\'a}ndez}, {Hartmann}, {Saucedo}, \& {Gordon}}]{calvet04}
{Calvet}, N., {Muzerolle}, J., {Brice{\~n}o}, C., {et~al.} 2004, \aj, 128, 1294

\bibitem[{{Cardelli} {et~al.}(1989){Cardelli}, {Clayton}, \&
  {Mathis}}]{cardelli89}
{Cardelli}, J.~A., {Clayton}, G.~C., \& {Mathis}, J.~S. 1989, \apj, 345, 245

\bibitem[{{Cauley} \& {Johns-Krull}(2014)}]{cauley14}
{Cauley}, P.~W., \& {Johns-Krull}, C.~M. 2014, \apj, 797, 112

\bibitem[{{Cauley} \& {Johns-Krull}(2015)}]{cauley15}
---. 2015, \apj, 810, 5

\bibitem[{{Cohen} \& {Kuhi}(1979)}]{cohen-kuhi79}
{Cohen}, M., \& {Kuhi}, L.~V. 1979, \apjs, 41, 743

\bibitem[{{Cutri} \& {et al.}(2013)}]{cutri13}
{Cutri}, R.~M., \& {et al.} 2013, VizieR Online Data Catalog, II/328

\bibitem[{{Davies} {et~al.}(1990){Davies}, {Evans}, {Bode}, \&
  {Whittet}}]{davies90}
{Davies}, J.~K., {Evans}, A., {Bode}, M.~F., \& {Whittet}, D.~C.~B. 1990,
  \mnras, 247, 517

\bibitem[{{Doering} \& {Meixner}(2009)}]{doering_meixner09}
{Doering}, R.~L., \& {Meixner}, M. 2009, \aj, 138, 780

\bibitem[{{Donehew} \& {Brittain}(2011)}]{donehew-brittain11}
{Donehew}, B., \& {Brittain}, S. 2011, \aj, 141, 46

\bibitem[{{Edwards} {et~al.}(1987){Edwards}, {Cabrit}, {Strom}, {Heyer},
  {Strom}, \& {Anderson}}]{edwards87}
{Edwards}, S., {Cabrit}, S., {Strom}, S.~E., {et~al.} 1987, \apj, 321, 473

\bibitem[{{Eisner} {et~al.}(2015){Eisner}, {Rieke}, {Rieke}, {Flaherty},
  {Stone}, {Arnold}, {Cortes}, {Cox}, {Hawkins}, {Cole}, {Zajac}, \&
  {Rudolph}}]{eisner15}
{Eisner}, J.~A., {Rieke}, G.~H., {Rieke}, M.~J., {et~al.} 2015, \mnras, 447,
  202

\bibitem[{{Espaillat} {et~al.}(2014){Espaillat}, {Muzerolle}, {Najita},
  {Andrews}, {Zhu}, {Calvet}, {Kraus}, {Hashimoto}, {Kraus}, \&
  {D'Alessio}}]{espaillat14}
{Espaillat}, C., {Muzerolle}, J., {Najita}, J., {et~al.} 2014, Protostars and
  Planets VI, 497

\bibitem[{{Fairlamb} {et~al.}(2015){Fairlamb}, {Oudmaijer},
  {Mendigut{\'{\i}}a}, {Ilee}, \& {van den Ancker}}]{fairlamb15}
{Fairlamb}, J.~R., {Oudmaijer}, R.~D., {Mendigut{\'{\i}}a}, I., {Ilee}, J.~D.,
  \& {van den Ancker}, M.~E. 2015, \mnras, 453, 976

\bibitem[{{Fairlamb} {et~al.}(2017){Fairlamb}, {Oudmaijer}, {Mendigutia},
  {Ilee}, \& {van den Ancker}}]{fairlamb17}
{Fairlamb}, J.~R., {Oudmaijer}, R.~D., {Mendigutia}, I., {Ilee}, J.~D., \& {van
  den Ancker}, M.~E. 2017, \mnras, 464, 4721

\bibitem[{{Finkenzeller} \& {Mundt}(1984)}]{finkenzeller&mundt84}
{Finkenzeller}, U., \& {Mundt}, R. 1984, \aaps, 55, 109

\bibitem[{{Fuente} {et~al.}(2003){Fuente}, {Rodr{\'\i}guez-Franco}, {Testi},
  {Natta}, {Bachiller}, \& {Neri}}]{fuente03}
{Fuente}, A., {Rodr{\'\i}guez-Franco}, A., {Testi}, L., {et~al.} 2003, \apjl,
  598, L39

\bibitem[{{Gaia Collaboration} {et~al.}(2016){Gaia Collaboration}, {Prusti,
  T.}, {de Bruijne, J. H. J.}, {Brown, A. G. A.}, {Vallenari, A.}, {Babusiaux,
  C.}, {Bailer-Jones, C. A. L.}, {Bastian, U.}, {Biermann, M.}, {Evans, D. W.},
  {Eyer, L.}, {Jansen, F.}, {Jordi, C.}, {Klioner, S. A.}, {Lammers, U.},
  {Lindegren, L.}, {Luri, X.}, {Mignard, F.}, {Milligan, D. J.}, {Panem, C.},
  {Poinsignon, V.}, {Pourbaix, D.}, {Randich, S.}, {Sarri, G.}, {Sartoretti,
  P.}, {Siddiqui, H. I.}, {Soubiran, C.}, {Valette, V.}, {van Leeuwen, F.},
  {Walton, N. A.}, {Aerts, C.}, {Arenou, F.}, {Cropper, M.}, {Drimmel, R.},
  {H\o{}g, E.}, {Katz, D.}, {Lattanzi, M. G.}, {O\'{}Mullane, W.}, {Grebel, E.
  K.}, {Holland, A. D.}, {Huc, C.}, {Passot, X.}, {Bramante, L.}, {Cacciari,
  C.}, {Casta\~neda, J.}, {Chaoul, L.}, {Cheek, N.}, {De Angeli, F.},
  {Fabricius, C.}, {Guerra, R.}, {Hern\'andez, J.}, {Jean-Antoine-Piccolo, A.},
  {Masana, E.}, {Messineo, R.}, {Mowlavi, N.}, {Nienartowicz, K.},
  {Ord\'o\~nez-Blanco, D.}, {Panuzzo, P.}, {Portell, J.}, {Richards, P. J.},
  {Riello, M.}, {Seabroke, G. M.}, {Tanga, P.}, {Th\'evenin, F.}, {Torra, J.},
  {Els, S. G.}, {Gracia-Abril, G.}, {Comoretto, G.}, {Garcia-Reinaldos, M.},
  {Lock, T.}, {Mercier, E.}, {Altmann, M.}, {Andrae, R.}, {Astraatmadja, T.
  L.}, {Bellas-Velidis, I.}, {Benson, K.}, {Berthier, J.}, {Blomme, R.},
  {Busso, G.}, {Carry, B.}, {Cellino, A.}, {Clementini, G.}, {Cowell, S.},
  {Creevey, O.}, {Cuypers, J.}, {Davidson, M.}, {De Ridder, J.}, {de Torres,
  A.}, {Delchambre, L.}, {Dell\'{}Oro, A.}, {Ducourant, C.}, {Fr\'emat, Y.},
  {Garc\'{\i}a-Torres, M.}, {Gosset, E.}, {Halbwachs, J.-L.}, {Hambly, N. C.},
  {Harrison, D. L.}, {Hauser, M.}, {Hestroffer, D.}, {Hodgkin, S. T.}, {Huckle,
  H. E.}, {Hutton, A.}, {Jasniewicz, G.}, {Jordan, S.}, {Kontizas, M.}, {Korn,
  A. J.}, {Lanzafame, A. C.}, {Manteiga, M.}, {Moitinho, A.}, {Muinonen, K.},
  {Osinde, J.}, {Pancino, E.}, {Pauwels, T.}, {Petit, J.-M.}, {Recio-Blanco,
  A.}, {Robin, A. C.}, {Sarro, L. M.}, {Siopis, C.}, {Smith, M.}, {Smith, K.
  W.}, {Sozzetti, A.}, {Thuillot, W.}, {van Reeven, W.}, {Viala, Y.}, {Abbas,
  U.}, {Abreu Aramburu, A.}, {Accart, S.}, {Aguado, J. J.}, {Allan, P. M.},
  {Allasia, W.}, {Altavilla, G.}, {\'Alvarez, M. A.}, {Alves, J.}, {Anderson,
  R. I.}, {Andrei, A. H.}, {Anglada Varela, E.}, {Antiche, E.}, {Antoja, T.},
  {Ant\'on, S.}, {Arcay, B.}, {Atzei, A.}, {Ayache, L.}, {Bach, N.}, {Baker, S.
  G.}, {Balaguer-N\'u\~nez, L.}, {Barache, C.}, {Barata, C.}, {Barbier, A.},
  {Barblan, F.}, {Baroni, M.}, {Barrado y Navascu\'es, D.}, {Barros, M.},
  {Barstow, M. A.}, {Becciani, U.}, {Bellazzini, M.}, {Bellei, G.}, {Bello
  Garc\'{\i}a, A.}, {Belokurov, V.}, {Bendjoya, P.}, {Berihuete, A.}, {Bianchi,
  L.}, {Bienaym\'e, O.}, {Billebaud, F.}, {Blagorodnova, N.}, {Blanco-Cuaresma,
  S.}, {Boch, T.}, {Bombrun, A.}, {Borrachero, R.}, {Bouquillon, S.}, {Bourda,
  G.}, {Bouy, H.}, {Bragaglia, A.}, {Breddels, M. A.}, {Brouillet, N.},
  {Br\"usemeister, T.}, {Bucciarelli, B.}, {Budnik, F.}, {Burgess, P.},
  {Burgon, R.}, {Burlacu, A.}, {Busonero, D.}, {Buzzi, R.}, {Caffau, E.},
  {Cambras, J.}, {Campbell, H.}, {Cancelliere, R.}, {Cantat-Gaudin, T.},
  {Carlucci, T.}, {Carrasco, J. M.}, {Castellani, M.}, {Charlot, P.}, {Charnas,
  J.}, {Charvet, P.}, {Chassat, F.}, {Chiavassa, A.}, {Clotet, M.}, {Cocozza,
  G.}, {Collins, R. S.}, {Collins, P.}, {Costigan, G.}, {Crifo, F.}, {Cross, N.
  J. G.}, {Crosta, M.}, {Crowley, C.}, {Dafonte, C.}, {Damerdji, Y.},
  {Dapergolas, A.}, {David, P.}, {David, M.}, {De Cat, P.}, {de Felice, F.},
  {de Laverny, P.}, {De Luise, F.}, {De March, R.}, {de Martino, D.}, {de
  Souza, R.}, {Debosscher, J.}, {del Pozo, E.}, {Delbo, M.}, {Delgado, A.},
  {Delgado, H. E.}, {di Marco, F.}, {Di Matteo, P.}, {Diakite, S.}, {Distefano,
  E.}, {Dolding, C.}, {Dos Anjos, S.}, {Drazinos, P.}, {Dur\'an, J.}, {Dzigan,
  Y.}, {Ecale, E.}, {Edvardsson, B.}, {Enke, H.}, {Erdmann, M.}, {Escolar, D.},
  {Espina, M.}, {Evans, N. W.}, {Eynard Bontemps, G.}, {Fabre, C.}, {Fabrizio,
  M.}, {Faigler, S.}, {Falc\~ao, A. J.}, {Farr\`as Casas, M.}, {Faye, F.},
  {Federici, L.}, {Fedorets, G.}, {Fern\'andez-Hern\'andez, J.}, {Fernique,
  P.}, {Fienga, A.}, {Figueras, F.}, {Filippi, F.}, {Findeisen, K.}, {Fonti,
  A.}, {Fouesneau, M.}, {Fraile, E.}, {Fraser, M.}, {Fuchs, J.}, {Furnell, R.},
  {Gai, M.}, {Galleti, S.}, {Galluccio, L.}, {Garabato, D.},
  {Garc\'{\i}a-Sedano, F.}, {Gar\'e, P.}, {Garofalo, A.}, {Garralda, N.},
  {Gavras, P.}, {Gerssen, J.}, {Geyer, R.}, {Gilmore, G.}, {Girona, S.},
  {Giuffrida, G.}, {Gomes, M.}, {Gonz\'alez-Marcos, A.}, {Gonz\'alez-N\'u\~nez,
  J.}, {Gonz\'alez-Vidal, J. J.}, {Granvik, M.}, {Guerrier, A.}, {Guillout,
  P.}, {Guiraud, J.}, {G\'urpide, A.}, {Guti\'errez-S\'anchez, R.}, {Guy, L.
  P.}, {Haigron, R.}, {Hatzidimitriou, D.}, {Haywood, M.}, {Heiter, U.},
  {Helmi, A.}, {Hobbs, D.}, {Hofmann, W.}, {Holl, B.}, {Holland, G.}, {Hunt, J.
  A. S.}, {Hypki, A.}, {Icardi, V.}, {Irwin, M.}, {Jevardat de Fombelle, G.},
  {Jofr\'e, P.}, {Jonker, P. G.}, {Jorissen, A.}, {Julbe, F.}, {Karampelas,
  A.}, {Kochoska, A.}, {Kohley, R.}, {Kolenberg, K.}, {Kontizas, E.}, {Koposov,
  S. E.}, {Kordopatis, G.}, {Koubsky, P.}, {Kowalczyk, A.}, {Krone-Martins,
  A.}, {Kudryashova, M.}, {Kull, I.}, {Bachchan, R. K.}, {Lacoste-Seris, F.},
  {Lanza, A. F.}, {Lavigne, J.-B.}, {Le Poncin-Lafitte, C.}, {Lebreton, Y.},
  {Lebzelter, T.}, {Leccia, S.}, {Leclerc, N.}, {Lecoeur-Taibi, I.}, {Lemaitre,
  V.}, {Lenhardt, H.}, {Leroux, F.}, {Liao, S.}, {Licata, E.}, {Lindstr\o{}m,
  H. E. P.}, {Lister, T. A.}, {Livanou, E.}, {Lobel, A.}, {L\"offler, W.},
  {L\'opez, M.}, {Lopez-Lozano, A.}, {Lorenz, D.}, {Loureiro, T.}, {MacDonald,
  I.}, {Magalh\~aes Fernandes, T.}, {Managau, S.}, {Mann, R. G.}, {Mantelet,
  G.}, {Marchal, O.}, {Marchant, J. M.}, {Marconi, M.}, {Marie, J.}, {Marinoni,
  S.}, {Marrese, P. M.}, {Marschalk\'o, G.}, {Marshall, D. J.},
  {Mart\'{\i}n-Fleitas, J. M.}, {Martino, M.}, {Mary, N.}, {Matijevic, G.},
  {Mazeh, T.}, {McMillan, P. J.}, {Messina, S.}, {Mestre, A.}, {Michalik, D.},
  {Millar, N. R.}, {Miranda, B. M. H.}, {Molina, D.}, {Molinaro, R.},
  {Molinaro, M.}, {Moln\'ar, L.}, {Moniez, M.}, {Montegriffo, P.}, {Monteiro,
  D.}, {Mor, R.}, {Mora, A.}, {Morbidelli, R.}, {Morel, T.}, {Morgenthaler,
  S.}, {Morley, T.}, {Morris, D.}, {Mulone, A. F.}, {Muraveva, T.}, {Musella,
  I.}, {Narbonne, J.}, {Nelemans, G.}, {Nicastro, L.}, {Noval, L.},
  {Ord\'enovic, C.}, {Ordieres-Mer\'e, J.}, {Osborne, P.}, {Pagani, C.},
  {Pagano, I.}, {Pailler, F.}, {Palacin, H.}, {Palaversa, L.}, {Parsons, P.},
  {Paulsen, T.}, {Pecoraro, M.}, {Pedrosa, R.}, {Pentik\"ainen, H.}, {Pereira,
  J.}, {Pichon, B.}, {Piersimoni, A. M.}, {Pineau, F.-X.}, {Plachy, E.}, {Plum,
  G.}, {Poujoulet, E.}, {Prsa, A.}, {Pulone, L.}, {Ragaini, S.}, {Rago, S.},
  {Rambaux, N.}, {Ramos-Lerate, M.}, {Ranalli, P.}, {Rauw, G.}, {Read, A.},
  {Regibo, S.}, {Renk, F.}, {Reyl\'e, C.}, {Ribeiro, R. A.}, {Rimoldini, L.},
  {Ripepi, V.}, {Riva, A.}, {Rixon, G.}, {Roelens, M.}, {Romero-G\'omez, M.},
  {Rowell, N.}, {Royer, F.}, {Rudolph, A.}, {Ruiz-Dern, L.}, {Sadowski, G.},
  {Sagrist\`a Sell\'es, T.}, {Sahlmann, J.}, {Salgado, J.}, {Salguero, E.},
  {Sarasso, M.}, {Savietto, H.}, {Schnorhk, A.}, {Schultheis, M.}, {Sciacca,
  E.}, {Segol, M.}, {Segovia, J. C.}, {Segransan, D.}, {Serpell, E.}, {Shih,
  I-C.}, {Smareglia, R.}, {Smart, R. L.}, {Smith, C.}, {Solano, E.}, {Solitro,
  F.}, {Sordo, R.}, {Soria Nieto, S.}, {Souchay, J.}, {Spagna, A.}, {Spoto,
  F.}, {Stampa, U.}, {Steele, I. A.}, {Steidelm\"uller, H.}, {Stephenson, C.
  A.}, {Stoev, H.}, {Suess, F. F.}, {S\"uveges, M.}, {Surdej, J.}, {Szabados,
  L.}, {Szegedi-Elek, E.}, {Tapiador, D.}, {Taris, F.}, {Tauran, G.}, {Taylor,
  M. B.}, {Teixeira, R.}, {Terrett, D.}, {Tingley, B.}, {Trager, S. C.},
  {Turon, C.}, {Ulla, A.}, {Utrilla, E.}, {Valentini, G.}, {van Elteren, A.},
  {Van Hemelryck, E.}, {van Leeuwen, M.}, {Varadi, M.}, {Vecchiato, A.},
  {Veljanoski, J.}, {Via, T.}, {Vicente, D.}, {Vogt, S.}, {Voss, H.}, {Votruba,
  V.}, {Voutsinas, S.}, {Walmsley, G.}, {Weiler, M.}, {Weingrill, K.}, {Werner,
  D.}, {Wevers, T.}, {Whitehead, G.}, {Wyrzykowski, L.}, {Yoldas, A.}, {Zerjal,
  M.}, {Zucker, S.}, {Zurbach, C.}, {Zwitter, T.}, {Alecu, A.}, {Allen, M.},
  {Allende Prieto, C.}, {Amorim, A.}, {Anglada-Escud\'e, G.}, {Arsenijevic,
  V.}, {Azaz, S.}, {Balm, P.}, {Beck, M.}, {Bernstein, H.-H.}, {Bigot, L.},
  {Bijaoui, A.}, {Blasco, C.}, {Bonfigli, M.}, {Bono, G.}, {Boudreault, S.},
  {Bressan, A.}, {Brown, S.}, {Brunet, P.-M.}, {Bunclark, P.}, {Buonanno, R.},
  {Butkevich, A. G.}, {Carret, C.}, {Carrion, C.}, {Chemin, L.}, {Ch\'ereau,
  F.}, {Corcione, L.}, {Darmigny, E.}, {de Boer, K. S.}, {de Teodoro, P.}, {de
  Zeeuw, P. T.}, {Delle Luche, C.}, {Domingues, C. D.}, {Dubath, P.}, {Fodor,
  F.}, {Fr\'ezouls, B.}, {Fries, A.}, {Fustes, D.}, {Fyfe, D.}, {Gallardo, E.},
  {Gallegos, J.}, {Gardiol, D.}, {Gebran, M.}, {Gomboc, A.}, {G\'omez, A.},
  {Grux, E.}, {Gueguen, A.}, {Heyrovsky, A.}, {Hoar, J.}, {Iannicola, G.},
  {Isasi Parache, Y.}, {Janotto, A.-M.}, {Joliet, E.}, {Jonckheere, A.}, {Keil,
  R.}, {Kim, D.-W.}, {Klagyivik, P.}, {Klar, J.}, {Knude, J.}, {Kochukhov, O.},
  {Kolka, I.}, {Kos, J.}, {Kutka, A.}, {Lainey, V.}, {LeBouquin, D.}, {Liu,
  C.}, {Loreggia, D.}, {Makarov, V. V.}, {Marseille, M. G.}, {Martayan, C.},
  {Martinez-Rubi, O.}, {Massart, B.}, {Meynadier, F.}, {Mignot, S.}, {Munari,
  U.}, {Nguyen, A.-T.}, {Nordlander, T.}, {Ocvirk, P.}, {O\'{}Flaherty, K. S.},
  {Olias Sanz, A.}, {Ortiz, P.}, {Osorio, J.}, {Oszkiewicz, D.}, {Ouzounis,
  A.}, {Palmer, M.}, {Park, P.}, {Pasquato, E.}, {Peltzer, C.}, {Peralta, J.},
  {P\'eturaud, F.}, {Pieniluoma, T.}, {Pigozzi, E.}, {Poels, J.}, {Prat, G.},
  {Prod\'{}homme, T.}, {Raison, F.}, {Rebordao, J. M.}, {Risquez, D.},
  {Rocca-Volmerange, B.}, {Rosen, S.}, {Ruiz-Fuertes, M. I.}, {Russo, F.},
  {Sembay, S.}, {Serraller Vizcaino, I.}, {Short, A.}, {Siebert, A.}, {Silva,
  H.}, {Sinachopoulos, D.}, {Slezak, E.}, {Soffel, M.}, {Sosnowska, D.},
  {Straizys, V.}, {ter Linden, M.}, {Terrell, D.}, {Theil, S.}, {Tiede, C.},
  {Troisi, L.}, {Tsalmantza, P.}, {Tur, D.}, {Vaccari, M.}, {Vachier, F.},
  {Valles, P.}, {Van Hamme, W.}, {Veltz, L.}, {Virtanen, J.}, {Wallut, J.-M.},
  {Wichmann, R.}, {Wilkinson, M. I.}, {Ziaeepour, H.}, \& {Zschocke,
  S.}}]{gaia16}
{Gaia Collaboration}, {Prusti, T.}, {de Bruijne, J. H. J.}, {et~al.} 2016,
  A\&A, 595, A1

\bibitem[{{Gaia Collaboration} {et~al.}(2018){Gaia Collaboration}, {Brown, A.
  G. A.}, {Vallenari, A.}, {Prusti, T.}, {de Bruijne, J. H. J.}, {Babusiaux,
  C.}, {Bailer-Jones, C. A. L.}, {Biermann, M.}, {Evans, D. W.}, {Eyer, L.},
  {Jansen, F.}, {Jordi, C.}, {Klioner, S. A.}, {Lammers, U.}, {Lindegren, L.},
  {Luri, X.}, {Mignard, F.}, {Panem, C.}, {Pourbaix, D.}, {Randich, S.},
  {Sartoretti, P.}, {Siddiqui, H. I.}, {Soubiran, C.}, {van Leeuwen, F.},
  {Walton, N. A.}, {Arenou, F.}, {Bastian, U.}, {Cropper, M.}, {Drimmel, R.},
  {Katz, D.}, {Lattanzi, M. G.}, {Bakker, J.}, {Cacciari, C.}, {Casta\~neda,
  J.}, {Chaoul, L.}, {Cheek, N.}, {De Angeli, F.}, {Fabricius, C.}, {Guerra,
  R.}, {Holl, B.}, {Masana, E.}, {Messineo, R.}, {Mowlavi, N.}, {Nienartowicz,
  K.}, {Panuzzo, P.}, {Portell, J.}, {Riello, M.}, {Seabroke, G. M.}, {Tanga,
  P.}, {Th\'evenin, F.}, {Gracia-Abril, G.}, {Comoretto, G.},
  {Garcia-Reinaldos, M.}, {Teyssier, D.}, {Altmann, M.}, {Andrae, R.}, {Audard,
  M.}, {Bellas-Velidis, I.}, {Benson, K.}, {Berthier, J.}, {Blomme, R.},
  {Burgess, P.}, {Busso, G.}, {Carry, B.}, {Cellino, A.}, {Clementini, G.},
  {Clotet, M.}, {Creevey, O.}, {Davidson, M.}, {De Ridder, J.}, {Delchambre,
  L.}, {Dell\'{}Oro, A.}, {Ducourant, C.}, {Fern\'andez-Hern\'andez, J.},
  {Fouesneau, M.}, {Fr\'emat, Y.}, {Galluccio, L.}, {Garc\'{\i}a-Torres, M.},
  {Gonz\'alez-N\'u\~nez, J.}, {Gonz\'alez-Vidal, J. J.}, {Gosset, E.}, {Guy, L.
  P.}, {Halbwachs, J.-L.}, {Hambly, N. C.}, {Harrison, D. L.}, {Hern\'andez,
  J.}, {Hestroffer, D.}, {Hodgkin, S. T.}, {Hutton, A.}, {Jasniewicz, G.},
  {Jean-Antoine-Piccolo, A.}, {Jordan, S.}, {Korn, A. J.}, {Krone-Martins, A.},
  {Lanzafame, A. C.}, {Lebzelter, T.}, {L\"offler, W.}, {Manteiga, M.},
  {Marrese, P. M.}, {Mart\'{\i}n-Fleitas, J. M.}, {Moitinho, A.}, {Mora, A.},
  {Muinonen, K.}, {Osinde, J.}, {Pancino, E.}, {Pauwels, T.}, {Petit, J.-M.},
  {Recio-Blanco, A.}, {Richards, P. J.}, {Rimoldini, L.}, {Robin, A. C.},
  {Sarro, L. M.}, {Siopis, C.}, {Smith, M.}, {Sozzetti, A.}, {S\"uveges, M.},
  {Torra, J.}, {van Reeven, W.}, {Abbas, U.}, {Abreu Aramburu, A.}, {Accart,
  S.}, {Aerts, C.}, {Altavilla, G.}, {\'Alvarez, M. A.}, {Alvarez, R.}, {Alves,
  J.}, {Anderson, R. I.}, {Andrei, A. H.}, {Anglada Varela, E.}, {Antiche, E.},
  {Antoja, T.}, {Arcay, B.}, {Astraatmadja, T. L.}, {Bach, N.}, {Baker, S. G.},
  {Balaguer-N\'u\~nez, L.}, {Balm, P.}, {Barache, C.}, {Barata, C.}, {Barbato,
  D.}, {Barblan, F.}, {Barklem, P. S.}, {Barrado, D.}, {Barros, M.}, {Barstow,
  M. A.}, {Bartholom\'e Mu\~noz, S.}, {Bassilana, J.-L.}, {Becciani, U.},
  {Bellazzini, M.}, {Berihuete, A.}, {Bertone, S.}, {Bianchi, L.}, {Bienaym\'e,
  O.}, {Blanco-Cuaresma, S.}, {Boch, T.}, {Boeche, C.}, {Bombrun, A.},
  {Borrachero, R.}, {Bossini, D.}, {Bouquillon, S.}, {Bourda, G.}, {Bragaglia,
  A.}, {Bramante, L.}, {Breddels, M. A.}, {Bressan, A.}, {Brouillet, N.},
  {Br\"usemeister, T.}, {Brugaletta, E.}, {Bucciarelli, B.}, {Burlacu, A.},
  {Busonero, D.}, {Butkevich, A. G.}, {Buzzi, R.}, {Caffau, E.}, {Cancelliere,
  R.}, {Cannizzaro, G.}, {Cantat-Gaudin, T.}, {Carballo, R.}, {Carlucci, T.},
  {Carrasco, J. M.}, {Casamiquela, L.}, {Castellani, M.}, {Castro-Ginard, A.},
  {Charlot, P.}, {Chemin, L.}, {Chiavassa, A.}, {Cocozza, G.}, {Costigan, G.},
  {Cowell, S.}, {Crifo, F.}, {Crosta, M.}, {Crowley, C.}, {Cuypers+, J.},
  {Dafonte, C.}, {Damerdji, Y.}, {Dapergolas, A.}, {David, P.}, {David, M.},
  {de Laverny, P.}, {De Luise, F.}, {De March, R.}, {de Martino, D.}, {de
  Souza, R.}, {de Torres, A.}, {Debosscher, J.}, {del Pozo, E.}, {Delbo, M.},
  {Delgado, A.}, {Delgado, H. E.}, {Di Matteo, P.}, {Diakite, S.}, {Diener,
  C.}, {Distefano, E.}, {Dolding, C.}, {Drazinos, P.}, {Dur\'an, J.},
  {Edvardsson, B.}, {Enke, H.}, {Eriksson, K.}, {Esquej, P.}, {Eynard Bontemps,
  G.}, {Fabre, C.}, {Fabrizio, M.}, {Faigler, S.}, {Falc\~ao, A. J.}, {Farr\`as
  Casas, M.}, {Federici, L.}, {Fedorets, G.}, {Fernique, P.}, {Figueras, F.},
  {Filippi, F.}, {Findeisen, K.}, {Fonti, A.}, {Fraile, E.}, {Fraser, M.},
  {Fr\'ezouls, B.}, {Gai, M.}, {Galleti, S.}, {Garabato, D.},
  {Garc\'{\i}a-Sedano, F.}, {Garofalo, A.}, {Garralda, N.}, {Gavel, A.},
  {Gavras, P.}, {Gerssen, J.}, {Geyer, R.}, {Giacobbe, P.}, {Gilmore, G.},
  {Girona, S.}, {Giuffrida, G.}, {Glass, F.}, {Gomes, M.}, {Granvik, M.},
  {Gueguen, A.}, {Guerrier, A.}, {Guiraud, J.}, {Guti\'errez-S\'anchez, R.},
  {Haigron, R.}, {Hatzidimitriou, D.}, {Hauser, M.}, {Haywood, M.}, {Heiter,
  U.}, {Helmi, A.}, {Heu, J.}, {Hilger, T.}, {Hobbs, D.}, {Hofmann, W.},
  {Holland, G.}, {Huckle, H. E.}, {Hypki, A.}, {Icardi, V.}, {Jan\ss{}en, K.},
  {Jevardat de Fombelle, G.}, {Jonker, P. G.}, {Juh\'asz, \'A. L.}, {Julbe,
  F.}, {Karampelas, A.}, {Kewley, A.}, {Klar, J.}, {Kochoska, A.}, {Kohley,
  R.}, {Kolenberg, K.}, {Kontizas, M.}, {Kontizas, E.}, {Koposov, S. E.},
  {Kordopatis, G.}, {Kostrzewa-Rutkowska, Z.}, {Koubsky, P.}, {Lambert, S.},
  {Lanza, A. F.}, {Lasne, Y.}, {Lavigne, J.-B.}, {Le Fustec, Y.}, {Le
  Poncin-Lafitte, C.}, {Lebreton, Y.}, {Leccia, S.}, {Leclerc, N.},
  {Lecoeur-Taibi, I.}, {Lenhardt, H.}, {Leroux, F.}, {Liao, S.}, {Licata, E.},
  {Lindstr\o{}m, H. E. P.}, {Lister, T. A.}, {Livanou, E.}, {Lobel, A.},
  {L\'opez, M.}, {Managau, S.}, {Mann, R. G.}, {Mantelet, G.}, {Marchal, O.},
  {Marchant, J. M.}, {Marconi, M.}, {Marinoni, S.}, {Marschalk\'o, G.},
  {Marshall, D. J.}, {Martino, M.}, {Marton, G.}, {Mary, N.}, {Massari, D.},
  {Matijevic, G.}, {Mazeh, T.}, {McMillan, P. J.}, {Messina, S.}, {Michalik,
  D.}, {Millar, N. R.}, {Molina, D.}, {Molinaro, R.}, {Moln\'ar, L.},
  {Montegriffo, P.}, {Mor, R.}, {Morbidelli, R.}, {Morel, T.}, {Morris, D.},
  {Mulone, A. F.}, {Muraveva, T.}, {Musella, I.}, {Nelemans, G.}, {Nicastro,
  L.}, {Noval, L.}, {O\'{}Mullane, W.}, {Ord\'enovic, C.}, {Ord\'o\~nez-Blanco,
  D.}, {Osborne, P.}, {Pagani, C.}, {Pagano, I.}, {Pailler, F.}, {Palacin, H.},
  {Palaversa, L.}, {Panahi, A.}, {Pawlak, M.}, {Piersimoni, A. M.}, {Pineau,
  F.-X.}, {Plachy, E.}, {Plum, G.}, {Poggio, E.}, {Poujoulet, E.}, {Prsa, A.},
  {Pulone, L.}, {Racero, E.}, {Ragaini, S.}, {Rambaux, N.}, {Ramos-Lerate, M.},
  {Regibo, S.}, {Reyl\'e, C.}, {Riclet, F.}, {Ripepi, V.}, {Riva, A.}, {Rivard,
  A.}, {Rixon, G.}, {Roegiers, T.}, {Roelens, M.}, {Romero-G\'omez, M.},
  {Rowell, N.}, {Royer, F.}, {Ruiz-Dern, L.}, {Sadowski, G.}, {Sagrist\`a
  Sell\'es, T.}, {Sahlmann, J.}, {Salgado, J.}, {Salguero, E.}, {Sanna, N.},
  {Santana-Ros, T.}, {Sarasso, M.}, {Savietto, H.}, {Schultheis, M.}, {Sciacca,
  E.}, {Segol, M.}, {Segovia, J. C.}, {S\'egransan, D.}, {Shih, I-C.},
  {Siltala, L.}, {Silva, A. F.}, {Smart, R. L.}, {Smith, K. W.}, {Solano, E.},
  {Solitro, F.}, {Sordo, R.}, {Soria Nieto, S.}, {Souchay, J.}, {Spagna, A.},
  {Spoto, F.}, {Stampa, U.}, {Steele, I. A.}, {Steidelm\"uller, H.},
  {Stephenson, C. A.}, {Stoev, H.}, {Suess, F. F.}, {Surdej, J.}, {Szabados,
  L.}, {Szegedi-Elek, E.}, {Tapiador, D.}, {Taris, F.}, {Tauran, G.}, {Taylor,
  M. B.}, {Teixeira, R.}, {Terrett, D.}, {Teyssandier, P.}, {Thuillot, W.},
  {Titarenko, A.}, {Torra Clotet, F.}, {Turon, C.}, {Ulla, A.}, {Utrilla, E.},
  {Uzzi, S.}, {Vaillant, M.}, {Valentini, G.}, {Valette, V.}, {van Elteren,
  A.}, {Van Hemelryck, E.}, {van Leeuwen, M.}, {Vaschetto, M.}, {Vecchiato,
  A.}, {Veljanoski, J.}, {Viala, Y.}, {Vicente, D.}, {Vogt, S.}, {von Essen,
  C.}, {Voss, H.}, {Votruba, V.}, {Voutsinas, S.}, {Walmsley, G.}, {Weiler,
  M.}, {Wertz, O.}, {Wevers, T.}, {Wyrzykowski, L.}, {Yoldas, A.}, {Zerjal,
  M.}, {Ziaeepour, H.}, {Zorec, J.}, {Zschocke, S.}, {Zucker, S.}, {Zurbach,
  C.}, \& {Zwitter, T.}}]{gaia18}
{Gaia Collaboration}, {Brown, A. G. A.}, {Vallenari, A.}, {et~al.} 2018, A\&A,
  616, A1

\bibitem[{{Garcia Lopez} {et~al.}(2006){Garcia Lopez}, {Natta}, {Testi}, \&
  {Habart}}]{garcialopez06}
{Garcia Lopez}, R., {Natta}, A., {Testi}, L., \& {Habart}, E. 2006, \aap, 459,
  837

\bibitem[{{Garufi} {et~al.}(2017){Garufi}, {Meeus}, {Benisty}, {Quanz},
  {Banzatti}, {Kama}, {Canovas}, {Eiroa}, {Schmid}, {Stolker}, {Pohl},
  {Rigliaco}, {M{\'e}nard}, {Meyer}, {van Boekel}, \& {Dominik}}]{garufi17}
{Garufi}, A., {Meeus}, G., {Benisty}, M., {et~al.} 2017, \aap, 603, A21

\bibitem[{{GRAVITY Collaboration} {et~al.}(2019){GRAVITY Collaboration},
  {Perraut}, {Labadie}, {Lazareff}, {Klarmann}, {Segura-Cox}, {Benisty},
  {Bouvier}, {Brandner}, {Caratti O Garatti}, {Caselli}, {Dougados}, {Garcia},
  {Garcia-Lopez}, {Kendrew}, {Koutoulaki}, {Kervella}, {Lin}, {Pineda},
  {Sanchez-Bermudez}, {van Dishoeck}, {Abuter}, {Amorim}, {Berger}, {Bonnet},
  {Buron}, {Cantalloube}, {Cl{\'e}net}, {Coud{\'e} Du Foresto}, {Dexter}, {de
  Zeeuw}, {Duvert}, {Eckart}, {Eisenhauer}, {Eupen}, {Gao}, {Gendron},
  {Genzel}, {Gillessen}, {Gordo}, {Grellmann}, {Haubois}, {Haussmann},
  {Henning}, {Hippler}, {Horrobin}, {Hubert}, {Jocou}, {Lacour}, {Le Bouquin},
  {L{\'e}na}, {M{\'e}rand}, {Ott}, {Paumard}, {Perrin}, {Pfuhl}, {Rabien},
  {Ray}, {Rau}, {Rousset}, {Scheithauer}, {Straub}, {Straubmeier}, {Sturm},
  {Vincent}, {Waisberg}, {Wank}, {Widmann}, {Wieprecht}, {Wiest}, {Wiezorrek},
  {Woillez}, \& {Yazici}}]{gravity19}
{GRAVITY Collaboration}, {Perraut}, K., {Labadie}, L., {et~al.} 2019, \aap,
  632, A53

\bibitem[{{Gregory} {et~al.}(2012){Gregory}, {Donati}, {Morin}, {Hussain},
  {Mayne}, {Hillenbrand}, \& {Jardine}}]{gregory12}
{Gregory}, S.~G., {Donati}, J.~F., {Morin}, J., {et~al.} 2012, \apj, 755, 97

\bibitem[{{Hartmann} {et~al.}(2016){Hartmann}, {Herczeg}, \&
  {Calvet}}]{hartmann16}
{Hartmann}, L., {Herczeg}, G., \& {Calvet}, N. 2016, \araa, 54, 135

\bibitem[{{Herbig}(1958)}]{herbig58}
{Herbig}, G.~H. 1958, \apj, 128, 259

\bibitem[{{Herbig}(1960)}]{herbig60}
---. 1960, \apjs, 4, 337

\bibitem[{{Hern{\'a}ndez} {et~al.}(2004){Hern{\'a}ndez}, {Calvet},
  {Brice{\~n}o}, {Hartmann}, \& {Berlind}}]{hernandez04}
{Hern{\'a}ndez}, J., {Calvet}, N., {Brice{\~n}o}, C., {Hartmann}, L., \&
  {Berlind}, P. 2004, \aj, 127, 1682

\bibitem[{{Hillenbrand} {et~al.}(1992){Hillenbrand}, {Strom}, {Vrba}, \&
  {Keene}}]{hillenbrand92}
{Hillenbrand}, L.~A., {Strom}, S.~E., {Vrba}, F.~J., \& {Keene}, J. 1992, \apj,
  397, 613

\bibitem[{{Joy}(1949)}]{joy49}
{Joy}, A.~H. 1949, \apj, 110, 424

\bibitem[{{Juh{\'a}sz} {et~al.}(2010){Juh{\'a}sz}, {Bouwman}, {Henning},
  {Acke}, {van den Ancker}, {Meeus}, {Dominik}, {Min}, {Tielens}, \&
  {Waters}}]{juhasz10}
{Juh{\'a}sz}, A., {Bouwman}, J., {Henning}, T., {et~al.} 2010, \apj, 721, 431

\bibitem[{{Kausch} {et~al.}(2015){Kausch}, {Noll}, {Smette}, {Kimeswenger},
  {Barden}, {Szyszka}, {Jones}, {Sana}, {Horst}, \& {Kerber}}]{kausch15}
{Kausch}, W., {Noll}, S., {Smette}, A., {et~al.} 2015, \aap, 576, A78

\bibitem[{{Kenyon} \& {Hartmann}(1995)}]{k&h95}
{Kenyon}, S.~J., \& {Hartmann}, L. 1995, \apjs, 101, 117

\bibitem[{{Khokhlov} {et~al.}(2017){Khokhlov}, {Miroshnichenko}, {Mennickent},
  {Cabezas}, {Zhanabaev}, {Reichart}, {Ivarsen}, {Haislip}, {Nysewander}, \&
  {LaCluyze}}]{khokhlov17}
{Khokhlov}, S.~A., {Miroshnichenko}, A.~S., {Mennickent}, R., {et~al.} 2017,
  \apj, 835, 53

\bibitem[{{Kim} {et~al.}(2016){Kim}, {Watson}, {Manoj}, {Forrest}, {Furlan},
  {Najita}, {Sargent}, {Hern{\'a}ndez}, {Calvet}, {Adame}, {Espaillat},
  {Megeath}, {Muzerolle}, \& {McClure}}]{kim16}
{Kim}, K.~H., {Watson}, D.~M., {Manoj}, P., {et~al.} 2016, \apjs, 226, 8

\bibitem[{{Kraus}(2009)}]{kraus09}
{Kraus}, M. 2009, \aap, 494, 253

\bibitem[{{Lazareff} {et~al.}(2017){Lazareff}, {Berger}, {Kluska}, {Le
  Bouquin}, {Benisty}, {Malbet}, {Koen}, {Pinte}, {Thi}, {Absil}, {Baron},
  {Delboulb{\'e}}, {Duvert}, {Isella}, {Jocou}, {Juhasz}, {Kraus}, {Lachaume},
  {M{\'e}nard}, {Millan-Gabet}, {Monnier}, {Moulin}, {Perraut}, {Rochat},
  {Soulez}, {Tallon}, {Thi{\'e}baut}, {Traub}, \& {Zins}}]{lazareff17}
{Lazareff}, B., {Berger}, J.~P., {Kluska}, J., {et~al.} 2017, \aap, 599, A85

\bibitem[{{Lee} {et~al.}(2016){Lee}, {Chen}, \& {Liu}}]{lee16}
{Lee}, C.-D., {Chen}, W.-P., \& {Liu}, S.-Y. 2016, \aap, 592, A130

\bibitem[{Lee {et~al.}(2017)Lee, Gullikson, \& Kaplan}]{igrinsplp}
Lee, J.-J., Gullikson, K., \& Kaplan, K. 2017, igrins/plp 2.2.0, , ,
  doi:10.5281/zenodo.845059

\bibitem[{{Lynden-Bell} \& {Pringle}(1974)}]{lynden-bell&pringle74}
{Lynden-Bell}, D., \& {Pringle}, J.~E. 1974, \mnras, 168, 603

\bibitem[{{Maaskant} {et~al.}(2013){Maaskant}, {Honda}, {Waters}, {Tielens},
  {Dominik}, {Min}, {Verhoeff}, {Meeus}, \& {van den Ancker}}]{maaskant13}
{Maaskant}, K.~M., {Honda}, M., {Waters}, L.~B.~F.~M., {et~al.} 2013, \aap,
  555, A64

\bibitem[{{Manoj} {et~al.}(2006){Manoj}, {Bhatt}, {Maheswar}, \&
  {Muneer}}]{manoj06}
{Manoj}, P., {Bhatt}, H.~C., {Maheswar}, G., \& {Muneer}, S. 2006, \apj, 653,
  657

\bibitem[{{Maravelias} {et~al.}(2018){Maravelias}, {Kraus}, {Cidale}, {Borges
  Fernandes}, {Arias}, {Cur{\'e}}, \& {Vasilopoulos}}]{maravelias18}
{Maravelias}, G., {Kraus}, M., {Cidale}, L.~S., {et~al.} 2018, \mnras, 480, 320

\bibitem[{{Mathis}(1990)}]{mathis90}
{Mathis}, J.~S. 1990, \araa, 28, 37

\bibitem[{{McClure}(2009)}]{mcclure09}
{McClure}, M. 2009, \apjl, 693, L81

\bibitem[{{Meeus} {et~al.}(2001){Meeus}, {Waters}, {Bouwman}, {van den Ancker},
  {Waelkens}, \& {Malfait}}]{meeus01}
{Meeus}, G., {Waters}, L.~B.~F.~M., {Bouwman}, J., {et~al.} 2001, \aap, 365,
  476

\bibitem[{{Meeus} {et~al.}(2012){Meeus}, {Montesinos}, {Mendigut{\'\i}a},
  {Kamp}, {Thi}, {Eiroa}, {Grady}, {Mathews}, {Sandell}, {Martin-Za{\"\i}di},
  {Brittain}, {Dent}, {Howard}, {M{\'e}nard}, {Pinte}, {Roberge}, {Vand
  enbussche}, \& {Williams}}]{meeus12}
{Meeus}, G., {Montesinos}, B., {Mendigut{\'\i}a}, I., {et~al.} 2012, \aap, 544,
  A78

\bibitem[{{Mehner} {et~al.}(2016){Mehner}, {de Wit}, {Groh}, {Oudmaijer},
  {Baade}, {Rivinius}, {Selman}, {Boffin}, \& {Martayan}}]{mehner16}
{Mehner}, A., {de Wit}, W.~J., {Groh}, J.~H., {et~al.} 2016, \aap, 585, A81

\bibitem[{{Mel'Nikov}(2001)}]{melnikov01}
{Mel'Nikov}, S.~Y. 2001, Astronomy Reports, 45, 686

\bibitem[{{Mendigut{\'\i}a}(2020)}]{mendigutia20}
{Mendigut{\'\i}a}, I. 2020, arXiv e-prints, arXiv:2005.01745

\bibitem[{{Mendigut{\'{\i}}a} {et~al.}(2011){Mendigut{\'{\i}}a}, {Calvet},
  {Montesinos}, {Mora}, {Muzerolle}, {Eiroa}, {Oudmaijer}, \&
  {Mer{\'{\i}}n}}]{mendigutia11b}
{Mendigut{\'{\i}}a}, I., {Calvet}, N., {Montesinos}, B., {et~al.} 2011, \aap,
  535, A99

\bibitem[{{Mendigut{\'{\i}}a} {et~al.}(2012){Mendigut{\'{\i}}a}, {Mora},
  {Montesinos}, {Eiroa}, {Meeus}, {Mer{\'{\i}}n}, \&
  {Oudmaijer}}]{mendigutia12}
{Mendigut{\'{\i}}a}, I., {Mora}, A., {Montesinos}, B., {et~al.} 2012, \aap,
  543, A59

\bibitem[{{Mendigut{\'\i}a} {et~al.}(2015){Mendigut{\'\i}a}, {Oudmaijer},
  {Rigliaco}, {Fairlamb}, {Calvet}, {Muzerolle}, {Cunningham}, \&
  {Lumsden}}]{mendigutia15}
{Mendigut{\'\i}a}, I., {Oudmaijer}, R.~D., {Rigliaco}, E., {et~al.} 2015,
  \mnras, 452, 2837

\bibitem[{{Miroshnichenko} \& {Corporon}(1999)}]{miroshnichenko99}
{Miroshnichenko}, A., \& {Corporon}, P. 1999, \aap, 349, 126

\bibitem[{{Miroshnichenko}(2006)}]{miroshnichenko06}
{Miroshnichenko}, A.~S. 2006, Astronomical Society of the Pacific Conference
  Series, Vol. 355, {Galactic B[e] Stars: A Review of 30 Years of
  Investigation}, ed. M.~{Kraus} \& A.~S. {Miroshnichenko}, 13

\bibitem[{{Miroshnichenko}(2007)}]{miroshnichenko07}
---. 2007, \apj, 667, 497

\bibitem[{{Miroshnichenko} {et~al.}(2001){Miroshnichenko}, {Levato},
  {Bjorkman}, \& {Grosso}}]{miroshnichenko01}
{Miroshnichenko}, A.~S., {Levato}, H., {Bjorkman}, K.~S., \& {Grosso}, M. 2001,
  \aap, 371, 600

\bibitem[{{Monnier} {et~al.}(2006){Monnier}, {Berger}, {Millan-Gabet}, {Traub},
  {Schloerb}, {Pedretti}, {Benisty}, {Carleton}, {Haguenauer}, {Kern},
  {Labeye}, {Lacasse}, {Malbet}, {Perraut}, {Pearlman}, \& {Zhao}}]{monnier06}
{Monnier}, J.~D., {Berger}, J.~P., {Millan-Gabet}, R., {et~al.} 2006, \apj,
  647, 444

\bibitem[{{Muratore} {et~al.}(2015){Muratore}, {Kraus}, {Oksala}, {Arias},
  {Cidale}, {Borges Fernandes}, \& {Liermann}}]{muratore15}
{Muratore}, M.~F., {Kraus}, M., {Oksala}, M.~E., {et~al.} 2015, \aj, 149, 13

\bibitem[{{Muzerolle} {et~al.}(2001){Muzerolle}, {Calvet}, \&
  {Hartmann}}]{muzerolle01}
{Muzerolle}, J., {Calvet}, N., \& {Hartmann}, L. 2001, \apj, 550, 944

\bibitem[{{Muzerolle} {et~al.}(2004){Muzerolle}, {D'Alessio}, {Calvet}, \&
  {Hartmann}}]{muzerolle04}
{Muzerolle}, J., {D'Alessio}, P., {Calvet}, N., \& {Hartmann}, L. 2004, \apj,
  617, 406

\bibitem[{{Muzerolle} {et~al.}(1998){Muzerolle}, {Hartmann}, \&
  {Calvet}}]{muzerolle98}
{Muzerolle}, J., {Hartmann}, L., \& {Calvet}, N. 1998, \aj, 116, 2965

\bibitem[{{Myers} {et~al.}(1987){Myers}, {Fuller}, {Mathieu}, {Beichman},
  {Benson}, {Schild}, \& {Emerson}}]{myers87}
{Myers}, P.~C., {Fuller}, G.~A., {Mathieu}, R.~D., {et~al.} 1987, \apj, 319,
  340

\bibitem[{{Ochsenbein} {et~al.}(2000){Ochsenbein}, {Bauer}, \&
  {Marcout}}]{vizier}
{Ochsenbein}, F., {Bauer}, P., \& {Marcout}, J. 2000, \aaps, 143, 23

\bibitem[{{Park} {et~al.}(2014){Park}, {Jaffe}, {Yuk}, {Chun}, {Pak}, {Kim},
  {Pavel}, {Lee}, {Oh}, {Jeong}, {Sim}, {Lee}, {Nguyen Le}, {Strubhar},
  {Gully-Santiago}, {Oh}, {Cha}, {Moon}, {Park}, {Brooks}, {Ko}, {Han}, {Nah},
  {Hill}, {Lee}, {Barnes}, {Yu}, {Kaplan}, {Mace}, {Kim}, {Lee}, {Hwang}, \&
  {Park}}]{park14}
{Park}, C., {Jaffe}, D.~T., {Yuk}, I.-S., {et~al.} 2014, in \procspie, Vol.
  9147, Ground-based and Airborne Instrumentation for Astronomy V, 91471D

\bibitem[{{Pecaut} \& {Mamajek}(2013)}]{pecaut-mamajek13}
{Pecaut}, M.~J., \& {Mamajek}, E.~E. 2013, \apjs, 208, 9

\bibitem[{Pedregosa {et~al.}(2011)Pedregosa, Varoquaux, Gramfort, Michel,
  Thirion, Grisel, Blondel, Prettenhofer, Weiss, Dubourg, Vanderplas, Passos,
  Cournapeau, Brucher, Perrot, \& Duchesnay}]{scikit-learn}
Pedregosa, F., Varoquaux, G., Gramfort, A., {et~al.} 2011, Journal of Machine
  Learning Research, 12, 2825

\bibitem[{{Pogodin} {et~al.}(2006){Pogodin}, {Malanushenko}, {Kozlova},
  {Tarasova}, \& {Franco}}]{pogodin06}
{Pogodin}, M.~A., {Malanushenko}, V.~P., {Kozlova}, O.~V., {Tarasova}, T.~N.,
  \& {Franco}, G.~A.~P. 2006, \aap, 452, 551

\bibitem[{{Reiter} {et~al.}(2018){Reiter}, {Calvet}, {Thanathibodee}, {Kraus},
  {Cauley}, {Monnier}, {Rubinstein}, {Aarnio}, \& {Harries}}]{reiter18}
{Reiter}, M., {Calvet}, N., {Thanathibodee}, T., {et~al.} 2018, \apj, 852, 5

\bibitem[{{Robinson} \& {Espaillat}(2019)}]{robinson19}
{Robinson}, C.~E., \& {Espaillat}, C.~C. 2019, \apj, 874, 129

\bibitem[{{Sartori} {et~al.}(2010){Sartori}, {Gregorio-Hetem}, {Rodrigues},
  {Hetem}, \& {Batalha}}]{sartori10}
{Sartori}, M.~J., {Gregorio-Hetem}, J., {Rodrigues}, C.~V., {Hetem}, Annibal,
  J., \& {Batalha}, C. 2010, \aj, 139, 27

\bibitem[{{Smette} {et~al.}(2015){Smette}, {Sana}, {Noll}, {Horst}, {Kausch},
  {Kimeswenger}, {Barden}, {Szyszka}, {Jones}, {Gallenne}, {Vinther},
  {Ballester}, \& {Taylor}}]{smette15}
{Smette}, A., {Sana}, H., {Noll}, S., {et~al.} 2015, \aap, 576, A77

\bibitem[{{Stephenson} \& {Sanduleak}(1977)}]{stephenson-sanduleak77}
{Stephenson}, C.~B., \& {Sanduleak}, N. 1977, Publications of the Warner \&
  Swasey Observatory, 2, 4

\bibitem[{{Strai{\v{z}}ys} {et~al.}(1999){Strai{\v{z}}ys}, {Corbally}, \&
  {Laugalys}}]{straizys99}
{Strai{\v{z}}ys}, V., {Corbally}, C.~J., \& {Laugalys}, V. 1999, Baltic
  Astronomy, 8, 355

\bibitem[{{Strai{\v{z}}ys} {et~al.}(2008){Strai{\v{z}}ys}, {Corbally}, \&
  {Laugalys}}]{straizys08}
---. 2008, Baltic Astronomy, 17, 125

\bibitem[{{Su{\'a}rez} {et~al.}(2006){Su{\'a}rez}, {Garc{\'\i}a-Lario},
  {Manchado}, {Manteiga}, {Ulla}, \& {Pottasch}}]{suarez06}
{Su{\'a}rez}, O., {Garc{\'\i}a-Lario}, P., {Manchado}, A., {et~al.} 2006, \aap,
  458, 173

\bibitem[{{Vacca} {et~al.}(2003){Vacca}, {Cushing}, \& {Rayner}}]{vacca03}
{Vacca}, W.~D., {Cushing}, M.~C., \& {Rayner}, J.~T. 2003, \pasp, 115, 389

\bibitem[{{Valenti} {et~al.}(2003){Valenti}, {Fallon}, \&
  {Johns-Krull}}]{valenti03}
{Valenti}, J.~A., {Fallon}, A.~A., \& {Johns-Krull}, C.~M. 2003, \apjs, 147,
  305

\bibitem[{{Valenti} {et~al.}(2000){Valenti}, {Johns-Krull}, \&
  {Linsky}}]{valenti00}
{Valenti}, J.~A., {Johns-Krull}, C.~M., \& {Linsky}, J.~L. 2000, \apjs, 129,
  399

\bibitem[{{van Boekel} {et~al.}(2005){van Boekel}, {Min}, {Waters}, {de Koter},
  {Dominik}, {van den Ancker}, \& {Bouwman}}]{vanboekel05}
{van Boekel}, R., {Min}, M., {Waters}, L.~B.~F.~M., {et~al.} 2005, \aap, 437,
  189

\bibitem[{{van den Ancker} {et~al.}(1998){van den Ancker}, {de Winter}, \&
  {Tjin A Djie}}]{vandenancker98}
{van den Ancker}, M.~E., {de Winter}, D., \& {Tjin A Djie}, H.~R.~E. 1998,
  \aap, 330, 145

\bibitem[{{Verhoeff} {et~al.}(2012){Verhoeff}, {Waters}, {van den Ancker},
  {Min}, {Stap}, {Pantin}, {van Boekel}, {Acke}, {Tielens}, \& {de
  Koter}}]{verhoeff12}
{Verhoeff}, A.~P., {Waters}, L.~B.~F.~M., {van den Ancker}, M.~E., {et~al.}
  2012, \aap, 538, A101

\bibitem[{{Vieira} {et~al.}(2003){Vieira}, {Corradi}, {Alencar}, {Mendes},
  {Torres}, {Quast}, {Guimar{\~a}es}, \& {da Silva}}]{vieira03}
{Vieira}, S.~L.~A., {Corradi}, W.~J.~B., {Alencar}, S.~H.~P., {et~al.} 2003,
  \aj, 126, 2971

\bibitem[{{Villebrun} {et~al.}(2019){Villebrun}, {Alecian}, {Hussain},
  {Bouvier}, {Folsom}, {Lebreton}, {Amard}, {Charbonnel}, {Gallet},
  {Haemmerl{\'e}}, {B{\"o}hm}, {Johns-Krull}, {Kochukhov}, {Marsden}, {Morin},
  \& {Petit}}]{villebrun19}
{Villebrun}, F., {Alecian}, E., {Hussain}, G., {et~al.} 2019, \aap, 622, A72

\bibitem[{{Vink} {et~al.}(2002){Vink}, {Drew}, {Harries}, \&
  {Oudmaijer}}]{vink02}
{Vink}, J.~S., {Drew}, J.~E., {Harries}, T.~J., \& {Oudmaijer}, R.~D. 2002,
  \mnras, 337, 356

\bibitem[{{Vink} {et~al.}(2005){Vink}, {Drew}, {Harries}, {Oudmaijer}, \&
  {Unruh}}]{vink05}
{Vink}, J.~S., {Drew}, J.~E., {Harries}, T.~J., {Oudmaijer}, R.~D., \& {Unruh},
  Y. 2005, \mnras, 359, 1049

\bibitem[{Vioque {et~al.}(2018)Vioque, D.~Oudmaijer, Baines, Mendigutía, \&
  Pérez-Martínez}]{vioque18}
Vioque, M., D.~Oudmaijer, R., Baines, D., Mendigutía, I., \& Pérez-Martínez,
  R. 2018, Astronomy \& Astrophysics, 620, doi:10.1051/0004-6361/201832870

\bibitem[{{Vioque} {et~al.}(2020){Vioque}, {Oudmaijer}, {Schreiner},
  {Mendigut{\'\i}a}, {Baines}, {Mowlavi}, \&
  {P{\'e}rez-Mart{\'\i}nez}}]{vioque20}
{Vioque}, M., {Oudmaijer}, R.~D., {Schreiner}, M., {et~al.} 2020, arXiv
  e-prints, arXiv:2005.01727

\bibitem[{{Virtanen} {et~al.}(2020){Virtanen}, {Gommers}, {Oliphant},
  {Haberland}, {Reddy}, {Cournapeau}, {Burovski}, {Peterson}, {Weckesser},
  {Bright}, {van der Walt}, {Brett}, {Wilson}, {Jarrod Millman}, {Mayorov},
  {Nelson}, {Jones}, {Kern}, {Larson}, {Carey}, {Polat}, {Feng}, {Moore}, {Vand
  erPlas}, {Laxalde}, {Perktold}, {Cimrman}, {Henriksen}, {Quintero}, {Harris},
  {Archibald}, {Ribeiro}, {Pedregosa}, {van Mulbregt}, \&
  {Contributors}}]{scipy}
{Virtanen}, P., {Gommers}, R., {Oliphant}, T.~E., {et~al.} 2020, Nature
  Methods, 17, 261

\bibitem[{{Welin}(1973)}]{welin73}
{Welin}, G. 1973, \aaps, 9, 183

\bibitem[{{Wheelwright} {et~al.}(2013){Wheelwright}, {Weigelt}, {Caratti o
  Garatti}, \& {Garcia Lopez}}]{wheelwright13}
{Wheelwright}, H.~E., {Weigelt}, G., {Caratti o Garatti}, A., \& {Garcia
  Lopez}, R. 2013, \aap, 558, A116

\bibitem[{{Wichittanakom} {et~al.}(2020){Wichittanakom}, {Oudmaijer},
  {Fairlamb}, {Mendigut{\'\i}a}, {Vioque}, \& {Ababakr}}]{wichittanakom20}
{Wichittanakom}, C., {Oudmaijer}, R.~D., {Fairlamb}, J.~R., {et~al.} 2020,
  \mnras, 493, 234

\bibitem[{{Zickgraf}(2003)}]{zickgraf03}
{Zickgraf}, F.~J. 2003, \aap, 408, 257

\bibitem[{{Zinnecker} \& {Yorke}(2007)}]{zinnecker_yorke07}
{Zinnecker}, H., \& {Yorke}, H.~W. 2007, \araa, 45, 481

\end{thebibliography}

\begin{appendix}\label{appendix}
For intermediate- and high-mass objects, it can be difficult to determine their evolutionary status. For instance, HBes share observational features with classical Be objects as well as post-main-sequence B[e] supergiants. For many objects, there is a debate about their evolutionary state. We list some known cases here for completeness, although we include them in the IGRINS analysis above.  

\begin{itemize}
    \item MWC 137 -- This object is included in the HAeBe samples of \citet{verhoeff12,cauley14,cauley15,ababakr17,vioque18,arun19}. However, it is included in the B[e] supergiants list of \citet{maravelias18} and the candidate B[e] supergiant list of \citet{kraus09}. \citet{mehner16} classifies this object as a post-main-sequence B[e] star based on optical and near-infrared color-magnitude and color-color diagrams. The authors detail observations of a jet coming from this system. \citet{muratore15} also determine that this is an evolved B[e] object. \citet{fuente03} find an upper limit on the disk mass of 0.007 solar masses based on data from the IRAM interferometer at the Plateau de Bure and the NRAO Very Large Array.
\end{itemize}

\begin{itemize}
    \item BD+41 3731 -- This object is considered a HAeBe by  \citet{alecian13} and \citet{reiter18}, but \citet{cauley14} suggest that it is not a pre-main-sequence object based on the featureless He I $\lambda$10830 spectrum and lack of H$\alpha$ emission.
\end{itemize}

\begin{itemize}
    \item HD 52721/GU CMa -- This object is largely considered a HAeBe including by \citet{alecian13,fairlamb15,reiter18,vioque20}. However, we note that \citet{cauley14} suggests that this may be a classical Be star based on its He I $\lambda$10830 spectrum and lack of H$\alpha$ emission paired with a lack of IR emission found by \citet{hillenbrand92}. 
\end{itemize}

\begin{itemize}
    \item HD 53367 -- This objects is considered a HAeBe by \citet{fairlamb15,reiter18,vioque18,vioque20}. \citet{cauley14} notes that \citet{hillenbrand92} found no IR excess for this object and \citet{pogodin06} find evidence that this is a binary system consisting of a main-sequence B0e star and a secondary that is most likely a 5 solar mass pre-main-sequence object, with circumstellar material near the companion. \citet{cauley14} finds that the H$\alpha$ is well-fit by a two-component Gaussian with different velocities, perhaps due to the binary nature of the object. 
\end{itemize}

\begin{itemize}
    \item LkHa 215 -- This object is included as a HAeBe in \citet{hernandez04,vink05,manoj06,alecian13,reiter18}. \citet{alonso-albi09} did not detect millimeter or centimeter flux for this object. They fit the spectral energy distribution to find a low disk mass with a surrounding envelope. \citet{acke-anker06} and \citet{hillenbrand92} classify this object as being a Group I object due to the IR excess. \citet{cauley14} find that the He I $\lambda$10830 spectrum is featureless and thus suggest this object may be a classical Be star. 
\end{itemize}

\begin{itemize}
    \item HD 85567 -- This object is included as a HAeBe in \citet{acke-anker06,acke09,fairlamb15,ababakr16,lazareff17,gravity19}. \citet{wheelwright13} analyze NIR interferometry of HD 85567 and conclude that it is a YSO and not a classical Be star. However, there has been some evidence that it is a FS CMa system \citet{miroshnichenko07,lee16,khokhlov17}. \citet{khokhlov17} and \citet{miroshnichenko01} conclude that this object cannot be an HAeBe object due to its location on the HR diagram, lack of far-IR excess, and no close-by star-forming regions. \citet{khokhlov17} suggest that this is a binary system with an undetected secondary component. \citet{baines06} consider HD 85567 to be a HAeBe but they do find evidence that it is in a binary system. 
\end{itemize}

\begin{itemize}
    \item Hen 3-847 -- This object is included in the HAeBe samples of \citet{valenti03,verhoeff12,fairlamb15,ababakr16,ababakr17,vioque18,arun19}. \citet{vioque20} rejects this object as being in the group of FS CMa objects. Hen 3-847 is given as an isolated B[e] star in \citet{vandenancker98}. \citet{miroshnichenko06} considers this object to be a B[e] with warm dust (B[e]WD).
\end{itemize}

\begin{itemize}
    \item MWC 342 -- This object is included in the HAeBe samples of \citet{eisner15,ababakr17,vioque18}. \citet{vioque20} rejects this object as being in the group of FS CMa objects. \citet{aret16} lists this object as a B[e] supergiant candidate. \citet{miroshnichenko07} considers this object a B[e] star in the FS CMa group and \citet{zickgraf03} lists it as an ``unclassified'' B[e]. \citet{miroshnichenko99} consider it a B[e] star with unknown evolutionary state. It is considered a YSO by \citet{monnier06}.
\end{itemize}

\begin{itemize}
    \item IRAS 06071+2925 -- This object is included in the HAeBe samples of \citet{cauley15,vioque18}. \citet{vioque20} rejects this object as being in the group of FS CMa objects and it is included in the sample of FS CMa systems in \citet{miroshnichenko07}. \citet{vieira03} presents data for this object, however its evolutionary status is unclear and it is listed as not having a clear association with a star-forming region. It is also presented in the HAeBe candidates list of \citet{sartori10}.
\end{itemize}

\end{appendix}

\end{document}